\documentclass[apj,numberedappendix]{emulateapj}

\usepackage{amsmath}
\usepackage{color}
\usepackage{hyperref}

\usepackage{color}

\usepackage[]{units}
\usepackage{hyphenat}

\slugcomment{To be submitted to Astrophysical Journal}

\usepackage{natbib}
\begin{document}

\title{The extragalactic background light, the Hubble constant, and anomalies: conclusions from 20 years of TeV gamma-ray observations}
\author{J.~Biteau$^1$ \& D.~A.~Williams$^2$}
\shorttitle{EBL, Hubble constant, and anomalies}
\shortauthors{Biteau \& Williams}

\affil{Santa Cruz Institute for Particle Physics and Department of Physics,\\ University of California at Santa Cruz, Santa Cruz, CA 95064, USA}

\email{$^1$ jbiteau@ucsc.edu}
\email{$^2$ daw@ucsc.edu}

\begin{abstract}
Ground-based observatories have been collecting $\unit[0.2-20]{TeV}$ gamma rays from blazars for about twenty years. These gamma rays can experience absorption along the line of sight due to interactions with the extragalactic background light (EBL). In this paper, we show that the gamma-ray optical depth can be reduced to the convolution product of an EBL kernel with the EBL intensity, assuming a particular form for the EBL evolution. We extract the absorption signal from the most extensive set of TeV spectra from blazars collected so far and unveil a broad-band EBL spectrum from mid-ultraviolet to far infrared. This spectrum is in good agreement with the accumulated emission of galaxies, constraining unresolved populations of sources. We propose a data-driven estimate of the Hubble constant based on the comparison of local and gamma-ray measurements of the EBL. After setting stringent upper-limits on the redshift of four TeV blazars, we investigate the 106 gamma-ray spectra in our sample and find no significant evidence for anomalies. The intrinsic TeV spectra are not harder than their GeV counterpart, and no spectral upturn is visible at the highest optical depths. Finally, we investigate a modification of the pair-creation threshold due to Lorentz invariance violation. A mild excess prevents us from ruling out an effect at the Planck energy, and we constrain for the first time the energy scale of the modification to values larger than sixty percent of the Planck energy.
\end{abstract}


\section{Introduction}
\label{Sec:Intro}

The universe is not as dark as we sometimes imagine; even its largest voids are filled with light. The most intense of these photon fields, the cosmic microwave background (CMB), covers the millimeter wavelength range and carries the relic radiation that escaped the epoch of recombination, less than half a million years after the Big Bang. At lower wavelength, from $0.1$ to $\unit[1000]{\mu m}$, the universe is populated by the light that stars and galaxies have emitted since the epoch of reionization ($z\lesssim10$). Part of the light initially radiated in the ultraviolet (UV) and optical (O) bands is directly observable in the cosmic optical background (COB, $\unit[0.1-8]{\mu m}$). The rest was absorbed by dust in the interstellar medium and around active galactic nuclei (AGN) and was subsequently reradiated at lower energies, in the infrared (IR), forming the cosmic infrared background (CIB, $\unit[8-1000]{\mu m}$). The sum of the COB and CIB, the extragalactic background light (EBL), thus carries the $13$ billion years' radiation history of the universe and is a critical observable for models of reionization, galaxy formation and evolution, as well as high-energy-astrophysics phenomena, as we discuss.

The main constraints on the EBL from observations in the UV-O-IR come in two flavors: direct observations, which tend to be contaminated by bright foregrounds such as the zodiacal light, and estimates from integrated galaxy counts, which sum the light emitted by known populations of sources \citep[e.g.,][]{2000MNRAS.312L...9M}. The latter do not include contributions from truly diffuse components or unobserved populations of sources, such as primordial stars (Pop. III) and miniquasars that could have initiated the reionization of the universe \cite[e.g.][]{2004ApJ...604..484M,	2004MNRAS.351L..71C}, or intra-halo light that was recently invoked to explain the near-IR anisotropies observed by CIBER \citep{2014Sci...346..732Z}.

Stringent constraints also come from observations of gamma rays, which are more energetic than the EBL photons by twelve orders of magnitude. The underlying process, described by \cite{REF::NIKISHOV::JETP1962} and \cite{1967PhRv..155.1404G,1967PhRv..155.1408G}, is the creation of electron-positron pairs in the interaction of gamma rays from extragalactic sources with the EBL photon field. The survival probability of a gamma ray, or gamma-ray absorption, is characterized by an exponential attenuation law, $\exp(-\tau)$, where the optical depth, $\tau$, depends on the redshift of the source and on the gamma-ray energy. 

The detection of the first distant gamma-ray sources led to the first observational constraints on gamma-ray absorption \citep{1992ApJ...390L..49S}. Extragalactic sources observed in gamma rays, e.g. by {\it Fermi}-LAT in the high-energy band (HE, $\unit[0.1-300]{GeV}$) or by e.g. H.E.S.S., MAGIC and VERITAS in the very-high-energy band (VHE, $\unit[0.1-30]{TeV}$) are mostly AGN belonging to the class of blazars. The non-thermal relativistic jets emitted by blazars are pointed along the line-of-sight, resulting in an enhancement of the gamma-ray flux and energy as observed on Earth. Subclasses of blazars include flat-spectrum radio quasars (FSRQs), low-frequency-peaked BL Lac objects (LBLs), intermediate-frequency-peaked BL Lac objects (IBLs), and high-frequency-peaked BL Lac objects (HBLs). The energy of the peak gamma-ray emission increases from \nohyphens{FSRQs} to HBLs and the UV-O-IR photon fields encountered by gamma rays in the jets appears to be more and more suppressed along this sequence \citep{1998MNRAS.301..451G}, making BL~Lacs ideal cosmological probes of the EBL \citep[the case of FSRQs remains debated, see e.g.][]{2007ApJ...665.1023R}.

The main difficulty in constraining the EBL from gamma-ray observations lies in the uncertainties on the spectrum emitted at the source as would be observed on Earth without absorption,  the intrinsic spectrum. If some curvature is present in the observed spectrum, one could wonder whether it is related to the emission processes occurring in the blazar's jet or to absorption by the EBL during the propagation of the gamma rays to the Earth. Upper-limits on the EBL have been obtained assuming no intrinsic curvature, either in the VHE or HE-VHE spectra of the sources, as e.g. in \cite{2012A&A...542A..59M} or \cite{2010ApJ...714L.157G}, and constraints have been placed with similar hypotheses by \cite{2011ApJ...733...77O} and \cite{2014ApJ...795...91S}. The intrinsic HE-VHE curvature is accounted for on a statistical basis in  \cite{2013A&A...554A..75S}, through a broad-band modeling of the source emission from X rays to VHE in \cite{2013ApJ...770...77D} following a method suggested by \cite{2010ApJ...715L..16M}, by extrapolating the unabsorbed part of the HE spectrum in \citet[{\it Fermi}-LAT Collaboration,][]{2012Sci...338.1190A}, and by leaving the VHE curvature as a free parameter in \cite{2013A&A...550A...4H}. In particular, the {\it Fermi}-LAT and H.E.S.S. collaborations detected the imprint of the absorption by the EBL at the 6 standard deviation ($\sigma$) level and at the $\unit[9]{\sigma}$ level, respectively. Their approach is based on the scaling of existing EBL models via a normalization factor that, if significantly different from zero, indicates the best level of absorption by the EBL compatible with the gamma-ray data. The present work aims to overcome this model-dependent approach and to provide a broad-band EBL spectrum.

We first show in Sec.~\ref{Sec:Theory} that the optical depth calculated from the EBL spectrum can largely be simplified. One of the three integrals in the relation is fully reducible in an analytical way. We further show that the EBL spectrum can be deconvoluted from the pair-creation smearing effect, under the assumption of a simplified evolution of the EBL (which we validate in Appendix~\ref{App:Sys}). Section~\ref{Sec:DataAn} presents the dataset studied. The results are discussed in Sec.~\ref{Sec:Results}, with Sec.~\ref{Sec:EBLSpec} focusing on the EBL spectrum and the room left for reionization sources or truly diffuse components. We propose a model-independent method to constrain the Hubble constant in Sec.~\ref{Sec:Hubble}. We investigate the redshifts of under-constrained sources in Sec.~\ref{Sec:Redshift}. And finally, we search for anomalies that could originate from axion-like particles and from Lorentz invariance violation in Sec.~\ref{Sec:ALP} and \ref{Sec:LIV}. More details on the systematic uncertainties, best-fit parameters, and gamma-ray residuals are provided in appendices.

\section{EBL optical depth}
\label{Sec:Theory}

The EBL optical depth to gamma rays of energy $E_0$ (in the lab frame) emitted by a source at redshift $z_0$ is given by:
\begin{align}
\tau(E_0,z_0) =& \int_0^{z_0}{\rm d}z\ \frac{\partial L}{\partial z}(z)   \int_0^\infty {\rm d}\epsilon \frac{\partial n}{\partial \epsilon}(\epsilon,z) \nonumber \\
&\int_{-1}^1 {\rm d}\mu\ \frac{1-\mu}{2}\sigma_{\gamma\gamma}[ \beta(E_0,z,\epsilon,\mu) ]
\label{Eq:Opt1}
\end{align}
where $\partial n / \partial \epsilon$ is the density of EBL photons per infinitesimal energy band at energy $\epsilon$ and redshift $z$. The other terms are defined in the following. 

Assuming a flat $\Lambda$CDM cosmology, with Hubble constant $H_0$, matter density $\Omega_M$ and dark energy density $\Omega_\Lambda$, the distance element is:
\begin{align}
\frac{\partial L}{\partial z} =& \frac{c}{H_0}\ \frac{1}{1 + z} \frac{1}{\sqrt{\Omega_\Lambda + \Omega_M(1+z)^3}} \equiv \frac{c}{H_0}\ \frac{\partial l}{\partial z}
\label{Eq:Dist}
\end{align}

The integration over the energy of the EBL photons in Eq.~\ref{Eq:Opt1} is performed above the pair-creation threshold energy \citep[see e.g. Eq.~8 in][]{1967PhRv..155.1404G} and this information is encoded in the Bethe-Heitler cross section, with $\beta\in[0,1]$:
\begin{equation}
\sigma_{\gamma\gamma}(\beta) = \frac{3\sigma_T}{16} (1-\beta^2) \left[-4\beta+ 2\beta^3 + (3-\beta^4)\ln\frac{1+\beta}{1-\beta} \right]
\label{Eq:CrossSec}
\end{equation}
with $\sigma_T$ the Thomson cross section and where
\begin{equation}
\beta^2 = 1- \frac{2 m_e^2 c^4}{E_0 \epsilon}\frac{1}{1+z}\frac{1}{1-\mu}
\label{Eq:Beta}
\end{equation}
with $m_e$ mass of the electron and where $\mu$ is the cosine of the angle between the gamma ray and the EBL photons.

Calling $\epsilon_0 =\epsilon/(1+z)$ the energy of the EBL photons as measured today, $\beta^2 \geq 0$ is equivalent to:
\begin{equation}
(1+z)^2\frac{E_0 \epsilon_0}{m_e^2 c^4} \geq 1
\label{Eq:Threshold}
\end{equation}
which is the pair-creation threshold condition.

The definition of the EBL optical depth in Eq.~\ref{Eq:Opt1} requires an integration over the distance, the energy of the EBL photons, and the angle between the EBL photons and the gamma ray. We show here that one integration can be reduced in a fully analytical way without any loss of generality. Using the change of variables\footnote{This differs from the change of variable of \cite{REF::NIKISHOV::JETP1962} and \cite{1967PhRv..155.1404G} who use $\mu \rightarrow s = (1-\beta^2)^{-1}$.} $\mu \rightarrow \beta$, the optical depth indeed reads:
\begin{align}
\tau(E_0,z_0) &= \frac{3}{4} \frac{\sigma_T c}{H_0}  \int_0^{z_0} {\rm d}z\ \frac{\partial l}{\partial z}(z) \int_0^\infty {\rm d}\epsilon\frac{\partial n}{\partial \epsilon}(\epsilon,z) \nonumber \\ &\ \frac{1}{(1+z)^2} \left(\frac{m_e^2 c^4}{E_0 \epsilon}\right)^2 P(\beta_{\rm max})
\end{align}
with $\displaystyle \beta_{\rm max}^2 =  1- \frac{m_e^2 c^4}{E_0 \epsilon}\frac{1}{1+z}$, and where the particle-physics kernel $P$ admits, after integration over $\beta$, the following analytical expression:
\begin{align}
P(x) =& \ln^2{2} - \frac{\pi^2}{6} + 2\ {\rm Li}_2\left(\frac{1-x}{2} \right) - \frac{x+x^3}{1-x^2}   \nonumber \\
+& \left(\ln(1+x)-2\ln2\right)\ln(1-x) \nonumber \\ 
+& \frac{1}{2}\left(\ln^2(1-x)-\ln^2(1+x)\right) \nonumber \\ 
+& \frac{1+x^4}{2(1-x^2)}\ln\frac{1+x}{1-x} 
\end{align}
where ${\rm Li}_2(x)$ is the dilogarithm.

This important result already eases the computation of the optical depth for a given cosmology and EBL evolution. Going further requires an approximation: as in \cite{1996ApJ...456..124M}, we assume that the evolution and spectrum of the EBL can be locally decoupled, i.e.
\begin{equation}
{\rm d}\epsilon \frac{\partial n}{\partial \epsilon}(\epsilon,z) = {\rm d}\epsilon_0 \frac{\partial n}{\partial \epsilon_0}(\epsilon_0,0) \times evol(z) 
\label{Eq:EBLspec}
\end{equation}
where $\displaystyle evol(z)$ parametrizes the EBL history. We show in Appendix~\ref{App:EBLevol} that this approximation mildly impacts the optical depth up to redshifts\footnote{This parametrization remains acceptable up to $z\sim0.8$ for optical depths smaller than $2-3$.} of $z\sim0.6$.

Changing the variable $\displaystyle \epsilon_0 \rightarrow e_0 = \ln(E_0 \epsilon_0 / m_e^2 c^4)$, the EBL optical depth can be written as the convolution product:
\begin{align}
\tau(E_0,z_0) &= \frac{3\pi\sigma_T}{H_0}\times \frac{E_0}{m_e^2 c^4}\times\nonumber \\ &\quad  \int_{-\infty}^\infty {\rm d}e_0\ \nu I_\nu\left(e_0 - \ln\frac{E_0}{m_e c^2}\right) \times K_{z_0}(e_0)     \nonumber \\
&= \frac{3\pi\sigma_T}{H_0} \times \frac{E_0}{m_e^2 c^4}\times \nu I_\nu \otimes K_{z_0}\left(\ln\frac{E_0}{m_e c^2}\right)
\label{Eq:Opt3}
\end{align}
where $\nu I_\nu = c/4\pi \times \epsilon_0^2 \partial n/\partial \epsilon_0$ is the EBL specific intensity at $z=0$ as a function of $\ln(\epsilon_0/m_e c^2) = \ln(h\nu/m_e c^2)$, and where the EBL kernel is:
\begin{align}
K_{z_0}(e_0) = \exp(-3 e_0)  &\int_0^{z_0} {\rm d}z\ \frac{\partial l}{\partial z}(z)\times\frac{evol(z)}{(1+z)^4} \nonumber\\ & \times P\left(\sqrt{1-\frac{\exp(-e_0)}{(1+z)^2}}\right) 
\label{Eq:PPKernel2}
\end{align}

Following \cite{2008IJMPD..17.1515R}, we parametrize the evolution of the EBL as $evol(z) = (1+z)^{3-f_{\rm evol}}$, where $f_{\rm evol}$ quantifies the impact of the sources. For example, $f_{\rm evol} = 0$ would correspond to a simple cosmological expansion of the photon field with no source term. Assuming a flat $\Lambda$CDM cosmology with $\Omega_\Lambda = 0.7$, $\Omega_M = 0.3$, $H_0 = \unit[70]{km\ s^{-1}\ Mpc^{-1}}$, we show in Appendix~\ref{App:EBLevol} that $f_{\rm evol} = 1.7$ best reproduces the evolutions of the EBL models of \cite{FR08} and \cite{G12}. These two state-of-the-art models from independent groups have been chosen to calibrate our approach, and we note that using a different model \citep[e.g.][closer to $f_{\rm evol} = 1.2$]{D11} has a minimal impact on our results, well below the uncertainties quoted in the following.

\begin{figure}
\hspace{-0.2cm}
\includegraphics[width=0.535\textwidth]{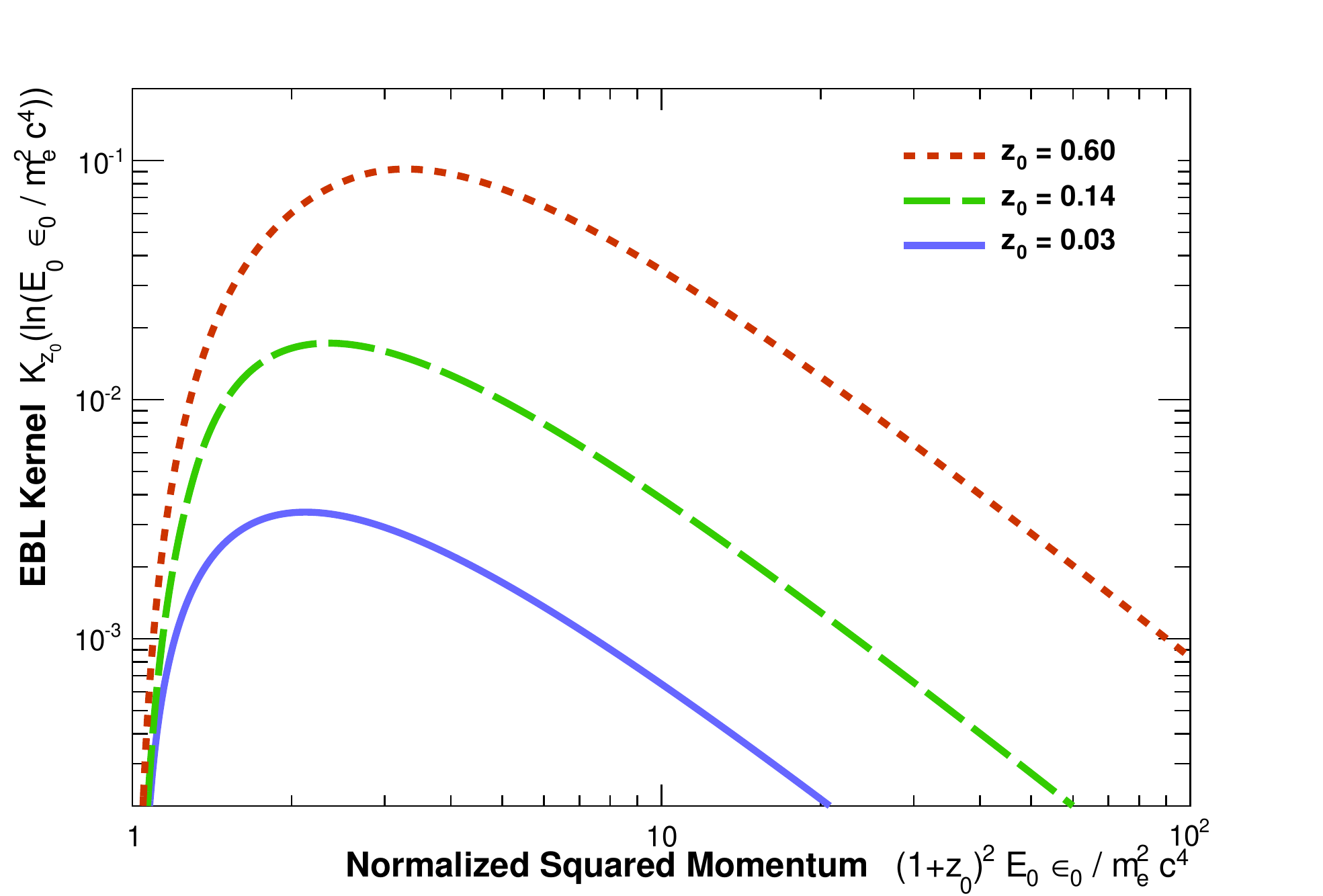}
\caption{EBL kernel, which yields the gamma-ray optical depth after convolution with the EBL intensity, as a function of the product of gamma-ray and EBL-photon energies in electron-mass units, in the lab frame.}
\label{fig:Kernel}
\end{figure}
The EBL kernel from Eq.~\ref{Eq:PPKernel2} is shown in Fig.~\ref{fig:Kernel} as a function of the squared momentum of the interacting photons (EBL and gamma ray), normalized to twice the squared electron mass. Below one, the pair creation threshold condition in Eq.~\ref{Eq:Threshold} is not satisfied, resulting in a null kernel. The kernel peaks between 2 and 4 times the threshold with a peak position and amplitude increasing with redshift. Note that for a fixed gamma-ray energy, the full width at half maximum of the EBL kernel almost spans an order of magnitude in EBL-photon energy (or equivalently in wavelength), whatever the redshift. This broadness argues against simple delta-function approximations to reconstruct the EBL spectrum, such as assumed in \cite{2014ApJ...795...91S}. This also implies that any spectral reconstruction of the EBL based on gamma-ray observations yields correlated flux estimates across the probed wavelength range, though these correlations can be fully taken into account in further analyses, as demonstrated e.g. in Sec.~\ref{Sec:Hubble}.

Having shown that the EBL optical depth can be written as the convolution product in Eq.~\ref{Eq:Opt3}, we measure in the following the broad-band EBL spectrum based on local constraints and on gamma-ray absorption observed in gamma-ray spectra of blazars.

\section{Dataset and analysis}
\label{Sec:DataAn}

\subsection{Local constraints}
\label{Sec:EBLData}

Direct constraints on the intensity of the EBL from UV-O-IR observations come in two flavors: lower limits derived from galaxy counts, where faint emitters or truly diffuse components can be missed; and upper limits derived from direct measurements, which are contaminated by bright foregrounds (zodiacal and Galactic light), at least below $\unit[100]{\mu m}$. In the following, we exploit the extensive bibliographic study of \cite{2013APh....43..112D}. We select state-of-the-art constraints from independent datasets, resulting in 27 upper limits and 25 lower limits (called local constraints in the following). For the latter, we select the estimates corrected for the completeness of the sample. These data are reproduced in Table~\ref{Tab:EBLdata}, with uncertainties combining statistical and systematic uncertainties. In addition to constraints listed by \cite{2013APh....43..112D}, we also include the lower limits at $\unit[3.6]{\mu m}$ and $\unit[4.5]{\mu m}$ from \cite{2013ApJ...769...80A}.

\begin{deluxetable*}{rcccr}
\tablecolumns{5} 
\tabletypesize{\scriptsize}
\tablecaption{Local constraints on the EBL spectrum used in this paper, largely extracted from \cite{2013APh....43..112D}.\label{Tab:EBLdata}}
\tablehead{\colhead{$\lambda$} & \colhead{Lower limit} & \colhead{Upper limit} & \colhead{Experiment} & \colhead{ \qquad  \qquad  \qquad  \qquad Reference  \qquad  \qquad  \qquad  \qquad}\\
\colhead{$\mu$m} & \colhead{\ nW m$^{-2}$ sr$^{-1}$} & \colhead{\ nW m$^{-2}$ sr$^{-1}$} & & }
\startdata
$	0.153	$ & $	1.03\pm0.15	$ & \nodata &	Galex	&\cite{2005ApJ...619L..11X}\\
$	0.1595	$ & $	3.75\pm1.25	$ & \nodata &	HST/WFPC2	&\cite{2000ApJ...542L..79G}\\
$	0.231	$ & $	2.25\pm0.32	$ & \nodata &	Galex	&\cite{2005ApJ...619L..11X}\\
$	0.2365	$ & $	3.6\pm0.5	$ & \nodata &	HST/WFPC2	&\cite{2000ApJ...542L..79G}\\
$	0.3	$ & \nodata & $	18\pm12	$ &	HST/WFPC2	&\cite{2007ApJ...666..663B}\\
$	0.3	$ & $	3.7\pm0.7	$ & \nodata &	HST+ground	&\cite{2001ApJ...559..592T}\\
$	0.4	$ & \nodata & $	26\pm10	$ &	dark cloud	&\cite{1990IAUS..139..257M}\\
$	0.44	$ & \nodata & $	7.9\pm4.0	$ &	Pioneer 10/11	&\cite{2011ApJ...736..119M}\\
$	0.45	$ & $	6.1\pm1.8	$ & \nodata &	HST+ground	&\cite{2001ApJ...559..592T}\\
$	0.5115	$ & \nodata & $	30\pm9	$ &	ground	&\cite{1979ApJ...232..333D}\\
$	0.55	$ & \nodata & $	55\pm27	$ &	HST/WFPC2	&\cite{2007ApJ...666..663B}\\
$	0.61	$ & $	7.4\pm1.5	$ & \nodata &	HST+ground	&\cite{2001ApJ...559..592T}\\
$	0.64	$ & \nodata & $	7.7\pm5.8	$ &	Pioneer 10/11	&\cite{2011ApJ...736..119M}\\
$	0.81	$ & $	9.3\pm1.6	$ & \nodata &	HST+ground	&\cite{2001ApJ...559..592T}\\
$	0.814	$ & \nodata & $	57\pm32	$ &	HST/WFPC2	&\cite{2007ApJ...666..663B}\\
$	1.25	$ & \nodata & $	21\pm15	$ &	COBE/DIRBE	&\cite{2007ApJ...666...34L}\\
$	1.25	$ & $	11.5\pm1.3	$ & \nodata &	HST+ground	&\cite{2001ApJ...559..592T}\\
$	1.25	$ & $	11.7\pm2.6	$ & \nodata &	Subaru	&\cite{2010ApJ...723...40K}\\
$	1.6	$ & $	11.5\pm1.5	$ & \nodata &	Subaru	&\cite{2010ApJ...723...40K}\\
$	2.12	$ & $	10.0\pm0.8	$ & \nodata &	Subaru	&\cite{2010ApJ...723...40K}\\
$	2.2	$ & \nodata & $	20\pm6	$ &	COBE/DIRBE	&\cite{2007ApJ...666...34L}\\
$	2.2	$ & $	9.0\pm1.2	$ & \nodata &	HST+ground	&\cite{2001ApJ...559..592T}\\
$	3.5	$ & \nodata & $	13.3\pm2.8	$ &	COBE/DIRBE	&\cite{2007ApJ...666...34L}\\
$	3.6	$ & $	5.6\pm1.0	$ & \nodata &	Spitzer/IRAC	&\cite{2013ApJ...769...80A}\\
$	4.5	$ & $	4.4\pm0.8	$ & \nodata &	Spitzer/IRAC	&\cite{2013ApJ...769...80A}\\
$	4.9	$ & \nodata & $	22\pm12	$ &	COBE/DIRBE	&\cite{2003ApJ...585..305A}\\
$	15	$ & $1.9\pm0.4$ & \nodata &	ISO/ISOCAM	&\cite{2010ApJ...716L..45H}\\
$	16	$ & $2.2\pm0.2$ & \nodata &	Spitzer	&\cite{2011AJ....141....1T}\\
$	24	$ & $	2.86_{-0.16}^{+0.19}	$ & \nodata &	Spitzer/MIPS	&\cite{2010AA...512A..78B}\\
$	60	$ & \nodata & $	28.1\pm7.2	$ &	COBE/DIRBE	&\cite{2000ApJ...544...81F}\\
$	65	$ & \nodata & $	12.5\pm9.3	$ &	Akari	&\cite{2011ApJ...737....2M}\\
$	70	$ & $	6.6_{-0.6}^{+0.7}	$ & \nodata &	Spitzer/MIPS	&\cite{2010AA...512A..78B}\\
$	90	$ & \nodata & $	22.3\pm5.0	$ &	Akari	&\cite{2011ApJ...737....2M}\\
$	100	$ & $	8.35\pm0.95	$ & \nodata &	Herschel/PACS	&\cite{2011AA...532A..49B}\\
$	140	$ & \nodata & $	12.6\pm6.0	$ &	COBE/FIRAS	&\cite{1998ApJ...508..123F}\\
$	140	$ & \nodata & $	20.1\pm3.6	$ &	Akari	&\cite{2011ApJ...737....2M}\\
$	140	$ & \nodata & $	15.0\pm5.9	$ &	COBE/DIRBE	&\cite{2007ApJ...667...11O}\\
$	160	$ & $	14.6_{-2.9}^{+7.1}	$ & \nodata &	Spitzer/MIPS	&\cite{2010AA...512A..78B}\\
$	160	$ & \nodata & $	13.7\pm6.1	$ &	COBE/FIRAS	&\cite{1998ApJ...508..123F}\\
$	160	$ & \nodata & $	13.7\pm4.0	$ &	Akari	&\cite{2011ApJ...737....2M}\\
$	160	$ & \nodata & $	14.4\pm2.4	$ &	Spitzer/MIPS	&\cite{2012AA...543A.123P}\\
$	240	$ & \nodata & $	10.9\pm4.3	$ &	COBE/FIRAS	&\cite{1998ApJ...508..123F}\\
$	240	$ & \nodata & $	12.7\pm1.6	$ &	COBE/DIRBE	&\cite{2007ApJ...667...11O}\\
$	250	$ & $	10.13_{-2.33}^{+2.60}	$ & \nodata &	Herschel/SPIRE	&\cite{2012AA...542A..58B}\\
$	250	$ & \nodata & $	10.3\pm4.0	$ &	COBE/FIRAS	&\cite{1998ApJ...508..123F}\\
$	350	$ & $	6.46_{-1.57}^{+1.74}	$ & \nodata &	Herschel/SPIRE	&\cite{2012AA...542A..58B}\\
$	350	$ & \nodata & $	5.6\pm2.1	$ &	COBE/FIRAS	&\cite{1998ApJ...508..123F}\\
$	500	$ & $	2.80_{-0.81}^{+0.93}	$ & \nodata &	Herschel/SPIRE	&\cite{2012AA...542A..58B}\\
$	500	$ & \nodata & $	2.4\pm0.9	$ &	COBE/FIRAS	&\cite{1998ApJ...508..123F}\\
$	850	$ & $	0.24\pm0.03	$ & \nodata &	SCUBA	&\cite{2010ApJ...721..424Z}\\
$	850	$ & \nodata & $	0.50\pm0.21	$ &	COBE/FIRAS	&\cite{1998ApJ...508..123F}
\enddata
\end{deluxetable*}

\subsection{Gamma-ray data}
\label{Sec:GammaRayData}

We study the gamma-ray spectra of blazars published by ground based instruments up to 2014. We have succeeded in obtaining the spectral points for 106 non-redundant spectra, listed in Table~\ref{Tab:GammaData}, from 38 sources. Representing more than $\unit[80]{\%}$ of the spectral data from blazars indexed in TeVCat,\footnote{http://tevcat.uchicago.edu/} this is the most complete compilation of VHE gamma-ray data ever used for a study of the EBL. We paid particular attention to redundant data published in multiple articles, and selected the most up-to-date ones when a reanalysis was performed or when new data were included in the study.

To perform cosmological studies, we therefore select the spectra of known-redshift BL~Lac objects, called the gamma-ray cosmology sample in the following. With a total of $\sim270,000$ gamma rays from 86 spectra, this sample carries most of the information on gamma-ray absorption. The remaining 20 spectra, contributing an additional $\sim 30,000$ gamma rays, are used in Sections \ref{Sec:Redshift} and \ref{Sec:ALP}.

\begin{figure}
\hspace{-0.2cm}
\includegraphics[width=0.535\textwidth]{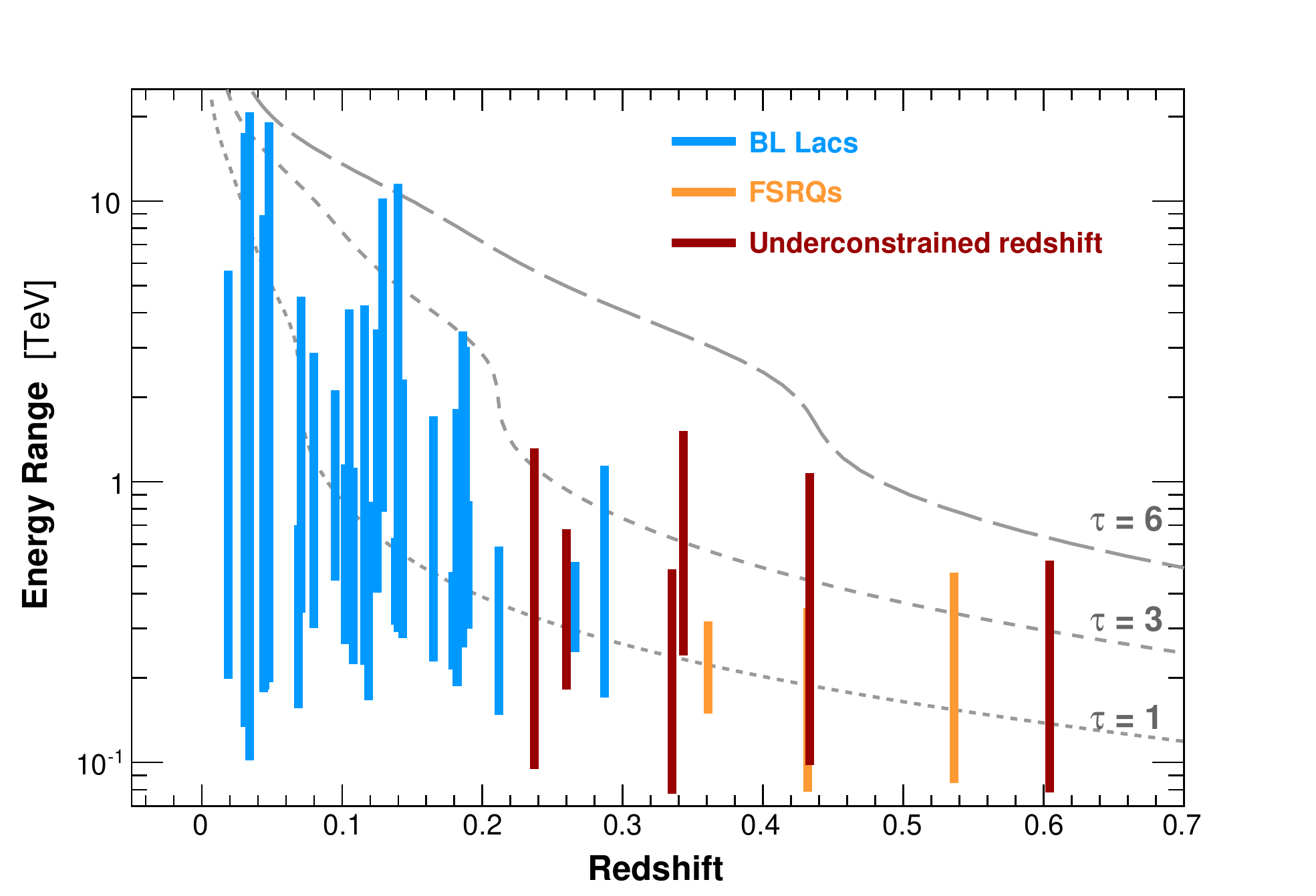}
\caption{Energy ranges spanned by the spectra from Table~\ref{Tab:GammaData} as a function of redshift. }
\label{fig:Ez}
\end{figure}

The energy ranges spanned by the gamma-ray spectra are shown in Fig.~\ref{fig:Ez}, together with the iso-optical-depth lines derived from our best-fit EBL spectrum (see Sec.~\ref{Sec:EBLSpec}). The full sample covers a redshift range up $z\sim0.604$ (PKS~1424+240) while the gamma-ray cosmology sample is limited to $z\leq0.287$ (1ES~0414+009).

Six objects have uncertain redshifts, shown as red lines in Fig.~\ref{fig:Ez} and indicated by a question mark in the third column of Table~\ref{Tab:GammaData}. They are discussed in more detail in Sec.~\ref{Sec:Redshift}.

We also associated a quasi-contemporaneous HE gamma-ray spectrum to each VHE spectrum whenever available. In particular, VHE observations performed after mid-2008 are contemporaneous with the sky survey of {\it Fermi}-LAT. Whenever HE best-fit spectral parameters were published in the same article as the VHE spectrum, we stored them for further analysis. The same procedure was followed using the 2FGL spectra \citep{2012ApJS..199...31N} for objects showing no significant variability in both gamma-ray bands, as long as the ground-based observations succeeded the launch of the satellite.

The quasi-contemporaneity of HE and VHE data motivates the use of the HE spectral shape to constrain the intrinsic VHE model. Photon indices have proven quite stable in the HE band except during flaring events, despite the rather high flux variability at all time scales that is characteristic of blazars \citep[see, e.g.,][]{2010ApJ...710.1271A}.

\subsection{Anaysis method}
\label{Sec:Ana}

\subsubsection{Test statistic}\label{Sec:TS}

We aim at finding the best EBL spectrum jointly accounting for local constraints and gamma-ray observations. We define the associated test statistic as:
\begin{equation}
\chi^2 = \chi^2_{\rm EBL} + \sum_{\rm \gamma ray\ spectra}\left(\chi^2_{\rm \gamma ray\ points} + \chi^2_{\rm HE-VHE}\right)
\label{Eq:EBLAndGammaChi2}
\end{equation} 

$\chi^2_{\rm \gamma ray\ points}$ is a measure of the quality of the fit of each gamma-ray spectrum:
\begin{equation}
\chi^2_{\rm \gamma ray\ points} = \sum_{i\ \in {\rm \gamma ray\ points}} \left(\frac{\phi_{\rm model}(E_i,z_0)-\phi_i}{\sigma_{\phi_i}}\right)^2
\end{equation} 
where $\phi_i$ and $\sigma_{\phi_i}$ are the measured flux and associated uncertainty, and where $\phi_{\rm model}$ is discussed in Sec.~\ref{Sec:IntModel}.

$\chi^2_{\rm HE-VHE}$ accounts for the constraints on the intrinsic spectral parameters. We impose, for contemporaneous observations in the HE and VHE bands, that the intrinsic VHE spectrum be softer than the HE spectrum:
\begin{equation}
\chi^2_{\rm HE-VHE} = \Theta(\Gamma_{\rm HE}-\Gamma)\left(\frac{\Gamma_{\rm HE}-\Gamma}{\sigma_{\Gamma_{\rm HE}}}\right)^2
\label{Eq:IntPrior}
\end{equation}
where $\Gamma_{\rm HE}$ and $\sigma_{\Gamma_{\rm HE}}$ are the photon index and associated uncertainty measured at HE, and where $\Gamma$ is the intrinsic VHE photon index. For curved spectra, the indices and uncertainties are computed at the intersection of the HE and VHE range.

$\chi^2_{\rm EBL}$ accounts for the lower limits and upper limits listed in Table~\ref{Tab:EBLdata}, $\displaystyle \left\{\nu I^i_\nu\ _{-\sigma^i_l}^{+\sigma^i_u}\right\}_i$. Only the constraining sides of the limits are considered, using again Heavyside functions:
\begin{align}
\chi^2_{\rm EBL} &= \sum_{i\in {\rm LL}} \Theta(\nu I^i_\nu - \nu I_\nu(\lambda_i))\times\left(\frac{\nu I_\nu(\lambda_i)-\nu I^i_\nu}{\sigma^i_l}\right)^2 \nonumber\\ &+ \sum_{i \in {\rm UL}} \Theta(\nu I_\nu(\lambda_i)-\nu I^i_\nu)\times\left(\frac{\nu I_\nu(\lambda_i)-\nu I^i_\nu}{\sigma^i_u}\right)^2
\label{Eq:EBLchi2}
\end{align}
where $\nu I_\nu(\lambda)$ is the model of the data, and where LL and UL denote the ensembles of lower and upper limits, respectively. 

The local constraints effectively taken into account in Eq.~\ref{Eq:EBLchi2} depend on the EBL-wavelength range probed by the gamma-ray data. The threshold condition in Eq.~\ref{Eq:Threshold} imposes $\lambda=hc/\epsilon_0<\lambda_{\rm max}$, where $\lambda_{\rm max}$ is the maximum wavelength associated with each spectrum, with:
\begin{align}\label{Eq:Lambda_max}
\lambda_{\rm max} &=   \frac{h}{m_e c} \times \frac{E_{\rm max,VHE}}{m_e c^2}\times (1+z_0)^2\\ \nonumber 
	& \sim \unit[95]{\mu m} \times\left(\frac{E_{\rm max,VHE}}{\unit[20]{TeV}}\right) \times(1+z_0)^2
\end{align}
The minimum wavelength is set {\it a posteriori} by successively adding free parameters to the model until no further improvement is found, as discussed in Sec.~\ref{Sec:NGauss}.

Expanding on the method of \cite{2013A&A...550A...4H}, the minimization of the $\chi^2$ in Eq.~\ref{Eq:EBLAndGammaChi2} in turn assumes, for each set of EBL parameters tested, a minimization over the parameters of the intrinsic spectra. This minimization is performed for each spectrum using the {\tt MIGRAD} method of {\tt MINUIT}. The minimization over the EBL parameters is performed in three steps, using successively {\tt SIMPLEX}, {\tt MIGRAD}, and {\tt HESSIAN}.

\subsubsection{Intrinsic spectral models}\label{Sec:IntModel}

The VHE gamma-ray spectra are modeled with
\begin{equation}
\phi_{\rm model}(E,z) = \phi_{\rm int}(E)\times \exp\left(-\tau(E,z)\right)
\label{Eq:SpectralModel}
\end{equation}

The intrinsic model, $\phi_{\rm int}(E)$, is determined iteratively as done by the \cite{2013A&A...550A...4H}. As a first step, the simplest two-parameter model is chosen: a power law (PWL in Table~\ref{Tab:GammaData}), $\phi_{\rm PWL}(E) = \phi_0 (E/E_0)^{-\Gamma}$, where $E_0$ is the reference energy, $\phi_0$ is the flux normalization and $\Gamma$ is the photon index. The reference energy is fixed, in this work, to the central value of the energy range of each spectrum, i.e. $E_0 = \sqrt{E_{\rm min,VHE}E_{\rm max,VHE}}$. To search for intrinsic curvature, we also test the three-parameter log parabola (LP), $\phi_{\rm LP}(E) = \phi_0 (E/E_0)^{-a - b \ln(E/E_0)}$, where $a$ is the photon index at $E_0$ and where $b$ is the curvature parameter, the exponential cutoff power law (EPWL), $\phi_{\rm EPWL}(E) = \phi_0 (E/E_0)^{-\Gamma}\exp(-E/E_{\rm cut})$, where $E_{\rm cut}$ is the cutoff energy. We also consider an exponential cutoff log parabola (ELP), $\phi_{\rm ELP}(E) = \phi_0 (E/E_0)^{-a - b \ln(E/E_0)}\exp(-E/E_{\rm cut})$, however none of the spectra is found to significantly prefer this model. We impose concave intrinsic spectra, i.e. $b\geq0$ and $E_{\rm cut}>0$.

For a given set of intrinsic models, we inspect the residuals of the individual spectra fixing the EBL parameters to their best-fit values. If a more complex intrinsic model is preferred at least at the $2\sigma$ level, it is chosen for the next iteration, until the set of models converges. To avoid any overrestriction of the parameter space, once the stable set of intrinsic models was found, we successively changed each intrinsic spectrum to its more complex counterpart, whichever was most preferred, and checked that it had no impact on other models nor on the best-fit EBL spectrum. This ensured the robustness of our method and of the choice of the intrinsic models.

\LongTables
\begin{deluxetable*}{lcccccr}
\tablecolumns{7} 
\tabletypesize{\scriptsize}
\tablecaption{Gamma-ray spectra used in this paper.\label{Tab:GammaData}}
\tablehead{\colhead{Source} & \colhead{Class} & \colhead{Redshift} & \colhead{Experiment} & \colhead{Obs. Period} & \colhead{Model} & \colhead{ \quad \qquad  \qquad Reference  \qquad  \qquad \quad}}
\startdata
	IC 310&HBL&0.019&	MAGIC&	2009-2010 (low)&PWL	& \cite{2014AA...563A..91A} \\
	\nodata&\nodata&\nodata&	MAGIC&	2009-2010 (high)&PWL	& \cite{2014AA...563A..91A} \\
	Markarian 421&HBL&0.031&	CAT&	1998&PWL	& \cite{2000AIPC..515..113P} \\
	\nodata&\nodata&\nodata&	HEGRA&	1999-2000&PWL	& \cite{2002AA...393...89A} \\
	\nodata&\nodata&\nodata&	HEGRA&	2000-2001&LP	& \cite{2002AA...393...89A} \\
	\nodata&\nodata&\nodata&	Tibet&	2000-2001 &PWL	& \cite{2003ApJ...598..242A} \\
	\nodata&\nodata&\nodata&	HESS&	2004&EPWL	& \cite{2005AA...437...95A} \\
	\nodata&\nodata&\nodata&	MAGIC&	2004-2005&EPWL	& \cite{2007ApJ...663..125A} \\
	\nodata&\nodata&\nodata&	MAGIC&	2006-04-22&PWL	& \cite{2010AA...519A..32A} \\
	\nodata&\nodata&\nodata&	MAGIC&	2006-04-24&PWL	& \cite{2010AA...519A..32A} \\
	\nodata&\nodata&\nodata&	MAGIC&	2006-04-25&PWL	& \cite{2010AA...519A..32A} \\
	\nodata&\nodata&\nodata&	MAGIC&	2006-04-26&PWL	& \cite{2010AA...519A..32A} \\
	\nodata&\nodata&\nodata&	MAGIC&	2006-04-27&EPWL	& \cite{2010AA...519A..32A} \\
	\nodata&\nodata&\nodata&	MAGIC&	2006-04-28&PWL	& \cite{2010AA...519A..32A} \\
	\nodata&\nodata&\nodata&	MAGIC&	2006-04-29&PWL	& \cite{2010AA...519A..32A} \\
	\nodata&\nodata&\nodata&	ARGO-YBJ&	2007-2010 (flux1)&PWL	& \cite{2011ApJ...734..110B} \\
	\nodata&\nodata&\nodata&	ARGO-YBJ&	2007-2010 (flux2)&LP	& \cite{2011ApJ...734..110B} \\
	\nodata&\nodata&\nodata&	ARGO-YBJ&	2007-2010 (flux3)&EPWL	& \cite{2011ApJ...734..110B} \\
	\nodata&\nodata&\nodata&	ARGO-YBJ&	2007-2010 (flux4)&PWL	& \cite{2011ApJ...734..110B} \\	
	\nodata&\nodata&\nodata&	VERITAS&	2008 (very low)&LP	& \cite{2011ApJ...738...25A} \\
	\nodata&\nodata&\nodata&	VERITAS&	2008 (low)&LP	& \cite{2011ApJ...738...25A} \\
	\nodata&\nodata&\nodata&	VERITAS&	2008 (mid)&LP	& \cite{2011ApJ...738...25A} \\
	\nodata&\nodata&\nodata&	VERITAS&	2008 (high A)&LP	& \cite{2011ApJ...738...25A} \\
	\nodata&\nodata&\nodata&	VERITAS&	2008 (high B)&EPWL	& \cite{2011ApJ...738...25A} \\
	\nodata&\nodata&\nodata&	VERITAS&	2008 (high C)&PWL	& \cite{2011ApJ...738...25A} \\
	\nodata&\nodata&\nodata&	VERITAS&	2008 (very high)&EPWL	& \cite{2011ApJ...738...25A} \\
	\nodata&\nodata&\nodata&	TACTIC&	2005-2006&PWL	& \cite{2015NIMPA.770...42S} \\
	\nodata&\nodata&\nodata&	TACTIC&	2008&PWL	& \cite{2010JPhG...37l5201C} \\
	\nodata&\nodata&\nodata&	TACTIC&	2009-2010&PWL	& \cite{2012JPhG...39d5201C} \\
	Markarian 501&HBL&0.034&	HEGRA&	1997&PWL	& \cite{2001AA...366...62A} \\
	\nodata&\nodata&\nodata&	TACTIC&	2005-2006&PWL	& \cite{2008JPhG...35f5202G} \\
	\nodata&\nodata&\nodata&	MAGIC&	2006&PWL	& \cite{2009ApJ...705.1624A} \\
	\nodata&\nodata&\nodata&	ARGO-YBJ&	2008-2011&PWL	& \cite{2012ApJ...758....2B} \\
	\nodata&\nodata&\nodata&	ARGO-YBJ&	2011 (high)&PWL	& \cite{2012ApJ...758....2B} \\
	1ES 2344+514&HBL&0.044&	Whipple&	1995 (dataset B)&PWL	& \cite{2005ApJ...634..947S} \\
	\nodata&\nodata&\nodata&	MAGIC&	2005-2006&PWL	& \cite{2007ApJ...662..892A} \\
	\nodata&\nodata&\nodata&	VERITAS&	2007 (low)&LP	& \cite{2011ApJ...738..169A} \\
	\nodata&\nodata&\nodata&	VERITAS&	2007 (high)&PWL	& \cite{2011ApJ...738..169A} \\
	\nodata&\nodata&\nodata&	MAGIC&	2008&PWL	& \cite{2013AA...556A..67A} \\
	Markarian 180&HBL&0.045&	MAGIC&	2006&PWL	& \cite{2006ApJ...648L.105A} \\
	1ES 1959+650&HBL&0.048&	HEGRA&	2002&PWL	& \cite{2003AA...406L...9A} \\
	\nodata&\nodata&\nodata&	Whipple&	2002&PWL	& \cite{2005ApJ...621..181D} \\
	\nodata&\nodata&\nodata&	MAGIC&	2006&PWL	& \cite{2008ApJ...679.1029T} \\
	\nodata&\nodata&\nodata&	VERITAS&	2007-2011&PWL	& \cite{2013ApJ...775....3A} \\
	BL Lacertae&IBL&0.069&	MAGIC&	2005&PWL	& \cite{2007ApJ...666L..17A} \\
	\nodata&\nodata&\nodata&	VERITAS&	2011&PWL	& \cite{2013ApJ...762...92A} \\
	PKS 2005-489&HBL&0.071&	HESS&	2004-2007&PWL	& \cite{2010AA...511A..52H} \\
	RGB J0152+017&HBL&0.08&	HESS&	2007&PWL	& \cite{2008AA...481L.103A} \\
	SHBL J001355.9-185406&HBL&0.095&	HESS&	2008-2011&PWL	& \cite{2013AA...554A..72H} \\
	W Comae&IBL&0.102&	VERITAS&	2008-01-04&PWL	& \cite{2008ApJ...684L..73A} \\
	1ES 1312-423&HBL&0.105&	HESS&	2004-2010&PWL	& \cite{2013MNRAS.434.1889H} \\
	VER J0521+211&IBL&0.108&	VERITAS&	2009-2010&PWL	& \cite{2013ApJ...776...69A} \\
	PKS 2155-304&HBL&0.116&	HESS&	2002-06&PWL	& \cite{2005AA...430..865A} \\
	\nodata&\nodata&\nodata&	HESS&	2002-10&PWL	& \cite{2005AA...430..865A} \\
	\nodata&\nodata&\nodata&	HESS&	2003-06&PWL	& \cite{2005AA...430..865A} \\
	\nodata&\nodata&\nodata&	HESS&	2003-07&PWL	& \cite{2005AA...430..865A} \\
	\nodata&\nodata&\nodata&	HESS&	2003-08&PWL	& \cite{2005AA...430..865A} \\
	\nodata&\nodata&\nodata&	HESS&	2003-09&PWL	& \cite{2005AA...430..865A} \\
	\nodata&\nodata&\nodata&	HESS&	2003-10-11&PWL	& \cite{2005AA...442..895A} \\
	\nodata&\nodata&\nodata&	HESS&	2005-2007&PWL	& \cite{2010AA...520A..83H} \\
	\nodata&\nodata&\nodata&	HESS&	2006-07/08&LP	& \cite{2013PhRvD..88j2003A} \\
	\nodata&\nodata&\nodata&	MAGIC&	2006-07/08&LP	& \cite{2012AA...544A..75A} \\
	\nodata&\nodata&\nodata&	HESS&	2006-08-02&PWL	& \cite{2012AA...539A.149H} \\
	\nodata&\nodata&\nodata&	HESS&	2006-08-03&PWL	& \cite{2012AA...539A.149H} \\
	\nodata&\nodata&\nodata&	HESS&	2008-08-09&PWL	& \cite{2009ApJ...696L.150A} \\
	B3 2247+381&HBL&0.119&	MAGIC&	2010&PWL	& \cite{2012AA...539A.118A} \\
	RGB J0710+591&HBL&0.125&	VERITAS&	2008-2009&PWL	& \cite{2010ApJ...715L..49A} \\
	H 1426+428&HBL&0.129&	HEGRA&	1999-2000&PWL	& \cite{2003AA...403..523A} \\
	\nodata&\nodata&\nodata&	HEGRA&	2002&PWL	& \cite{2003AA...403..523A} \\
	1ES 0806+524&HBL&0.138&	VERITAS&	2006-2008&PWL	& \cite{2009ApJ...690L.126A} \\
	1ES 0229+200&HBL&0.14&	HESS&	2005-2006&PWL	& \cite{2007AA...475L...9A} \\
	\nodata&\nodata&\nodata&	VERITAS&	2009-2012&PWL	& \cite{2014ApJ...782...13A} \\
	1RXS J101015.9-311909&HBL&0.143&	HESS&	2006-2010&PWL	& \cite{2012AA...542A..94H} \\
	H 2356-309&HBL&0.165&	HESS&	2004&PWL	& \cite{2010AA...516A..56H} \\
	\nodata&\nodata&\nodata&	HESS&	2005&PWL	& \cite{2010AA...516A..56H} \\
	\nodata&\nodata&\nodata&	HESS&	2006&PWL	& \cite{2010AA...516A..56H} \\
	RX J0648.7+1516&HBL&0.179&	VERITAS&	2010&PWL	& \cite{2011ApJ...742..127A} \\
	1ES 1218+304&HBL&0.182&	VERITAS&	2007&PWL	& \cite{2009ApJ...695.1370A} \\
	\nodata&\nodata&\nodata&	VERITAS&	2008-2009&PWL	& \cite{2010ApJ...709L.163A} \\
	1ES 1101-232&HBL&0.186&	HESS&	2004-2005&PWL	& \cite{2007AA...470..475A} \\
	1ES 0347-121&HBL&0.188&	HESS&	2006-09-12&PWL	& \cite{2007AA...473L..25A} \\
	RBS 0413&HBL&0.19&	VERITAS&	2008-2009&PWL	& \cite{2012ApJ...750...94A} \\
	1ES 1011+496&HBL&0.212&	MAGIC&	2007&PWL	& \cite{2007ApJ...667L..21A} \\
	1ES 1215+303&HBL&0.237?&	VERITAS&	2008-2012&PWL	& \cite{2013ApJ...779...92A} \\
	\nodata&\nodata&\nodata&	MAGIC&	2011&PWL	& \cite{2012AA...544A.142A} \\
	S5 0716+714&IBL&0.26?&	MAGIC&	2007-2008&PWL	& \cite{2009ApJ...704L.129A} \\
	PKS 0301-243&HBL&0.266&	HESS&	2009-2010&PWL	& \cite{2013AA...559A.136H} \\
	1ES 0414+009&HBL&0.287&	HESS&	2005-2009&PWL	& \cite{2012AA...538A.103H} \\
	\nodata&\nodata&\nodata&	VERITAS&	2008-2011&PWL	& \cite{2012ApJ...755..118A} \\
	3C 66A&IBL&0.335?&	MAGIC&	2009-2010&PWL	& \cite{2011ApJ...726...58A} \\
	PKS 0447-439&HBL&0.343?&	HESS&	2009-2010&PWL	& \cite{2013AA...552A.118H} \\
	PKS 1510-089&FSRQ&0.361&	HESS&	2009&PWL	& \cite{2013AA...554A.107H} \\
	4C +21.35&FSRQ&0.432&	MAGIC&	2010-06-17&PWL	& \cite{2011ApJ...730L...8A} \\
	PG 1553+113&HBL&0.433?&	HESS&	2005-2006&PWL	& \cite{2008AA...477..481A} \\
	\nodata&\nodata&\nodata&	MAGIC&	2007&PWL	& \cite{2012ApJ...748...46A} \\
	\nodata&\nodata&\nodata&	MAGIC&	2008&PWL	& \cite{2012ApJ...748...46A} \\
	\nodata&\nodata&\nodata&	MAGIC&	2009&PWL	& \cite{2012ApJ...748...46A} \\
	\nodata&\nodata&\nodata&	VERITAS&	2010-2012&PWL	& \cite{2015ApJ...799....7A} \\
	\nodata&\nodata&\nodata&	HESS&	2012&PWL	& \cite{2015ApJ...802...65A} \\
	3C 279&FSRQ&0.536&	MAGIC&	2006&PWL	& \cite{2008Sci...320.1752M} \\
	PKS 1424+240&IBL&0.604?&	VERITAS&	2009&PWL	& \cite{2014ApJ...785L..16A} \\
	\nodata&\nodata&\nodata&	VERITAS&	2011&PWL	& \cite{2014ApJ...785L..16A} \\
	\nodata&\nodata&\nodata&	VERITAS&	2013&PWL	& \cite{2014ApJ...785L..16A} \\
	\nodata&\nodata&\nodata&	MAGIC&	2009&PWL	& \cite{2014AA...567A.135A} \\
	\nodata&\nodata&\nodata&	MAGIC&	2010&PWL	& \cite{2014AA...567A.135A} \\
	\nodata&\nodata&\nodata&	MAGIC&	2011&PWL	& \cite{2014AA...567A.135A}
\enddata
\end{deluxetable*}

\subsubsection{Deconvolution method}\label{Sec:NGauss}

Similarly to the splines used in \cite{2007A&A...471..439M}, we chose a spectral model for the EBL that is a linear function of its parameters. As shown in Eq.~\ref{Eq:Opt3}, the pertinent variable is the logarithm of the EBL-photon energy, or equivalently the logarithm  of the EBL wavelength $l = \ln(\lambda/\lambda_{\rm ref})$, where $\lambda_{\rm ref}$ is a reference wavelength, set e.g. to $\unit[1]{\mu m}$. We model the EBL intensity as a sum of Gaussians of this variable, with fixed widths and peak positions:

\begin{align}
\nu I_\nu(l) &= \sum_i a_i \exp\left(-\frac{(l-l_i)^2}{2\sigma_l^2} \right) \nonumber \\
& \equiv \sum_i a_i \mathcal{N}(l;l_i,\sigma_l)
\label{Eq:GaussianModel}
\end{align}

The peak positions are logarithmically spaced:
\begin{equation}
\log(\lambda_i/\lambda_{\rm ref}) \equiv l_i = l_0 + i\times\Delta_l
\end{equation}

The width of the Gaussian functions depends on the distance between anchor points $\Delta_l$. We impose that the sum of two consecutive Gaussians of unity amplitude be equal to one right in between the two anchor points, i.e.
\begin{equation}
\Delta_l = \sigma_l\times 2\sqrt{2\ln2}
\label{Eq:DeltaE}
\end{equation}

The Gaussians leak into the neighboring bins and the average flux $\nu I_\nu^i$ in a bin centered on  $l_i$ is then
\begin{align}
\nu I_\nu^i &= \sum_j a_j \frac{1}{\Delta_l}\int_{l_i - \Delta_l/2}^{l_i + \Delta_l/2}{\rm d}l\exp\left(-\frac{(l-l_j)^2}{2\sigma_l^2}\right) \nonumber\\&		\equiv \sum_j a_j w_{ij}
\label{Eq:MatrixRelation}
\end{align}
with
\begin{align}
w_{ij} = \frac{\sqrt{\pi}}{4\sqrt{\ln2}}  &\left[{\rm erf}\left((i-j+\frac{1}{2})\times 2\sqrt{\ln2}\right) \right. \nonumber \\ &- \left. {\rm erf}\left((i-j-\frac{1}{2})\times 2\sqrt{\ln2}\right) \right]
\end{align}

The parameters $a_i$ are left free to vary. Once best-fit values, uncertainties, and correlations have been determined, the binned average flux $\nu I_\nu^i$ can easily be derived from the linear relation in Eq.~\ref{Eq:MatrixRelation}. This linearity permits the propagation of uncertainties for Gaussian distributions of the weight $a_i$, fully accounting for the correlation terms. This justifies the use of the {\tt HESSIAN} method in Sec.~\ref{Sec:TS}, which yields such symmetric Gaussian uncertainties.

Given the linearity of the model with respect to its parameters $a_i$, the optical depth can be rewritten as:
\begin{equation}
\tau(E_0,z_0) = \sum_i a_i\ t_i(E_0,z_0)
\end{equation}
where

\begin{align}
t_i(E_0,z_0) &=  \frac{3\pi\sigma_T}{H_0} \times \frac{E_0}{m_e^2 c^4}\nonumber \\ &\times \mathcal{N}(\cdot;0, \sigma_l) \otimes K_{z_0}\left(e_i+\ln\frac{E_0}{m_e c^2}\right)
\end{align}
where $e_i=\ln\frac{hc/\lambda_i}{m_ec^2}$.

One can compute the weights $t_i(E_0,z_0)$ in the very beginning of the fitting procedure, further reducing the computation expense. For a set of about 90 spectra and associated models, the full fitting procedure of the EBL spectrum takes about ten seconds of CPU time on a $\unit[3]{GHz}$ core, highlighting the significance of the analytical work shown in this section and in Sec.~\ref{Sec:Theory}.

\section{Results}
\label{Sec:Results}

\subsection{EBL spectrum}
\label{Sec:EBLSpec}

The best-fit spectral models of the intrinsic spectra are listed in column 6 of Table~\ref{Tab:GammaData}. Most of the spectra of the gamma-ray cosmology sample (71/86) are best described by PWL models. The other fifteen spectra, modeled by LP and EPWL models, correspond either to intensive campaigns in low states of prominent sources \citep[2004-05 MAGIC campaign of Markarian~421, large-zenith-angle H.E.S.S. observations of Markarian~421, 2007 VERITAS campaign on 1ES~2344+514, 2007-2010 campaign of ARGO-YBG on Markarian~421:][respectively]{2007ApJ...663..125A,2005AA...437...95A,2011ApJ...738..169A,2011ApJ...734..110B}, to observations of flares \citep[high state of Markarian~421 observed by HEGRA in 2000-2001 and MAGIC in 2004-2005, major outburst of PKS~2155-304 in 2006 observed by MAGIC and H.E.S.S.:][respectively]{2002AA...393...89A,2010AA...519A..32A,2012AA...544A..75A,2013PhRvD..88j2003A}, or to both \citep[2006-2008 campaign on Markarian~421 by VERITAS, including flares:][]{2011ApJ...727..129A}. In such cases, enhanced statistics at the highest energies enable the probe of intrinsic curvature. No spectrum is preferentially modeled with an ELP model.

\begin{figure}
\hspace{-0.2cm}
\includegraphics[width=0.535\textwidth]{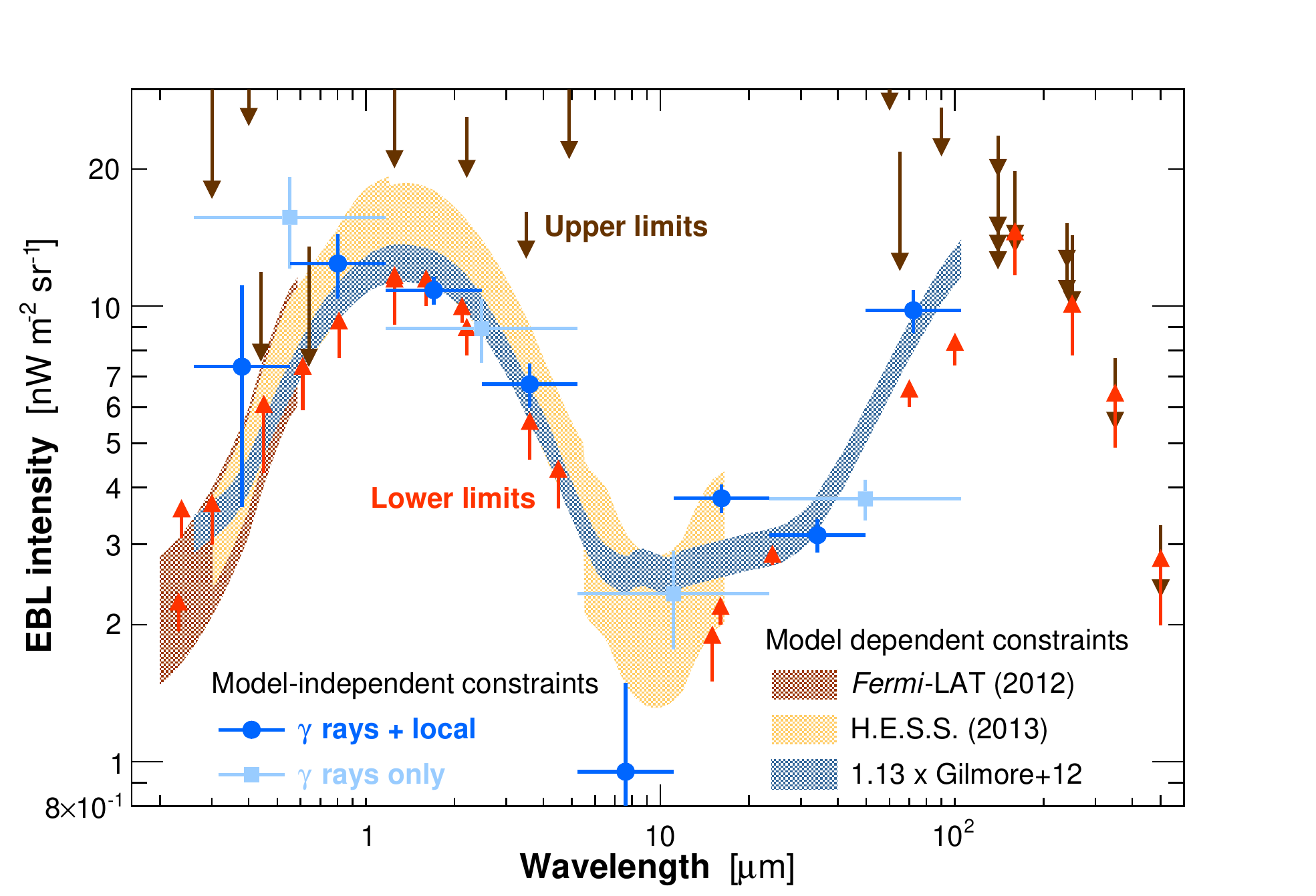}
\caption{EBL intensity at $z=0$ as a function of wavelength. The best-fit spectra derived in this work are shown with light blue (gamma rays only, four-point spectrum) and blue points (gamma rays + direct constraints, eight-point spectrum). Lower and upper limits are shown with orange upward-going and dark-brown downward-going arrows, respectively. For comparison with the work of \cite{2012Sci...338.1190A} and \cite{2013A&A...550A...4H}, the $\unit[1]{\sigma}$ (stat + sys) contour of the best-fit scaled-up model \citep{G12} is shown as filled blue region, using a scaling factor of $1.13$ as shown in Table~\ref{Tab:ResEBL}.}
\label{fig:3}
\end{figure}

The 86 spectra probe the wavelength range $\unit[0.26-105]{\mu m}$, for a bin size $\Delta_l=0.75$. The maximum wavelength corresponds to the pair-creation threshold, as described in Eq.~\ref{Eq:Lambda_max}. A smaller minimum wavelength would result in an underconstrained EBL intensity in the first bin. A smaller binning does not significantly improve the quality of the fit. With a total of 630 points and 187 free parameters for the intrinsic spectra, the best-fit model results in a test statistic of $(\sum\chi^2_{\rm \gamma ray\ points} + \chi^2_{\rm HE-VHE})/ndf = 340.1 / 443$. The small value of the reduced $\chi^2$ is not surprising, as the correlations between gamma-ray spectral points are not accounted for when fitting such archival data (the gamma-ray community is only starting to publish covariance matrices for spectral points). The uncertainties are also assumed to be Gaussian (underlying assumption for the $\chi^2$ test), while a full treatment at the event level would account for the Poisson statistics of the events from background and signal regions. 

The constraint from the hardness of the HE spectra associated with the VHE observations, proves {\it a posteriori} to play a minor role, $\sum \chi^2_{\rm HE-VHE} < 0.1$, indicating that there is no tension with the assumption of broad-band concavity in the intrinsic spectra. No tension is found with the local EBL constraints either, with $\chi^2_{\rm EBL}=2.4$ to which $n_{\rm EBL}=7$ local constraints contribute. Both the local EBL constraint and the hardness constraint thus barely impact the best-fit estimate of the EBL spectrum, but they nonetheless play a significant role when the spectrum departs from the best-fit point, thus impacting the uncertainties on the EBL. For the binning and the wavelength range probed here, the Gaussian-sum model admits eight parameters, resulting in a total test statistic of $\chi^2/ndf = 342.5/442$. The number of degrees of freedom accounts for eight free parameters to model the EBL and seven local points constraining this model, in addition to the $443$ degrees of freedom accounting for the gamma-ray spectra. 

Using the same intrinsic models and a null absorption, the gamma-ray spectra best-fit test statistic is $\sum(\chi^2_{\rm \gamma ray\ points} + \chi^2_{\rm HE-VHE})= 489.1$. With eight additional free parameters, the Gaussian-sum model is preferred by the gamma-ray data at the $\sqrt{2}{\rm\ erfc}^{-1}[\mathcal{P}_8(\Delta\chi^2=489.1-340.1)] = \unit[11]{\sigma}$ level, where $\mathcal{P}_8$ is the $\chi^2$ probability for eight degrees of freedom. This significance level may not readily be interpreted as the detection significance of the EBL signature, because the intrinsic models chosen for the likelihood ratio test are based on the best-fit EBL spectrum. Conservatively choosing instead the intrinsic models based on fits directly to the observed spectra assuming no EBL absorption results in eight intrinsic models being changed to a more complex or an equally complex model. The best-fit EBL model is then preferred at the $\unit[10.3]{\sigma}$ level, with no significant impact on the best-fit parameters. The cumulative EBL effect therefore leaves a strong imprint in the gamma-ray spectra, at a high significance level in between the two values given above. The EBL intensities computed from the best-fit parameters, as in Eq.~\ref{Eq:MatrixRelation}, are listed in Table~\ref{Tab:Int6} and shown in Fig.~\ref{fig:3} (blue points). The overall systematic uncertainty on the EBL level is estimated in Appendix~\ref{App:Sys} to be on the order of $\unit[7-8]{\%}$.

\begin{deluxetable}{cccc}
\tablecolumns{4} 
\tabletypesize{\scriptsize}
\tablecaption{Best fit EBL intensities.\label{Tab:Int6}}
\tablehead{\colhead{$\lambda$} & \colhead{$\lambda_{\rm min}$} & \colhead{$\lambda_{\rm max}$} & \colhead{$\nu I_\nu$}\\ 
\colhead{$\mu$m} &\colhead{$\mu$m} &\colhead{$\mu$m} & \colhead{\ nW m$^{-2}$ sr$^{-1}$} }
\startdata
$	0.38	$ & $	0.26	$ & $	0.55	$ & $	7.4	\pm	3.7	$\\
$	0.80	$ & $	0.55	$ & $	1.2	$ & $	12.4	\pm	1.9	$\\
$	1.7	$ & $	1.2	$ & $	2.5	$ & $	10.8	\pm	0.8	$\\
$	3.6	$ & $	2.5	$ & $	5.2	$ & $	6.7	\pm	0.7	$\\
$	7.6	$ & $	5.2	$ & $	11	$ & $	0.95	\pm	0.53	$\\
$	16	$ & $	11	$ & $	23	$ & $	3.79	\pm	0.26	$\\
$	34	$ & $	23	$ & $	50	$ & $	3.14	\pm	0.25	$\\
$	72	$ & $	50	$ & $	105	$ & $	9.8	\pm	1.1	$
\enddata
\end{deluxetable}
Theoretical and empirical EBL intensities can similarly be compared to local constraints and gamma-ray data. Table~\ref{Tab:ResEBL} lists the test statistics obtained with a null EBL, the best-fit eight-point EBL spectrum, and the models of \cite{G12} (G12), \cite{FR08} (F08), \cite{D11} (D11), \cite{2006ApJ...648..774S} (S06) \cite{2015ApJ...805...33K} (KS14), \cite{2010ApJ...712..238F} (F10), and \cite{2010A&A...515A..19K} (KD10). All the EBL models are preferred at the $\unit[10-12]{\sigma}$ level to a null EBL (column 7), as computed from the improvement in the quality of the fit of the gamma-ray data (column 4). The EBL models can also be compared to the best-fit eight-point spectrum with a likelihood ratio test, assuming that the models are nested. Such an approach is justified by a difference in optical depth between a model and its eight-point Gaussian-sum approximation smaller than $\unit[2]{\%}$ (see Table~\ref{Tab:Sys} of Appendix~\ref{App:GaussSum}). The difference between the best-fit eight-point spectrum and the models of G12, F08, and D11 corresponds to a small $\unit[0.5-1.5]{\sigma}$ preference for the Gaussian-sum description (column 8). The EBL shapes predicted by S06, KS14, F10, and K10 are disfavored at the $\geq\unit[3-6]{\sigma}$ level with respect to our best-fit spectrum. We have not included here the models of \cite{2012ApJ...752..113H} and \cite{2012ApJ...761..128S}, which estimate the EBL only up to the M ($\sim\unit[5]{\mu m}$) and I ($\sim\unit[0.8]{\mu m}$) photometric bands, respectively, and are thus not relevant for most of the TeV spectra listed in Table~\ref{Tab:GammaData}. We note nonetheless that their results are within $\unit[1-2]{\sigma}$ of the estimates derived in the present work at small wavelengths. Following the approach of the EBL studies by the H.E.S.S. and {\it Fermi}-LAT collaborations \citep{2013A&A...550A...4H,2012Sci...338.1190A}, we estimate the best-fit normalization factor of the EBL models, as listed in column 9.\footnote{Renormalizing theoretical or empirical EBL models is not physically motivated, since different components contributing to the spectrum need not scale in the same way. Nonetheless, measuring the best-fit scaling factor for a model is a straightforward way to quantify the model's compatibility with the experimental gamma-ray data.} The best-fit normalizations of the models listed in Table~\ref{Tab:ResEBL} are all larger than $1$, though only at the $\unit[1-2]{\sigma}$ level for the models of G12, F08, D11, and S06. This is not surprising as most of these models were designed to match sets of lower limits based on less complete surveys than the most recent studies, and as such tend to yield lower estimates of the EBL intensity. We compare in Fig.~\ref{fig:3} the scaled-up model of G12 with constraints derived from the measurements of H.E.S.S. below $z<0.2$ and {\it Fermi}-LAT between $0.5<z<1.6$. The H.E.S.S. and {\it Fermi}-LAT $\unit[1]{\sigma}$ confidence contours include the statistical and systematic uncertainties on the measurement of the parameter normalizing the model of F08, in the wavelength ranges relevant to these studies \citep[see][]{2013A&A...550A...4H,2013sf2a.conf..303B}. A good agreement on the level of EBL is found between these works and our results.

We investigate more closely the origin of the differences between the gamma-ray based EBL spectrum and the models. Good agreement between our results and all of the models in Table~\ref{Tab:ResEBL} is found below $\unit[5]{\mu m}$. The models of S06, KS14, F10, and KD10 are thus disfavored because of the rather low level of CIB they predict. The models of G12, F08, and D11 tend to predict a level of EBL that is higher than the gamma-ray estimate around $\unit[8]{\mu m}$, while lower around $\unit[16]{\mu m}$. The $\sim\unit[16]{\mu m}$ point from the gamma-ray data is found to be $\sim\unit[3]{\sigma}_{\rm stat+sys}$ above the galaxy-count estimate, but the lack of an observation around $\unit[8]{\mu m}$ prevents a direct comparison at this wavelength\footnote{A linear interpolation in $\log(\lambda)$ between the $4.5$ and $\unit[16]{\mu m}$ points from galaxy counts yields a level of EBL $\sim\unit[3]{\sigma}_{\rm stat+sys}$ higher than the gamma-ray estimate.}. These hints of deviations in the $\unit[5-20]{\mu m}$ region, which could be indicative of the signature of polycyclic aromatic hydrocarbons, should be taken with caution as neighboring points from the gamma-ray estimates are correlated. Using the covariance matrix shown in Appendix~\ref{App:BestFitAndCov}, the correlation coefficient between the $\sim\unit[8]{\mu m}$ and $\sim\unit[16]{\mu m}$ points is negative ($\unit[-16]{\%}$), showing that a decrease in the latter would result in an increase of the former, which would considerably reduce the deviations. A firm conclusion on the signature of polycyclic aromatic hydrocarbons cannot be drawn at this stage, as indicated by the small difference in quality between the model-independent fit of this work and the model-dependent fits (column 2 in Table~\ref{Tab:ResEBL}). Increased gamma-ray statistics from local sources ($z\lesssim0.1$) and improved constraints from galaxy counts in the $\unit[5-20]{\mu m}$ region would definitely help in deciding the matter.

\begin{deluxetable*}{lccccccccccc}
\tablecolumns{12} 
\tabletypesize{\small}
\tablecaption{Test statistics of the EBL models.\label{Tab:ResEBL}}
\tablehead{\colhead{Model}&\colhead{$\chi^2$}&\colhead{$ndf$}&\colhead{$\chi^2_\gamma$}&\colhead{$\chi^2_{\rm EBL}$}&\colhead{$n_{\rm EBL}$}&\colhead{$\sigma(\rm\cdot \neq no\ EBL)$}\tablenotemark{a}&\colhead{$\sigma(\rm this\ work\neq \cdot)$}\tablenotemark{b}&\colhead{$\alpha$}		&\colhead{$\chi^2_\gamma(\alpha)$}&\colhead{$\chi^2_{\rm EBL}(\alpha)$}&\colhead{$n_{\rm EBL}(\alpha)$}}
\startdata
No EBL	& $	489.1	$ & $	443	$ & $	489.1	$ & 	\nodata	& 	\nodata	& 	\nodata	& \nodata & 	\nodata			& 	\nodata	& 	\nodata	& 	\nodata	\\
This work	& $	342.5	$ & $	442	$ & $	340.1	$ & $	2.4	$ & $	7	$ & $	11.0	$ &	\nodata &	\nodata  &	\nodata &	\nodata&\nodata	\\
G12	& $	353.5	$ & $	455	$ & $	345.2	$ & $	8.4	$ & $	12	$ & $	12.0	$ & $	0.5	$ & $	1.13	\pm	0.07	$ & $	346.0	$ & $	1.9	$ & $	4	$\\
F08	& $	366.1	$ & $	451	$ & $	350.2	$ & $	15.9	$ & $	9	$ & $	11.8	$ & $	1.3	$ & $	1.05	\pm	0.07	$ & $	352.1	$ & $	13.6	$ & $	6	$\\
D11	& $	370.0	$ & $	453	$ & $	351.0	$ & $	19.0	$ & $	10	$ & $	11.8	$ & $	1.5	$ & $	1.16	\pm	0.05	$ & $	356.9	$ & $	3.5	$ & $	5	$\\
S06	& $	392.7	$ & $	445	$ & $	382.0	$ & $	10.7	$ & $	3	$ & $	10.3	$ & $	5.0	$ & $	1.18	\pm	0.08	$ & $	380.0	$ & $	7.5	$ & $	3	$\\
KS14	& $	401.4	$ & $	456	$ & $	362.1	$ & $	39.3	$ & $	13	$ & $	11.3	$ & $	3.0	$ & $	1.44	\pm	0.07	$ & $	361.2	$ & $	0.5	$ & $	4	$\\
F10	& $	424.0	$ & $	449	$ & $	390.2	$ & $	33.8	$ & $	7	$ & $	9.9	$ & $	5.7	$ & $	1.48	\pm	0.07	$ & $	384.6	$ & $	3.6	$ & $	3	$\\
KD10	& $	433.0	$ & $	457	$ & $	366.7	$ & $	66.3	$ & $	14	$ & $	11.1	$ & $	3.5	$ & $	1.52	\pm	0.14	$ & $	346.5	$ & $	0.2	$ & $	1	$
\enddata
\tablenotetext{a}{Significance with which the gamma-ray data prefer the model to the absence of EBL.}
\tablenotetext{b}{Significance with which the gamma-ray data prefer the EBL spectrum from this work to the model.\vspace{0.2cm}}
\end{deluxetable*}

To estimate the total brightness of the EBL, we perform a fit similar to that described above, extending the wavelength range to $\unit[0.1-1000]{\mu m}$ and using twelve points, corresponding roughly to $\Delta_l\sim0.75$. The fit is underconstrained below $\unit[0.25]{\mu m}$ and the EBL estimate in the lowest wavelength bin is compatible with zero at the $\unit[1]{\sigma}$ level. The brightness of the COB in $\unit[0.1-8]{\mu m}$ is estimated to be $\unit[36\pm11]{nW\ m^{-2}\ sr^{-1}}$, with a large uncertainty coming from the region $\lambda<\unit[0.25]{\mu m}$. Above $\unit[105]{\mu m}$, the gamma-ray data do not constrain the fit but the CIB is tightly constrained by local measurements, yielding a $\unit[8-1000]{\mu m}$ CIB brightness compatible with that of the COB, with a value of $\unit[25.9\pm3.4]{nW\ m^{-2}\ sr^{-1}}$. The total brightness of the EBL is measured with a $\sim\unit[20]{\%}$ accuracy as $\unit[62\pm12]{nW\ m^{-2}\ sr^{-1}}$, which is equivalent to $\unit[6.5\pm1.2]{\%}$ of the brightness of the CMB. This is compatible at the $\unit[1]{\sigma}$ level with previous estimations based on galaxy counts \citep{2006AA...451..417D} and on models (see e.g. Table 1 in G12, Table 4 in D11, and Table 2 in F10).

We also show in Fig.~\ref{fig:3} the best-fit EBL spectrum derived taking into account only the gamma-ray data (light-blue points). A larger binning ($\Delta_l = 1.5$) results because of the loss of information. Gamma-ray data show a spectrum of the EBL from mid-UV up to far IR that is in good agreement with estimates based on galaxy counts. The $\unit[11]{\sigma}$ deviation from a null EBL intensity and the fact that EBL reconstructed from gamma-ray observations follows the expected spectrum are pieces of evidence against scenarios in which the VHE emission of blazars would primarily come from ultra-high-energy cosmic rays (UHECR) reprocessed along the line of sight \citep[e.g.][]{2010APh....33...81E}. Quite a cosmic conspiracy would be needed to explain how secondary gamma rays from UHECR carry the very same imprint as that expected from absorption of primary gamma rays. A more quantitative study of the impact of this process on the EBL reconstruction is left for future work.

To quantify the compatibility between galaxy counts and the four-point EBL spectrum based only on gamma-ray data (light-blue points in Fig.~\ref{fig:3}), we compute the difference between these estimates in the wavelength range probed by the gamma-ray data. In order to account for the correlations between the gamma-ray based EBL points as well as for the broadness of the wavelength range including multiple points from galaxy counts, we build the following likelihood for the EBL excess, $\Delta\nu I_\nu$, marginalizing over the EBL parameters and taking into account multiple galaxy counts estimates for a single gamma-ray-based EBL point:

\begin{align}
\mathcal{L}(\Delta\nu I_\nu) =& \int{\rm d}A\ {\rm e}^{-\frac{1}{2}\sum_{i\in{\rm LL}}\left(\frac{\nu I_\nu(A,\lambda_i)-\nu I^i_\nu-\Delta\nu I_\nu}{\sigma^i_l}\right)^2}\nonumber \\
& \times{\rm e}^{-\frac{1}{2}\left[A - A_0\right]^{\rm T}V_{A_0}^{-1}\left[A - A_0\right]}
\label{Eq:Excess}
\end{align}
where $A_0$ is the vector of the best-fit EBL parameters and where $V_{A_0}$ is their covariance matrix, provided for reference in Appendix~\ref{App:BestFitAndCov}. $\nu I_\nu(\lambda_i)$ and $\nu I^i_\nu$ are the EBL specific intensity based on gamma rays and galaxy counts, respectively. The integration is performed numerically through a Monte Carlo probe of the parameter space within $\pm3\sigma$ of the best-fit $A_0$. 

\begin{figure}
\hspace{-0.2cm}
\includegraphics[width=0.535\textwidth]{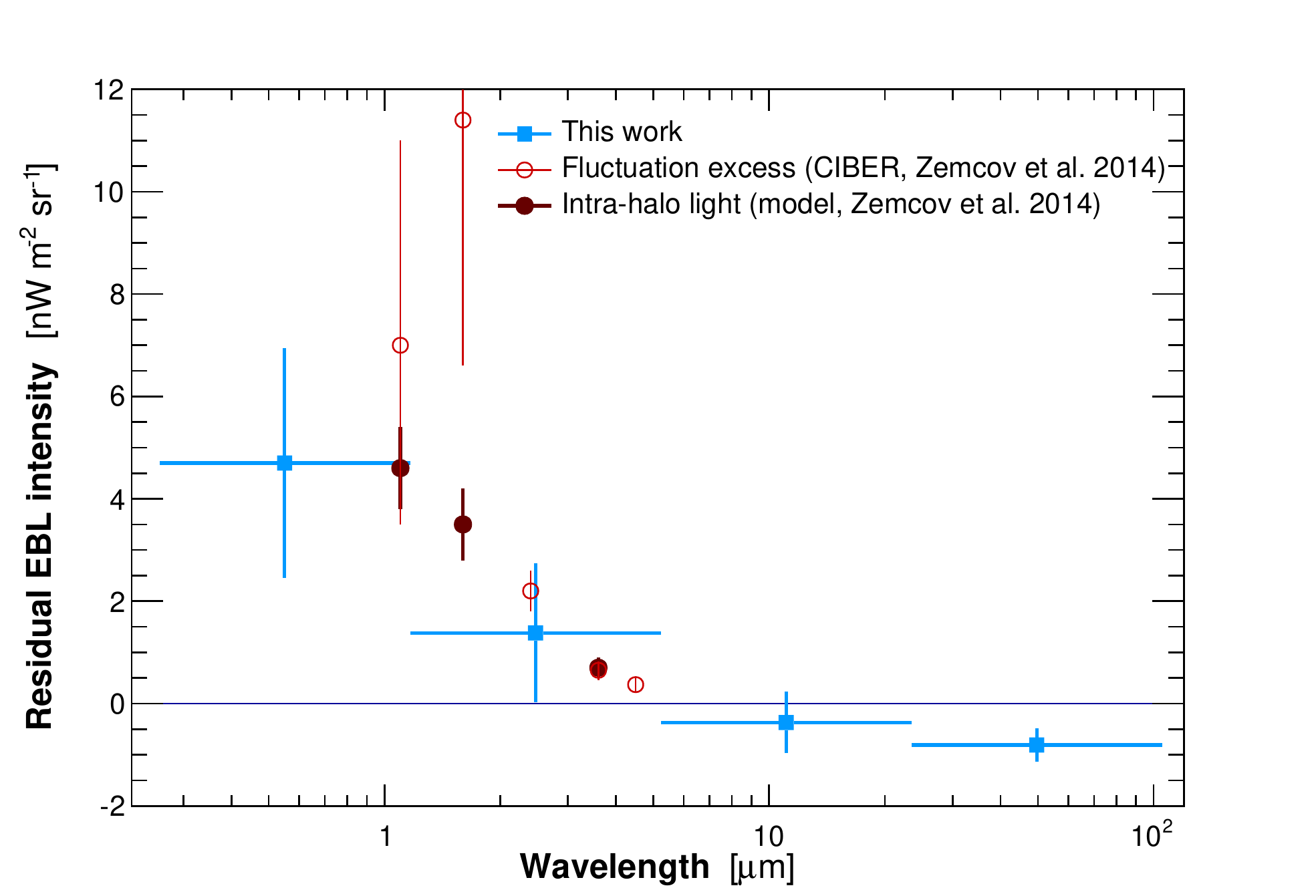}
\caption{Difference between the best-fit EBL spectrum derived from gamma-ray spectra only and the EBL estimates based on galaxy counts (blue squares). The anisotropy excess measured at $\unit[1.1]{\mu m}$ and $\unit[1.6]{\mu m}$ and a model of the IHL \citep[both from][]{2014Sci...346..732Z}  are shown with red-empty and dark-red filled points, respectively.}
\label{fig:EBLres}
\end{figure}

We measure an overall excess $\Delta\nu I_\nu = \unit[-0.8\pm0.4]{nW\ m^{-2}\ sr^{-1}}$, with a mild significance at the $\unit[2.0]{\sigma}$ level based on a likelihood ratio test. The slightly low gamma-ray estimate of the EBL is mostly due to the wavelength range above $\unit[25]{\mu m}$, with values in the same units of $4.7\pm2.2$, $1.4\pm1.4$, $-0.4\pm0.6$, and $-0.8\pm0.3$, in the wavelength ranges $\unit[0.26-1.2]{\mu m}$, $\unit[1.2-5.2]{\mu m}$, $\unit[5.2-23]{\mu m}$, and $\unit[23-105]{\mu m}$, respectively, as shown in Fig.~\ref{fig:EBLres}. The coarse binning of the EBL spectrum based only on gamma-ray data is possibly responsible for the mildly negative excess, as no tension is observed in the eight-point spectrum when taking into account the local constraints. We note that the model developed for the excess of anisotropy at $\unit[1.1]{\mu m}$ and $\unit[1.6]{\mu m}$ detected with CIBER by \cite{2014Sci...346..732Z} \citep[see also][for a critical review]{2015ApJ...804...99K}, explained as originating from stars stripped from their parent galaxies during mergers, cannot be ruled out by our results given the uncertainties from both approaches. Little room is left for contributions from other unknown populations of sources, especially above $\unit[1.2]{\mu m}$. This wavelength range is of particular interest for the study of the sources of reionization and we exclude significant contributions from miniquasars and Pop. III stars as presented e.g. in \cite{2004MNRAS.351L..71C}. We also exclude more exotic scenarios where Pop. III stars would experience a long ``Dark Star" phase powered by WIMPs \citep[see in particular Fig. 3 in][]{2012ApJ...745..166M}.

\subsection{Hubble constant}
\label{Sec:Hubble}

The possibility to constrain the expansion rate of the universe using gamma-ray observation of distant sources was first proposed by \cite{1994ApJ...423L...1S}. The idea is rather simple: measurements of the EBL optical depth, $\tau$, are proportional to $\nu I_\nu/H_0$, as shown in Eq.~\ref{Eq:Opt3}, while direct observations provide estimates of the EBL intensity $\nu I_\nu$, the combination of which thus constrains $H_0$.

Such an approach has been pursued e.g. by \cite{2008MNRAS.389..919B} who fixed the EBL intensity within estimates based on galaxy counts. Following a similar approach, we assume here that the EBL intensity can be described by the lower limits in Table~\ref{Tab:EBLdata}. This assumption follows the results from Sec.~\ref{Sec:EBLSpec}, showing a rather good agreement between gamma-ray data and galaxy counts. 

We hereafter use solely gamma-ray data to estimate $\nu I_\nu/H_0$. Calling $H$ the true value of the Hubble constant and $H_0 = \unit[70]{km\ s^{-1}\ Mpc^{-1}}$ the value used to derive the EBL parameters, $A_0$, we define the marginalized likelihood over the EBL parameters:
\begin{align}
\mathcal{L}(H) = \int{\rm d}A\ {\rm e}^{-\frac{1}{2}\left[\frac{H_0}{H}A - A_0\right]^{\rm T}V_{A_0}^{-1}\left[\frac{H_0}{H}A - A_0\right]}\times {\rm e}^{-\frac{\chi^2_{EBL}(A)}{2}}
\label{Eq:Hubble}
\end{align}
where $\chi^2_{EBL}(A)$ assesses the compatibility of the set of parameters $A$ with local constraints on the EBL, as described in Sec.~\ref{Sec:TS}. 

\begin{figure}
\hspace{-0.2cm}
\includegraphics[width=0.535\textwidth]{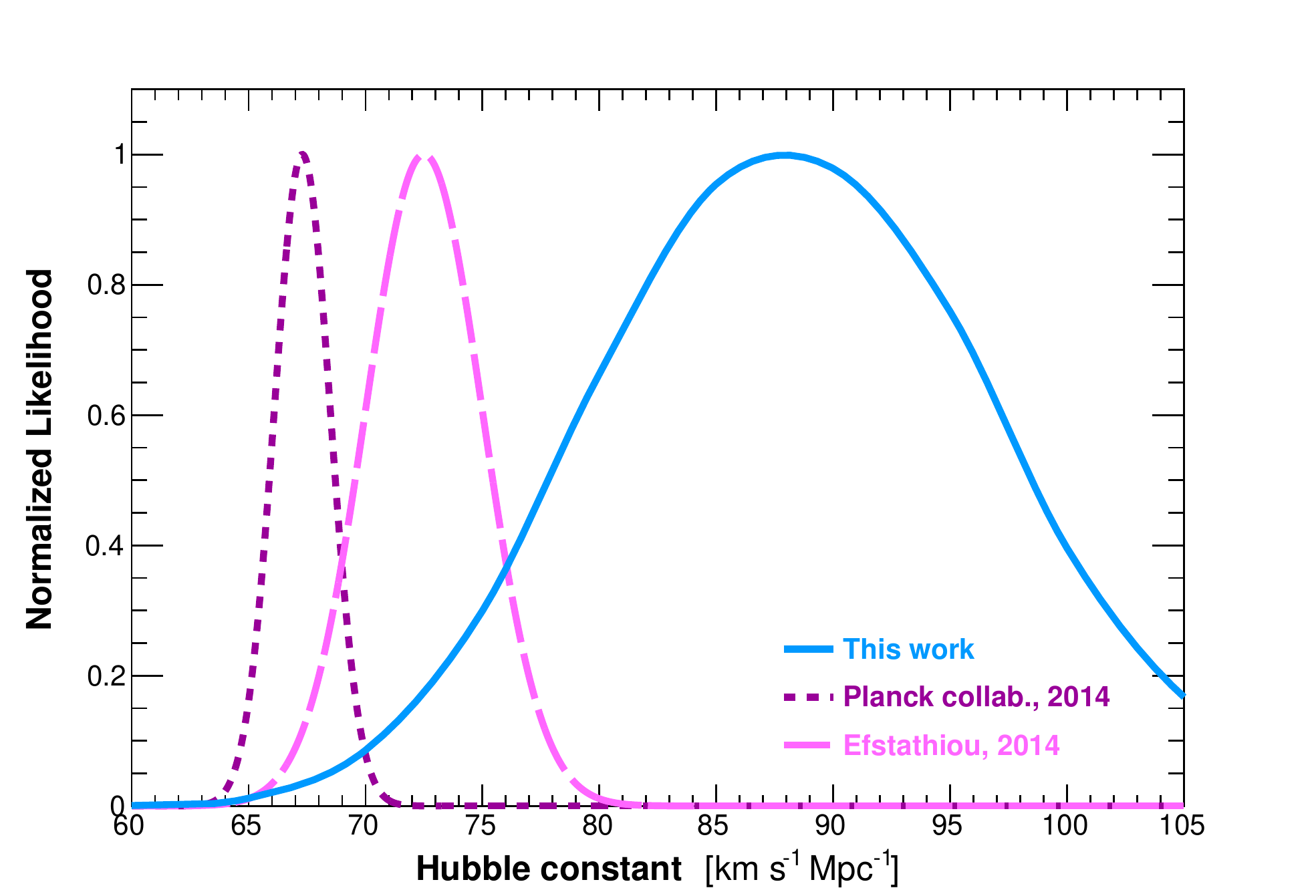}
\caption{Likelihood distribution of the Hubble constant. The estimate based on the comparison of gamma-ray data and local constraints is shown in solid blue. For reference, this estimate is compared to the best-fit values based on CMB observations and the local cosmic distance ladder.}
\label{fig:4}
\end{figure}

The marginalized likelihood shown in Fig.~\ref{fig:4} yields an estimate of $H_0 =  \unit[88 \pm 8_{\rm stat} \pm 13_{\rm sys}\ ]{km\ s^{-1}\ Mpc^{-1}}$. The systematic uncertainty is propagated from the optical depth ($\unit[7-8]{\%}$, see Appendix~\ref{App:Sys}), with $H_0 \propto 1/\tau$, and then added in quadrature to the bias expected from an excess of intensity with respect to galaxy counts. An offset of  $\unit[-0.8]{nW\ m^{-2}\ sr^{-1}}$, as determined in Sec.\ref{Sec:EBLSpec}, yields an estimate of the Hubble constant $\unit[11]{km\ s^{-1}\ Mpc^{-1}}$ lower than the best-fit value. Accounting for both statistical and systematics uncertainties on this measurement, no tension larger than $\unit[1.4]{\sigma}$ is observed with constraints based on the cosmic distance ladder \citep{2014MNRAS.440.1138E} or CMB-based measurements \citep{2014A&A...571A...1P}. 

\cite{2013ApJ...771L..34D} noticed that not only the distance term but also the density of photons used in the optical depth computation could depend on the Hubble constant for a given set of galaxy observations. Within such a formalism, the evolution parameter $f_{\rm evol}$ would be a function of $H_0$, with a typical span of $0.5<f_{\rm evol}<2.5$ for $0.5<H_0/\unit[100]{km\ s^{-1}\ Mpc^{-1}}<0.9$ \citep{PrivateDominguez}. Marginalizing the likelihood over this range of evolution parameters yields mildly larger uncertainties ($H_0 =  \unit[88 \pm 13_{\rm stat} \pm 13_{\rm sys}\ ]{km\ s^{-1}\ Mpc^{-1}}$), but does not affect our conclusions. We note that the result of \cite{2013ApJ...771L..34D}, $H_0 =  \unit[71\ ^{+ 4.6}_{- 5.6 {\rm (stat)}}\ ^{+ 7.2}_{- 13.8 {\rm (sys)}}]{km\ s^{-1}\ Mpc^{-1}}$, exploiting the cosmic gamma-ray horizon \citep{2005APh....23..608B} and a fixed EBL spectral shape at $z=0$ remains the most competitive gamma-ray estimate of the Hubble constant.

\subsection{Source redshifts}
\label{Sec:Redshift}

A similar approach to that devised in Sec.~\ref{Sec:Hubble} can be used to constrain the distance of unknown redshift sources. In the following, we use gamma-ray absorption as a guide for selecting possibly conflicting spectroscopic estimates. We assume a Hubble constant fixed to its nominal value $H_0 = \unit[70]{km\ s^{-1}\ Mpc^{-1}}$ and we describe the EBL by its best-fit eight-point spectrum obtained in Sec.~\ref{Sec:EBLSpec} (combined gamma-ray and local constraints). The best-fit parameters $A_0$ and covariance matrix $V_{A_0}$ for the eight-parameter spectrum are provided in Appendix~\ref{App:BestFitAndCov}.

We marginalize over the EBL parameters within uncertainties and define the following likelihood for a given set of spectra from a single source:
\begin{align}
\mathcal{L}(z) = \int&{\rm d}A\ {\rm e}^{-\frac{1}{2}\left[A - A_0\right]^{\rm T}V_{A_0}^{-1}\left[A - A_0\right]} \nonumber \\ &\times {\rm e}^{-\frac{1}{2}\sum\left(\chi^2_{\rm \gamma ray\ points} + \chi^2_{\rm HE-VHE}\right)(A,z)}
\label{Eq:z}
\end{align}
where the sum is over the spectra of a single source. Note that the spectra used in this sum must be different from the ones used in the estimation of the EBL parameters $A_0$, to avoid double counting the same datasets. The unknown-redshift sources are not included in the gamma-ray cosmology sample used to determine $A_0$, which justifies our approach. All the intrinsic spectra studied in this section are well matched by intrinsic PWL models.

\begin{figure}
\hspace{-0.2cm}
\includegraphics[width=0.535\textwidth]{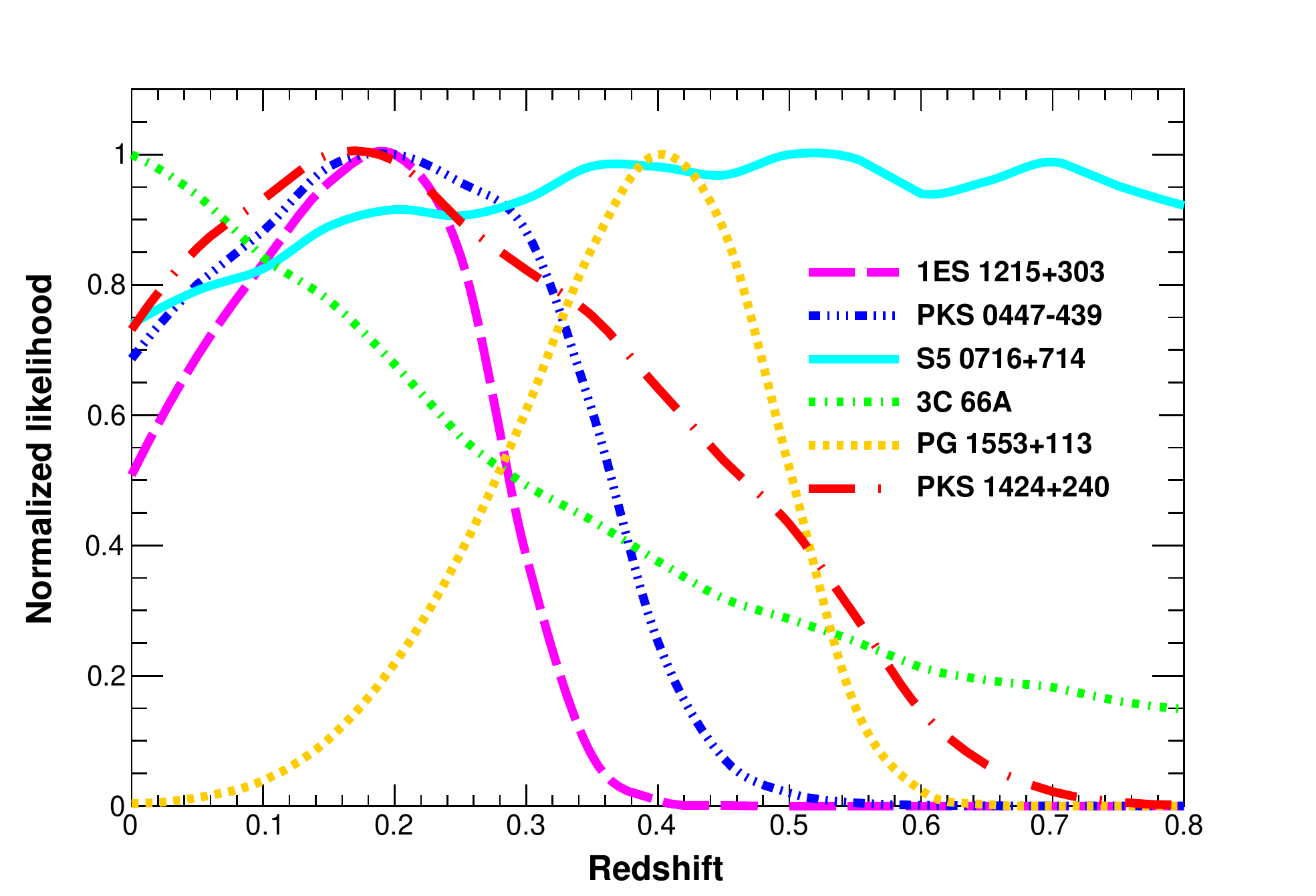}
\caption{Likelihood distribution of the redshift of six TeV blazars. Constraints are obtained after marginalization over the best-fit EBL parameters.}
\label{fig:5}
\end{figure}

Six sources with underconstrained distances are listed in Table~\ref{Tab:GammaData}. The likelihood distributions of the redshifts for the six sources are shown in Fig.~\ref{fig:5}.

\begin{itemize}
\item {1ES 1215+303}: for which we have gathered two spectra from MAGIC and VERITAS. Two spectroscopic estimates of the redshift of this HBL can be found in the literature: $z=0.130$ \citep{1998A&A...334..459B} and $z=0.237$ \citep{1993ApJS...84..109L}. \cite{2003ApJS..148..275A} show a spectrum and list the redshift as $0.130$, but it is not clear whether that redshift is supported by the spectrum or taken from the literature. As shown in Fig.~\ref{fig:5}, the likelihood profile for this source has a maximum at $z\sim0.2$. Because this estimate is compatible with zero, we provide an upper limit at the 99\% confidence level of $z<0.35$, and use in the following $z=0.237$.\\

\item {PKS 0447-439}: for which we have gathered  one spectrum from H.E.S.S. A spectroscopic estimate of $z=0.205$ was claimed by \cite{1998AJ....115.1253P}, based though on a rather weak feature that was not confirmed by further optical observations \citep[e.g.][]{2014A&A...565A..12P}. \cite{2015A&A...574A.101M} recently estimated the redshift of this source to be $z=0.343$, based on the observation of neighbouring galaxies possibly belonging to the same cluster. The H.E.S.S. spectrum has been used by several teams to constrain the distance of this source, with claimed measurements of $z=0.16\pm0.05$ and $z=0.20\pm0.05$ \citep[respectively]{2012A&A...543A.111P,2014PASJ...66...12Z}. Such small uncertainties, related to limiting assumptions on the EBL model or on the properties of the intrinsic spectra of TeV blazars, are not confirmed by the \cite{2013AA...552A.118H}, which provides an upper limit at the 95\% level of $z<0.59$. Our profile for PKS 0447-439 peaks at  $z\sim0.2$ and we obtain $z<0.45$ at the 99\% confidence level. In the following, we use $z=0.343$ for this source, noting that $z=0.205$ yields compatible results given the broadness of the likelihood profile.\\

\item {S5 0716+714}: for which we have gathered one spectrum from MAGIC. \cite{2013ApJ...764...57D} detected a system on the line of sight and have also set an upper limit based on the non-detection of lines farther away, constraining the redshift of this IBL within $0.232\leq z<0.322$. This range is consistent with the spectroscopy of three galaxies possibly hosted by the same cluster as the source, around $z\sim0.26$ \citep{2006ARep...50..802B}. In our study, no conclusion can be drawn for S5~0716+714, which shows a very broad maximum around $z\sim0.5$. In the following, we use the cluster-based $z=0.26$ as the redshift estimate of this source.\\

\item {3C 66A}: for which we have gathered one spectrum from MAGIC. \cite{2013ApJ...766...35F} detected an absorber on the line of sight and have constrained the maximum distance of this IBL based on the non-detection of farther lines, finding $0.335\leq z<0.41$. \cite{2010arXiv1006.4401Y} studied the gamma-ray emission of the object and set an upper limit at $z<0.58$. For 3C~66A, the likelihood in Fig.~\ref{fig:5} peaks at a rather low redshift  ($z\sim0$) and indicates a distance fully compatible with zero within uncertainties. No upper limit on the redshift of 3C~66A can be obtained given the broadness of the distribution. In the following, we use the spectroscopic lower-limit of $z=0.335$, which is only in mild tension with the likelihood profile (at the $\unit[1]{\sigma}$ level)\\

\item {PG 1553+113}: for which we have gathered two spectra from H.E.S.S. and three from MAGIC. \cite{2010ApJ...720..976D} constrained the redshift of this HBL through spectroscopic observations to $0.433\leq z<0.58$, though the lower limit is based on a single line. The authors obtained $z\geq0.395$ based on more numerous absorbers. Based on gamma-ray observations, \cite{2010arXiv1006.4401Y} determined $z<0.78$. Fixing the EBL absorption to the model of \cite{FR08}, the H.E.S.S. collaboration \citep{2015ApJ...802...65A} measured $z= 0.49 \pm 0.04$, or $z<0.56$ at the $\unit[95]{\%}$ confidence level, compared to $z<0.60$ and $z<0.62$ from the MAGIC and VERITAS collaborations, respectively \citep[][]{2014arXiv1408.1975M,2015ApJ...799....7A}. Our study accounts for the uncertainties on the EBL and thus yields a more conservative gamma-ray estimate of the redshift of this source, with most likely value $z=0.41_{-0.11}^{+0.08}$, preferred at the $\unit[3.4]{\sigma}$ level to a null redshift. For comparison with the work of \cite{2015ApJ...802...65A}, we obtain a $\unit[95]{\%}$ upper-limit on the redshift of the source $z<0.53$ ($z<0.58$ at $\unit[99]{\%}$). In the following, we set the redshift of PG 1553+113 to $z=0.433$.\\

\item {PKS 1424+240}: for which we have gathered three spectra from VERITAS and three from MAGIC. \cite{2013ApJ...768L..31F} determined a spectroscopic lower limit on the redshift of this object, $z\geq0.604$. \cite{2011arXiv1101.4098P} claim a measurement of $z=0.24\pm0.05$, with small uncertainties coming from the same limiting assumptions as for PKS 0447-439. \cite{2010arXiv1006.4401Y} provide a more robust upper-limit at $z<1.19$. As for 3C~66A, the likelihood profile for PKS 1424+240 is fully compatible with zero redshift, but an upper limit can be set at $z<0.64$ at the 99\% confidence level. In the following, we use the spectroscopic lower-limit of $z=0.604$, which is in slight tension ($\unit[2.4]{\sigma}$) with the likelihood profile.
\end{itemize}

The four constraints on redshifts derived in this section are the most stringent gamma-ray upper limits to date for these sources.

\subsection{Axion-like particles}
\label{Sec:ALP}
  
With a best-fit EBL spectrum at hand and redshift estimates for all sources, we can search for any siginificant residual, which might be indicative of new physics. Though we restricted the study to a gamma-ray cosmology sample of spectra for the determination of the EBL and of the Hubble constant, we study here all the spectra listed in Table~\ref{Tab:GammaData}, with spectral models provided in column 5.

We investigate in this section deviations from classical absorption resulting from coupling of gamma rays with axion-like particles. Gamma rays from blazars could convert into such hypothetical particles (cousins of the QCD axion with free mass and coupling) and could then convert back into gamma rays within the Galactic magnetic field. Several locations have been studied for the initial conversion, be it within the magnetic field of the emission region, in the vicinity of the source (host, jet galaxy, cluster) or in the intergalactic medium \citep[see e.g.][]{2007PhRvL..99w1102H, 2007PhRvD..76l1301D, 2009PhRvD..79l3511S, 2012PhRvD..86h5036T}. An observable effect would be a  gamma-ray spectral hardening at the highest optical depth, which has been weakly observed by several analyses \citep{2011JCAP...11..020D,2012JCAP...02..033H, 2013PhRvD..87c5027M, 2014JETPL.100..355R}. We note nonetheless that a large fraction of the parameter space corresponding to these hints has been excluded by \cite{2013PhRvD..88j2003A} using the lack of point-to-point fluctuations in the spectrum of the bright blazar PKS~2155-304.

The spectral points and best-fit models are shown in Appendix~\ref{App:IndivSpectra}. We show in Fig.~\ref{fig:6} the residuals to the best-fit models as a function of optical depth for these 737 points. All of the spectra are well represented by our models. The residuals are scattered around an average of $0.05\pm 0.03$ with an rms of $0.73\pm0.02$ and are well represented by a normal distribution, with a $p$ value of $\unit[22]{\%}$ following an Anderson-Darling test. The rms is significantly smaller than $1$, which is consistent with the small $\chi^2$ found in Sec.~\ref{Sec:EBLSpec}. 

\begin{figure}
\hspace{-0.2cm}
\includegraphics[width=0.535\textwidth]{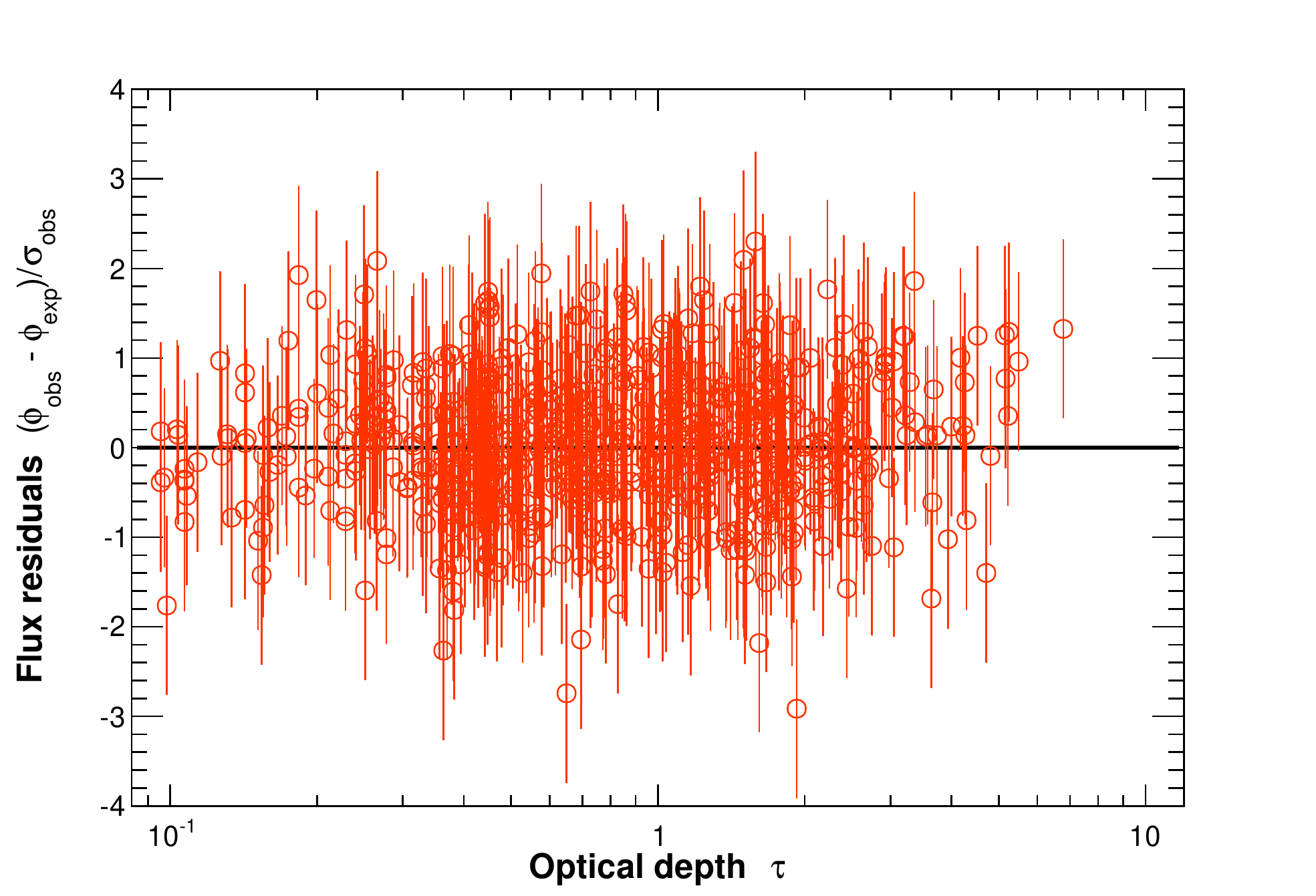}
\caption{Residuals to the best-fit models for the 106 spectra (737 points) studied in this paper, as a function of optical depth.}
\label{fig:6}
\end{figure}

In the following, we search for a hardening of the TeV spectra at the highest optical depths as in \cite{2012JCAP...02..033H}, while leaving model-dependent constraints on the axion-like particles' coupling to gamma rays to future studies. \cite{2012JCAP...02..033H} model a sample of 50 spectra and assume an EBL intensity fixed to the model of \cite{2010A&A...515A..19K}. Noting an rms of the flux residuals smaller than one, and deviations from a normal distribution based on an Anderson-Darling test, \cite{2012JCAP...02..033H} chose to ignore uncertainties on the data and to study the following quantity:
\begin{equation} 
R =\frac{\phi_i - \phi_{{\rm model},i}}{\phi_i + \phi_{{\rm model},i}}
\end{equation} 
where $\phi_i$ is the observed flux for the point $i$ and $\phi_{{\rm model},i}$ is the expected flux resulting from the model.

Comparing the distribution of $R$ in a reference sample composed of points at $1<\tau\leq2$ and in a search sample at $\tau>2$, the authors find a $\unit[4.2]{\sigma}$ discrepancy, with an average value of $R\sim0.25$ at large optical depth (estimated from the cumulative distribution function shown in Fig.~3 of their publication). Defining the flux-enhancement factor as $FE= \phi/\phi_{\rm model}$, the flux of TeV blazars would then exceed the expectations by $FE = (1+R)/(1-R) \sim 1.7$ above an optical depth of $\tau>2$. They explain this $\sim70\%$ increase in flux as an hint for mixing of gamma rays with axion-like particles. Performing the same test with our best-fit EBL spectrum and our larger dataset, we find a slightly larger discrepancy of $\unit[4.5]{\sigma}$, for an average $R\sim0.1$. However, we argue in the following that this test is flawed, because it neglects the uncertainties on the measured flux.

We measure the average flux enhancement $FE$ in various optical-depth bins, using a sample of gamma-ray spectra twice as large as any other studied. The residuals show a normal distribution, whose parameters are largely independent of redshift, energy, and optical depth. We thus include the uncertainties in the computation of the flux enhancement, weighting the relative contributions of the spectral points, as:
\begin{align}
\label{Eq:FE} 
FE &=\langle \frac{\phi} {\phi_{\rm model} } \rangle_i \nonumber \\
&= \frac{\sum_i \phi_i\phi_{{\rm model},i}/\sigma^2_{\phi,i} }{\sum_i \phi^2_{{\rm model},i}/\sigma^2_{\phi,i}}
\end{align} 
which is the usual $\chi^2$-based weighted average, propagating the uncertainty on the observed flux $\sigma_{\phi,i}$. Note that scaling the uncertainties on the observed flux up or down affects the errors on the flux enhancement estimate, but not the mean.

\begin{figure}
\hspace{-0.2cm}
\includegraphics[width=0.535\textwidth]{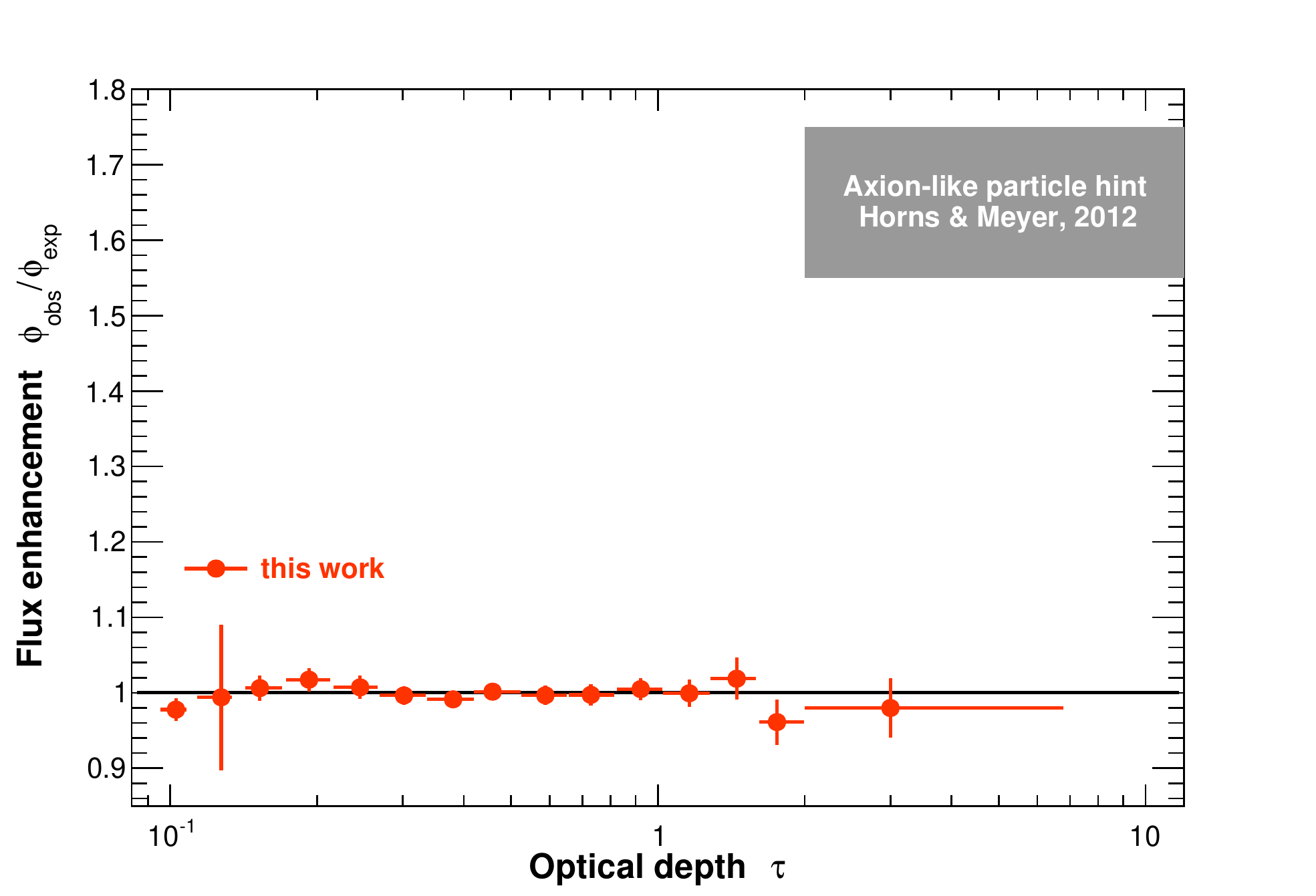}
\caption{Flux enhancement, defined by the ratio of observed and expected fluxes, as a function optical depth. The shaded gray region is the flux enhancement implied by the results of \cite{2012JCAP...02..033H}.}
\label{fig:7}
\end{figure}

The flux enhancement is shown as a function of optical depth in Fig.~\ref{fig:7}. The 106 spectra constrain the flux enhancement at optical depths larger than $\tau>2$ to $FE=0.98\pm0.04$, for an average optical depth $\tau=3.0$. This value, consistent with 1, shows no deviation from expectations. A flux enhancement of more than $\sim\unit[40]{\%}$ is excluded beyond the $\unit[5]{\sigma}$ level, taking only statistical uncertainties into account. Even assuming a systematic bias of $\unit[10]{\%}$ at large optical depth (see Appendix~\ref{App:Sys}), the flux enhancement corresponding to the results of \cite{2012JCAP...02..033H} (gray shaded region in Fig.~\ref{fig:7}) remains excluded at the $\unit[5]{\sigma}$ level. 

Applying the statistical test of \cite{2012JCAP...02..033H} to our dataset with our EBL spectrum therefore shows a more significant effect ($\unit[4.5]{\sigma}$ compared to $4.2$), albeit with smaller amplitude than they observe ($R \sim 0.1$ compared to $\sim 0.25$). However, we find little indication that the uncertainties on the individual flux measurements are unreliable, and when they are taken into account, we do not find a flux enhancement at large optical depths. The highest flux enhancements at optical depth above 2 in the sample studied here are obtained for the last points of the spectra of 1ES~1959+650 (Whipple, 2002), at $\tau=5.2$, 3C~279 (MAGIC, 2006), at $\tau=4.2$, and 1ES~0229+200 (H.E.S.S., 2005-2006), at $\tau=6.7$. They show flux enhancement of $13$, $14$ and even $23$, respectively, but with uncertainties on the order of $\unit[100]{\%}$ that strongly bias the test devised by \cite{2012JCAP...02..033H}. The most likely source of anomaly seen by these authors seems then to be the choice of statistical test rather than a specific dataset (their set of spectra is to a large extent included in Table~\ref{Tab:GammaData}). We note that \cite{2013PhRvD..87c5027M} studied the residuals normalized to uncertainties, as shown in Fig.~\ref{fig:6}, and found an average flux enhancement significant at the $\unit[4.4]{\sigma}$ level based on a $t$ test and using the EBL model of \cite{2010A&A...515A..19K}. Performing the same test with our sample yields a significance of $\unit[1.1]{\sigma}$ for the model of \cite{2010A&A...515A..19K} and $\unit[2.6]{\sigma}$ for the EBL spectrum derived in Sec.~\ref{Sec:EBLSpec}. The slightly different response of this test with respect to that shown in Fig.~\ref{fig:7} is possibly due to the weighting of the points in the averaging process ($\propto 1/\sigma_{\phi,i}$ for the $t$ test and $\propto 1/\sigma^2_{\phi,i}$ for a $\chi^2$-based average in Eq.~\ref{Eq:FE}). 

We conclude, based on the largest gamma-ray sample studied so far, that current VHE gamma-ray observations do not show a detectable flux enhancement at large optical depths and find little motivation for a lower limit on the coupling of axion-like particles with photons as reported in \cite{2013PhRvD..87c5027M}.

\subsection{Lorentz invariance violation}
\label{Sec:LIV}

We investigate in this section the compatibility of the 86 spectra in the gamma-ray cosmology sample with a quantum-gravity induced effect such as proposed by \cite{1999ApJ...518L..21K,2000PhRvD..62e3010A,2000PhLB..493....1P,2001PhRvD..63l4025E,2001PhRvD..64c6005A,2001APh....16...97S}. These authors conducted their investigations following the publication of the spectrum of Markarian~501 observed by HEGRA in 1997 \citep{1999AA...349...11A}, but the limited constraints on the EBL and on blazars' spectra, which were only a handful in 2000, prevented a quantitative constraint.  This spectrum is not included in our study because it was updated in \cite{2001AA...366...62A}.

We model the effect of a Lorentz invariance violation (LIV) adopting the formalism of \cite{2008PhRvD..78l4010J}. The starting point consists in a leading-order modification of the special relativistic relation $E^2 = p^2c^2 +m^2c^4$, where $E$ and $p$ are the energy and momentum of a particle of mass $m$. The effect should become significant around a quantum-gravity energy scale $E_{\rm QG}$, a correction of order $n=1,2$ yielding:

\begin{equation}
\label{Eq:ModDisp}
E^2 = p^2c^2 + m^2c^4 \pm E^2\left(\frac{E}{E_{\rm QG}}\right)^n
\end{equation}  
so that the norm of the momentum four-vector, $E^2 - p^2c^2$, is not a Lorentz invariant any more.

Equation~\ref{Eq:ModDisp} alters the kinematics of electron-positron pair creation, as in the collision of a TeV gamma ray interacting with an EBL photon. One needs to modify the special-relativistic threshold $\epsilon_{thr} > m_e^2 c^4 / E$, with $m_e$ the mass of the electron and where $\epsilon$ and $E$ are the energy of the two photons. The original spectrum of Markarian~501 published by HEGRA hinted at a gamma-ray absorption lower than predicted by classical interaction with the EBL, which corresponds to the subluminal case (``-" correction in Eq.~\ref{Eq:ModDisp}) that we study in the following. A superluminal correction (``+" in Eq.~\ref{Eq:ModDisp}) would yield either a lower threshold and pair creation on the CMB \citep{2008PhRvD..78l4010J} or photon decay if the dynamics are favorable \citep[e.g.][]{2010MPLA...25.3251S}, in which cases the absorption would be larger than classical. The investigation of the superluminal scenario, possibly requiring assumptions on the dynamics of the process, is left for future studies.

In the subluminal case, the energy-momentum conservation yields the modified pair-creation threshold:
\begin{equation}
\label{Eq:thr1}
\epsilon_{thr} = \frac{m_e^2c^4}{E}+\frac{1-2^{-n}}{4}\left(\frac{E}{E_{\rm QG}}\right)^n E
\end{equation}  
if Eq.~\ref{Eq:ModDisp} is applied to the photons and leptons. The term $1-2^{-n}$ should be replaced by $1$ if only the photons are affected, which is equivalent to considering a quantum-gravity energy scale twice as large for $n=1$. One can rewrite Eq.~\ref{Eq:thr1} as:
\begin{align}
\label{Eq:thr2}
\epsilon_{thr} &= \frac{m_e^2c^4}{E}\times\left[1+\left(\frac{E}{E_{\gamma, \rm LIV}}\right)^{n+2}\right]
\end{align}  
where
\begin{align}
\label{Eq:Eglin}
E_{\gamma, \rm LIV} &= \left[(2m_ec^2)^2E_{\rm QG}^{n}/(1-2^{-n})\right]^{1/(n+2)}\nonumber \\
&= \unit[29.4]{TeV}\times\left(\frac{E_{\rm QG}}{E_{\rm Planck}}\right)^{1/3}{\rm ,\ for\ }n=1
\end{align}  
with $E_{\rm Planck} = \sqrt{\hbar c^5/G} = \unit[1.22 \times 10^{28}]{eV}$. A leading quadratic correction, $n=2$, yields $E_{\gamma, \rm LIV} = \unit[120]{PeV}\times\sqrt{ E_{\rm QG}/E_{\rm Planck}}$, out of reach of current experiments for $E_{\rm QG}\sim E_{\rm Planck}$. We focus in the following on the linear case, $n=1$, but we derive the modified EBL optical depth in the general case as:
\begin{align}
\tau(E_0,z_0) =& \frac{3}{4} \frac{\sigma_T c}{H_0}  \int_0^{z_0} {\rm d}z\ \frac{\partial l}{\partial z}(z) \int_0^\infty {\rm d}\epsilon\frac{\partial n}{\partial \epsilon}(\epsilon,z) \nonumber \\ & \left(\frac{\epsilon_{thr}}{\epsilon}\right)^2 P\left(\sqrt{1-\frac{\epsilon_{thr}}{\epsilon}}\right)
\end{align}
where we follow the notations of Sec.~\ref{Sec:Theory}.

Assuming that the energy and redshift dependences of the EBL photon field can be decoupled and using the same changes of variable as in Sec.~\ref{Sec:Theory}, the LIV-affected EBL optical depth can be expressed as:
\begin{align}
\tau(E_0,z_0) =& \frac{3\pi\sigma_T}{H_0}\times \frac{E_0}{m_e^2 c^4}\times  \int_{-\infty}^\infty {\rm d}e_0\ \exp(-3e_0)
\nonumber\\
&\times\nu I_\nu\left(e_0 - \ln\frac{E_0}{m_e c^2}\right) \int_0^{z_0}{\rm d}z \frac{\partial l}{\partial z} \frac{evol(z)}{(1+z)^4}\nonumber\\
&\times \left[1+\left(\frac{(1+z)E_0}{E_{\gamma, \rm LIV}}\right)^{n+2}\right]P(\beta_{\rm max})
\end{align}
where
\begin{equation}
\beta_{\rm max}^2 = 1-\frac{\exp(-e_0)}{(1+z)^2} \left[1+\left(\frac{(1+z)E_0}{E_{\gamma, \rm LIV}}\right)^{n+2}\right]
\end{equation}

\begin{figure}
\hspace{-0.2cm}
\includegraphics[width=0.535\textwidth]{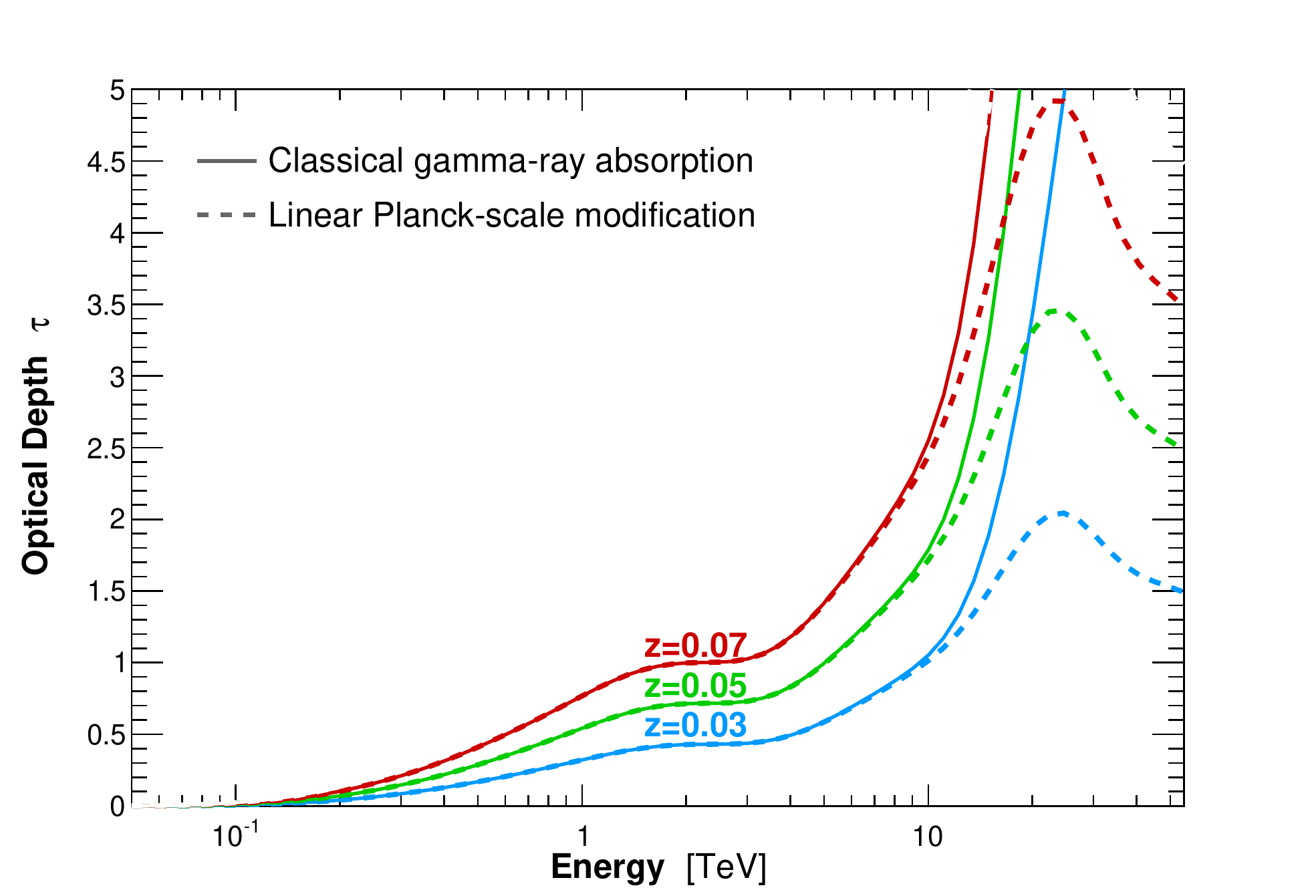}
\caption{EBL optical depth at three redshifts, in the classical case (solid lines) and in the case of a LIV modification at the Planck scale (dashed lines).}
\label{fig:OptDepth}
\end{figure}

We compare the LIV-affected optical depth to the classical one in Fig.~\ref{fig:OptDepth}, using the eight-point specific intensity of the EBL obtained in Sec.~\ref{Sec:EBLSpec}. The main effect of LIV on gamma-ray absorption is a reduction of the optical depth above $\unit[10]{TeV}$, largely independent of redshift. 

The approach developed in Sec.~\ref{Sec:Ana} consists in finding the best EBL spectrum jointly accounting for the absorption signature seen in gamma rays and for local constraints on the EBL. We add here an extra free parameter that is the energy scale $E_{\rm QG}$ at which LIV modifications of the pair-creation threshold take place. We vary $E_{\rm QG}$ and compute the best-fit $\chi^2$ accounting for both gamma-ray spectra and local constraints on the EBL. We define the test statistic $TS = \chi^2(E_{\rm QG},A_{\rm QG}) - \chi^2(\infty,A_{\rm \infty})$, where the latter $\chi^2$ is measured at the best-fit EBL level in the classical case, and where $A$ denotes the EBL parameters, left free to vary for each quantum-gravity scale. 

\begin{figure}
\hspace{-0.2cm}
\includegraphics[width=0.535\textwidth]{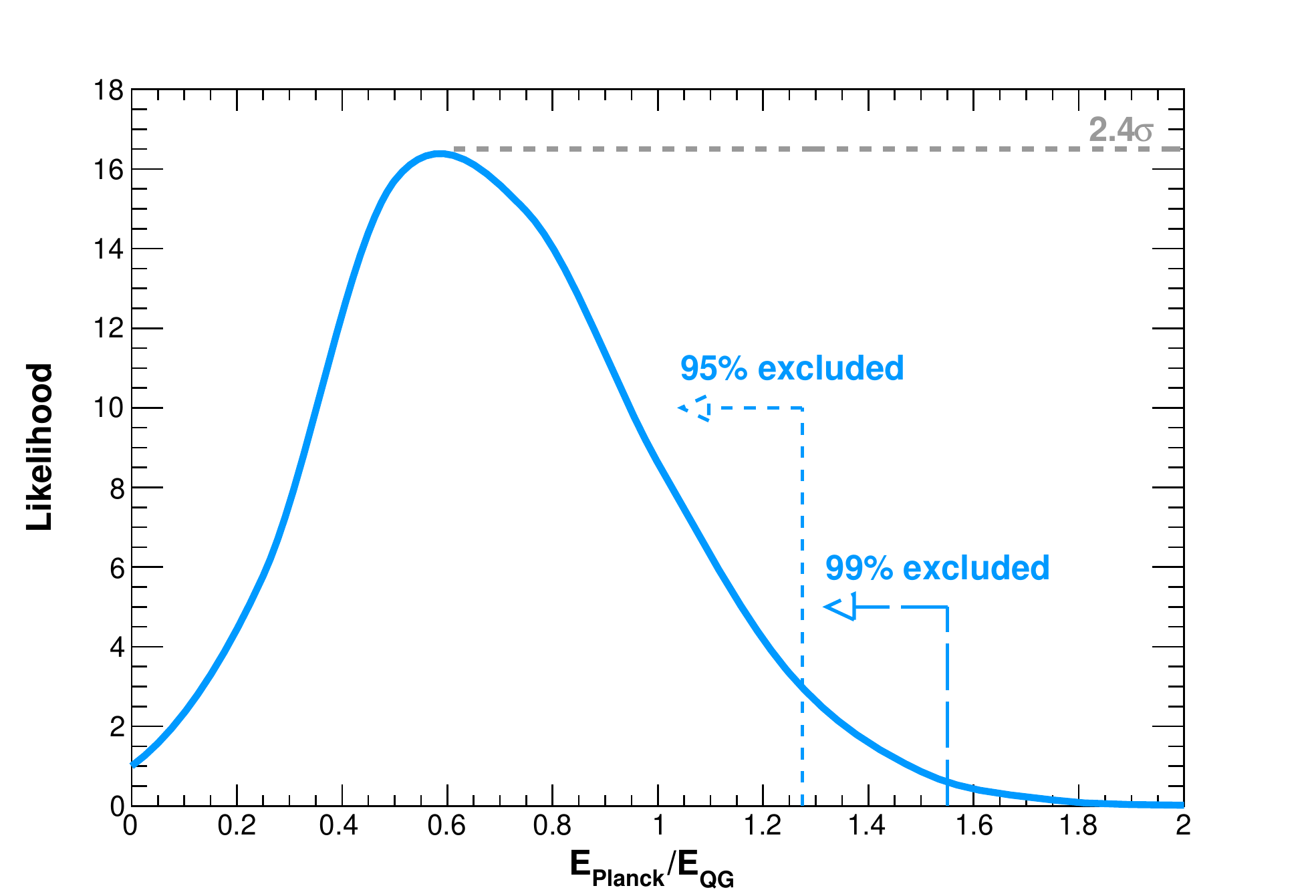}
\caption{Likelihood profile of the quantum-gravity energy scale, leaving the eight-point EBL spectrum free.}
\label{fig:LikelihoodLIV}
\end{figure}

Figure~\ref{fig:LikelihoodLIV} shows the likelihood profile, ${\cal L} = \exp(-TS/2)$, as a function of the inverse of $E_{\rm QG}$ normalized to the Planck energy. Interestingly, a slight excess can be seen around the the Planck scale, though with a significance of only $\unit[2.4]{\sigma}$. Our study is performed using a sample of 86 spectra that includes the very observation of Markarian~501 by HEGRA that triggered substantial theoretical developments on modifications of the pair-creation threshold. The original spectrum, which covers the energy range $\unit[0.56-22]{TeV}$, was reported in \cite{1999AA...349...11A}. The data were subsequently reanalyzed by \cite{2001AA...366...62A} with an improved energy resolution, yielding a compatible spectrum covering the energy range $\unit[3.3-21]{TeV}$, which is the spectrum included in our gamma-ray cosmology sample. Using instead the original spectrum enhances the Planck-scale excess to the $\unit[4]{\sigma}$ level. The smaller energy coverage of the higher-resolution spectrum results in a less constrained intrinsic emission and the observed spectrum has a somewhat sharper rollover at the highest energies, explaining the decrease in significance from $\unit[4]{\sigma}$ down to $\unit[2.4]{\sigma}$. Given the low significance of the excess, we provide lower limits on the quantum-gravity energy scale of $E_{\rm QG}>0.78\times E_{\rm Planck}$ ($\unit[95]{\%}$) and $E_{\rm QG}>0.65\times E_{\rm Planck}$ ($\unit[99]{\%}$). A $\unit[10]{\%}$ systematic uncertainty on the energy scale of gamma-ray instruments similarly impacts the quantum-gravity energy estimate, and we quote a robust lower limit accounting for the systematic uncertainties of $E_{\rm QG}>0.6\times E_{\rm Planck}$.

Other constraints on LIV have been derived from observations of high-energy sources. The synchrotron emission by electrons in the Crab nebula has been used, e.g. by \cite{2003Natur.424.1019J}, to constrain a linear subluminal term (the same as constrained here) to a quantum gravity scale seven orders of magnitude above the Planck scale. Some authors nonetheless argue that this test is not only kinematics but also depends on the dynamics of process, which are far from being understood within quantum-gravity phenomenology \citep{2013LRR....16....5A}. A suppressed cross section would then mitigate the constraints from the Crab nebula. A more pristine test comes from observations of gamma-ray bursts (GRBs) by {\it Fermi}, assuming that LIV affects the propagation time of photons in an energy-dependent way. The most recent constraints have been derived by \cite{2013PhRvD..87l2001V} who obtain, after taking into account possible internal delays in the source, a lower limit $E_{\rm QG} \geq 2\times E_{\rm Planck}$ from a single GRB (090510), and $E_{\rm QG} \geq 0.1\times E_{\rm Planck}$ from three others. 

One could conclude from the GRB~090510 limit than any modification of the pair-creation threshold is ruled out up to $2\times E_{\rm Planck}$, but this would be overlooking the fact that time-delay and absorption observations constrain two different processes. The underlying theory might indeed preserve the speed of light, not impacting time delay observations, but at the same time affect the dispersion relation of particles, leaving an imprint in blazars' spectra \citep[i.e. $v = \partial E / \partial p$ would no longer be valid, see][]{2013LRR....16....5A}. High-statistics observations above $\unit[10]{TeV}$ will be required to further study quantum-gravity effects, and the formalism we have developed in this section will prove useful to further constrain LIV with gamma-ray observations of blazars. 

\section{Conclusions}

A wealth of data has been published by ground-based gamma-ray observatories over the past two decades. We have compiled the most extensive set of gamma-ray spectra from VHE blazars to date, with 106 spectra from 38 objects, corresponding to a total of about 300,000 gamma rays. This is twice the size of any sample studied before. 

Our first result is purely analytical. We have discovered that the triple integral relating the gamma-ray optical depth to the EBL intensity can be reduced to a double integral without any loss of generality. We have further shown that, assuming a decoupling of the evolution of the EBL and of its rest-frame spectrum, the gamma-ray optical depth is the convolution of the EBL intensity with the EBL kernel.

This analytical work significantly reduces the complexity of the spectral reconstruction of the EBL based on gamma-ray observations. The decoupling approximation introduces rather small ($\lesssim\unit[5]{\%}$) systematic errors for sources in the local universe. Using a joint spectral analysis of a subsample of 86 spectra, we deconvolve the intrinsic emission of the sources from the imprint of the EBL spectrum. The reconstructed EBL intensity is preferred at the $\unit[11]{\sigma}$ level to the absence of gamma-ray absorption, and we reconstruct an eight-point spectrum covering the wavelength range $\unit[0.26-105]{\mu m}$, from mid-UV to far IR. The spectrum of the EBL based on gamma-ray observations is in good agreement with estimates based on galaxy counts, with uncertainties that leave some room for contributions from e.g. intra-halo light \citep{2014Sci...346..732Z}, while constraining the emission  of reionization sources such as Pop. III stars or miniquasars. The brightnesses of the COB and CIB are measured to be $\unit[36\pm11]{nW\ m^{-2}\ sr^{-1}}$ and $\unit[25.9\pm3.4]{nW\ m^{-2}\ sr^{-1}}$, respectively.  Once integrated between $\unit[0.1-1000]{\mu m}$, the EBL is $\unit[6.5]{\%}$ of the CMB, with an overall uncertainty on this number of $\unit[20]{\%}$. 

Our third result is a gamma-ray measurement of the Hubble constant, $H_0 =  \unit[88 \pm 8_{\rm stat} \pm 13_{\rm sys}\ ]{km\ s^{-1}\ Mpc^{-1}}$, that is both model independent and based on a significant number of gamma-ray spectra. Such constraints on the expansion rate of the universe are independent from measurements based on the CMB and the cosmic ladder of distances.

We measure no significant flux enhancement at large optical depths and we rule out at the $\unit[5]{\sigma}$ level the ``pair-production anomaly" as obtained by \cite{2012JCAP...02..033H}. This suggests that the level of mixing of axion-like particles with TeV photons, if any, is below previous estimates based on this effect. We would also like to correct two misconceptions sometimes circulating in the literature. (i) The best-fit EBL spectrum based on gamma-ray observations is not significantly lower than the minimum EBL level. (ii) The intrinsic gamma-ray spectra reconstructed after deabsorption from the EBL effect are not too hard with respect to expectations. All the intrinsic spectra in our fits are softer than measured at lower energy when contemporaneous data are available. This diminishes the motivation for scenarios where axion-like particles impact gamma-ray absorption, or where secondaries from UHECR contribute to the gamma-ray signal observed on Earth.

Finally, we constrain the impact of a Lorentz invariance violation on gamma-ray absorption. A modified dispersion relation, with a correction scaling as the ratio of the gamma-ray energy and the Planck energy, alters the threshold of pair creation and results in milder absorption of $>\unit[10]{TeV}$ gamma rays from blazars. Our formalism takes into account both the spectrum and the evolution of the EBL. It also enables a quantitative analysis of blazars' spectra in the presence of LIV. A weak $\unit[2.4]{\sigma}$ excess prevents us from ruling out a modification at the Planck energy, but we rule out for the first time any effect below $0.6 \times E_{\rm Planck}$ at the $\unit[99]{\%}$ confidence level.

The successes of gamma-ray cosmology \citep[see][for a review]{2013sf2a.conf..303B} are far from being closed by this work. Major progress in UV and IR observations of distant galaxies has been achieved in recent years, in particular with Galex, Spitzer, and Herschel. Further achievements are expected from the James Webb Space Telescope \citep[JWST,][]{2006SSRv..123..485G} during the next decade between $0.6$ and $\unit[27]{\mu m}$, with improved constraints from galaxy counts in the optical and mid-IR. The combination of next-generation local constraints with the tremendous gamma-ray sensitivity of the Chenrekov Telescope Array \citep[CTA,][]{2013APh....43....3A} above $\unit[30]{GeV}$ could significantly refine the binning of the EBL spectrum. The theoretical limit is dictated by the energy resolution of gamma-ray telescopes, about $\unit[10]{\%}$ in gamma-ray energy or equivalently $\unit[10]{\%}$ in EBL wavelength. Such fine spectroscopy could clearly show spectral signatures from polycyclic aromatic hydrocarbons, from intra-halo light, or from the sources of the reionization of the universe. The combined constraints of CTA and JWST will also be crucial for the measurement of the Hubble constant based on gamma-ray absorption.

Further developments of analyses at the event level (binned or unbinned), such as 3ML or GammaLib \citep{2013arXiv1307.5560K}, combined with full likelihood spectral techniques such as developed in \cite{2001A&A...374..895P}, could increase the statistical power of the approach developed in this paper. Such methods are in principle not affected by overestimated uncertainties from published spectra, where correlations are neglected, and would open the possibility of a broad-band spectral fit, e.g. combining the data from {\it Fermi}-LAT, Cherenkov telescopes, and HAWC \citep{2013APh....50...26A}. The subfield of gamma-ray cosmology focusing on the fate of the pairs produced by gamma-ray absorption, and how they could be impacted by the intergalactic magnetic field \citep[see][]{2013A&ARv..21...62D,2014arXiv1410.7717C}, will greatly benefit from these ongoing developments. Finally, we are working on the extension of the method that we have presented here to higher redshifts ($z>1$), where a simple parametrization of the EBL evolution starts to fail. If successful, significant improvements in the spectrum of the EBL below $\unit[1]{\mu m}$ can be expected from the analysis of {\it Fermi}-LAT blazars.

\acknowledgments

We would like to thank the corresponding authors of the many papers listed in Table 2 for providing the data points corresponding to their published results.  We thank the referee whose work significantly benefited this paper. We would also like to thank Joel Primack, Michael Dine, Stefano Profumo, Alberto Dominguez, Matthieu Bethermin, and Abelardo Moralejo for valuable discussions about this work, Amy Furniss for helping in the collection of the VERITAS spectra, and Steve Fegan for suggesting the use of a Gaussian-sum model for the EBL spectrum. We gratefully acknowledge support from the U.S. National Science Foundation award PHY-1229792.

\bibliography{ebl_hubble_anomalies.bbl}

\appendix
\section{Systematic uncertainties}\label{App:Sys}

We discuss in Sec.~\ref{App:EBLevol} and Sec.~\ref{App:GaussSum} the approximations to the spectrum and evolution of the EBL used in this paper. We estimate in Sec.~\ref{App:FulSys} the systematic uncertainty arising from these approximations as well as those coming from the modeling of the intrinsic spectra and the possible biases in the energy scale.

\subsection{EBL evolution}\label{App:EBLevol}

\begin{figure*}
\hspace{-0.2cm} \includegraphics[width=0.535\textwidth]{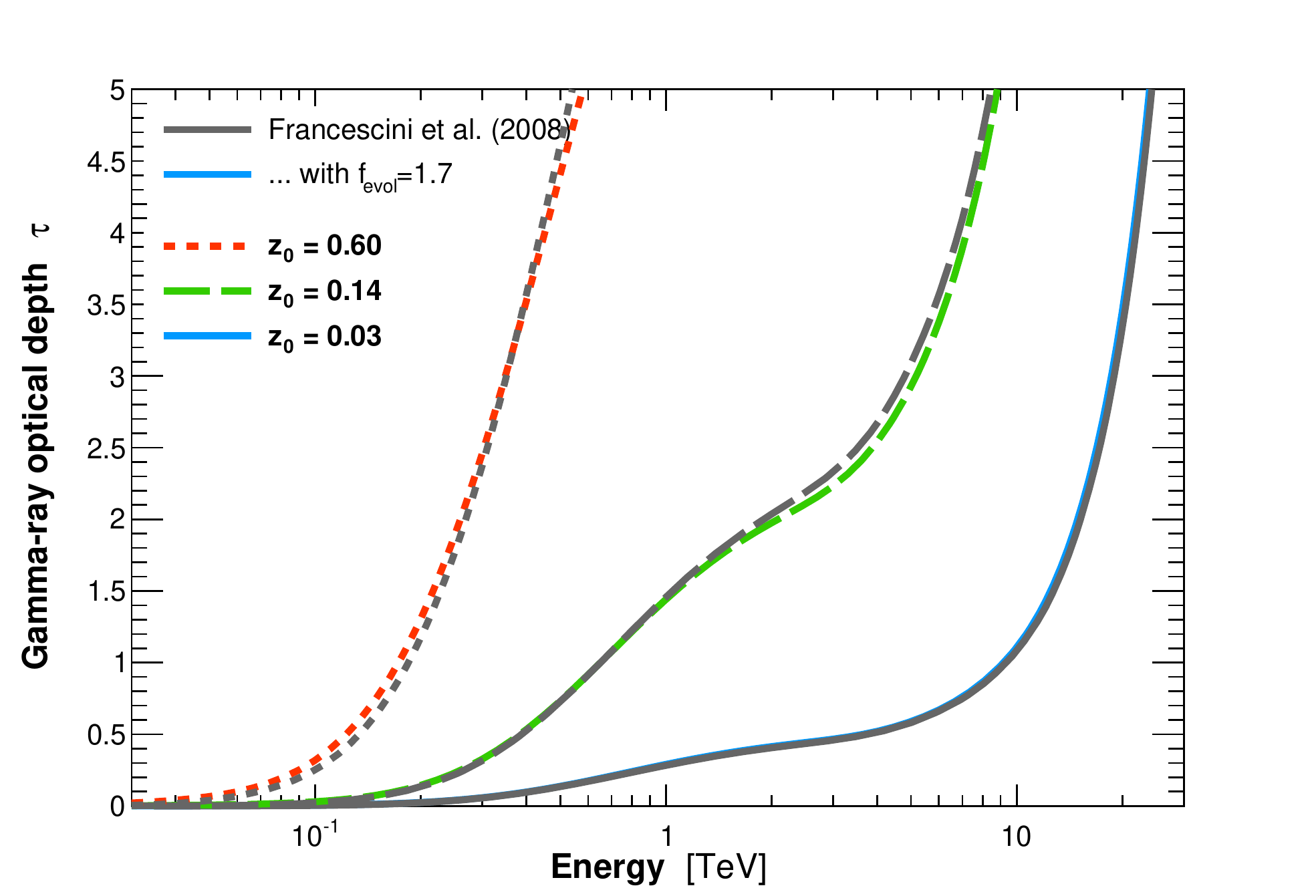}
\hspace{-0.2cm} \includegraphics[width=0.535\textwidth]{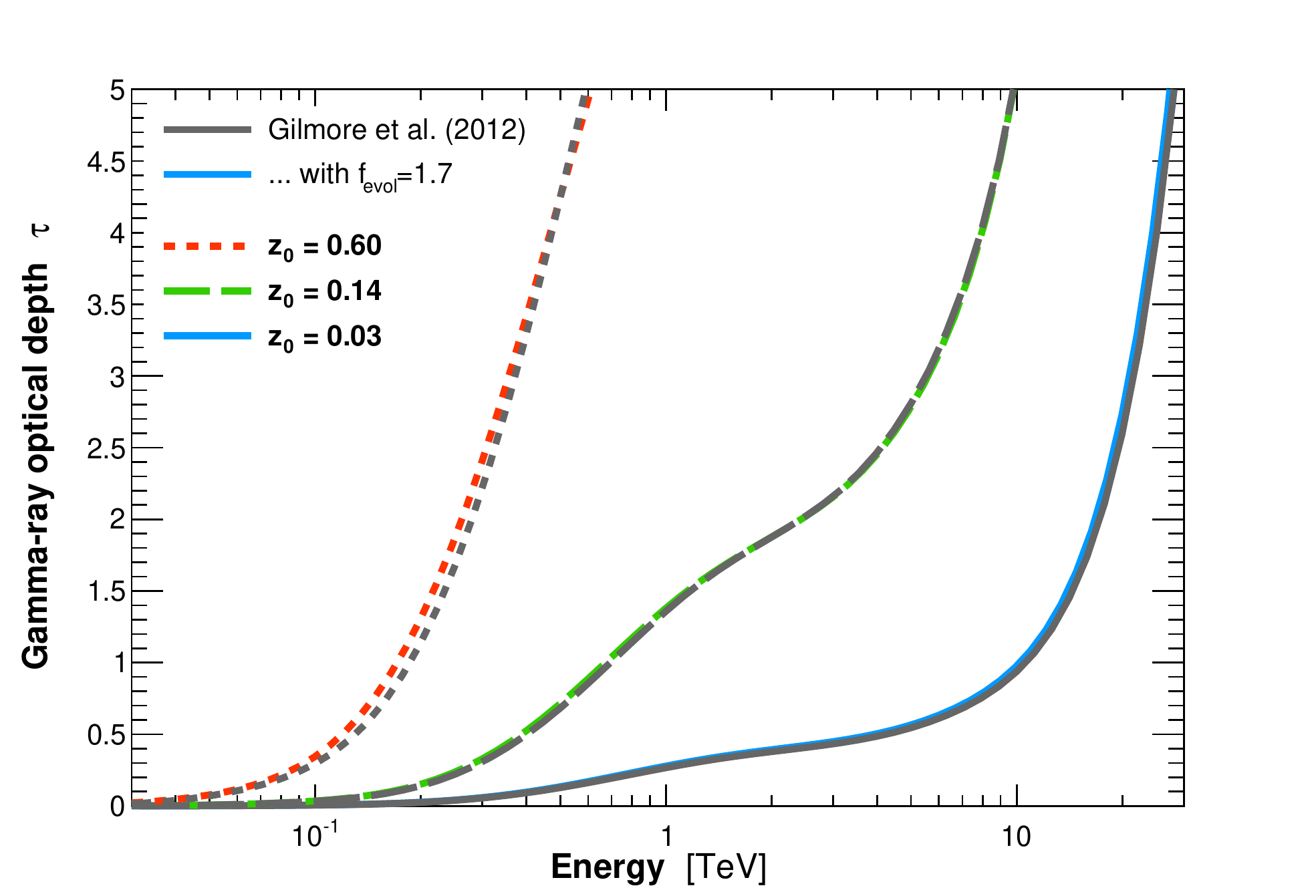}
\caption{Attenuation curves for three different redshifts, from $z=0.03$ up to $z=0.6$. Gray curves are directly extracted from the publications of \cite{FR08} ({\it left}) and \cite{G12} ({\it right}). Colored curves are based on the same $z=0$ EBL density as in the publications but assume a template EBL evolution with $f_{\rm evol} = 1.7$.}
\label{fig:Tau17F}
\end{figure*}

\begin{figure*}
\hspace{-0.2cm} \includegraphics[width=0.535\textwidth]{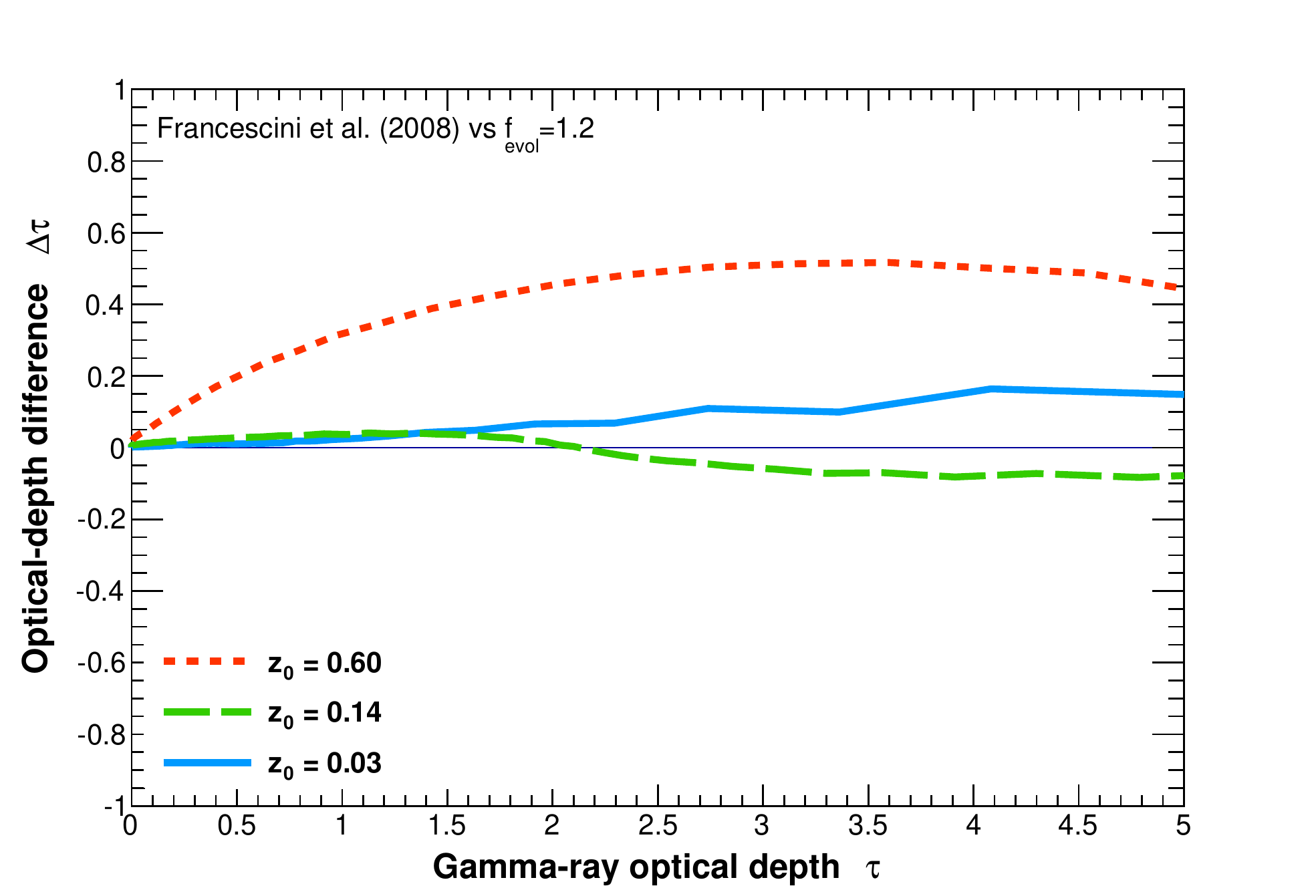}
\hspace{-0.2cm} \includegraphics[width=0.535\textwidth]{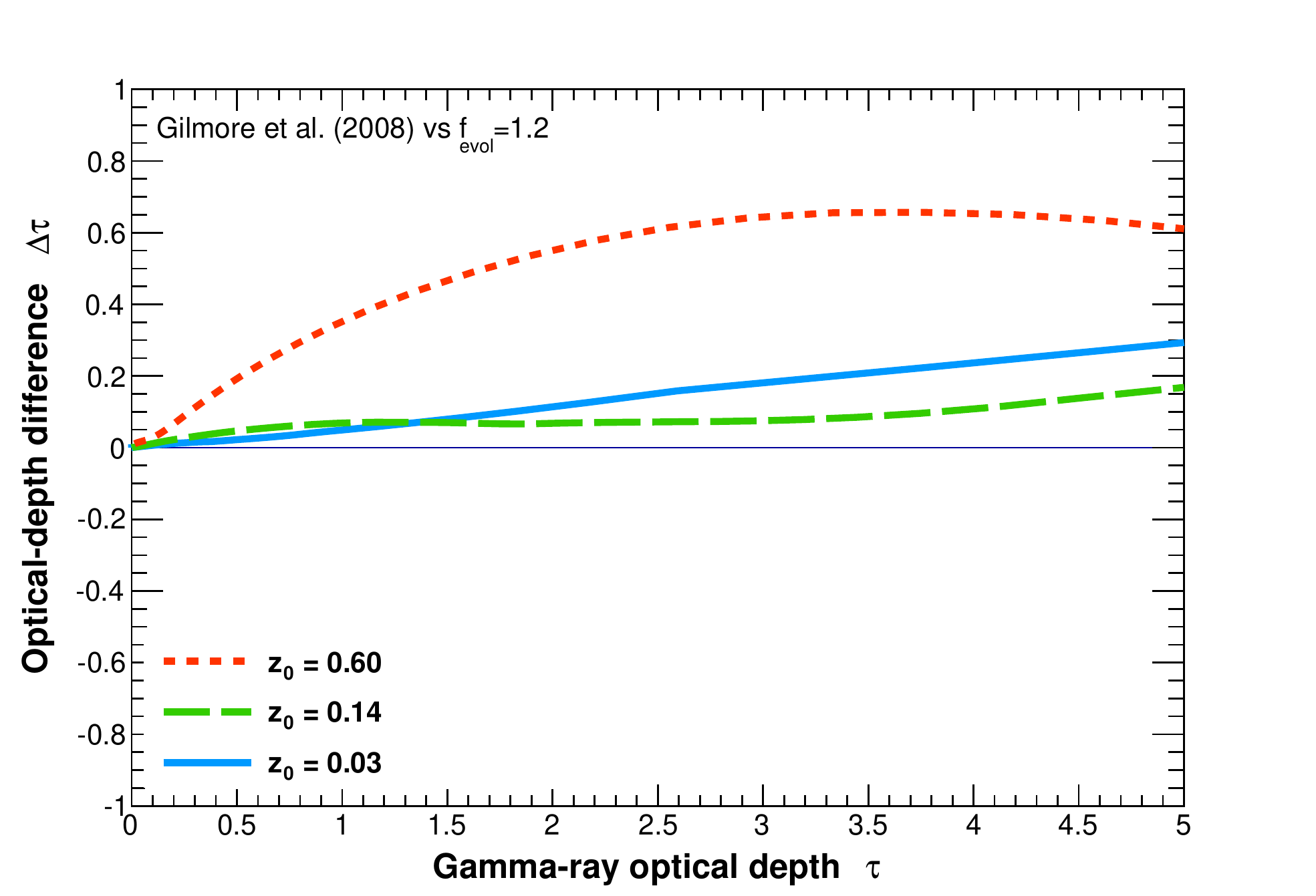}
\hspace{-0.2cm} \includegraphics[width=0.535\textwidth]{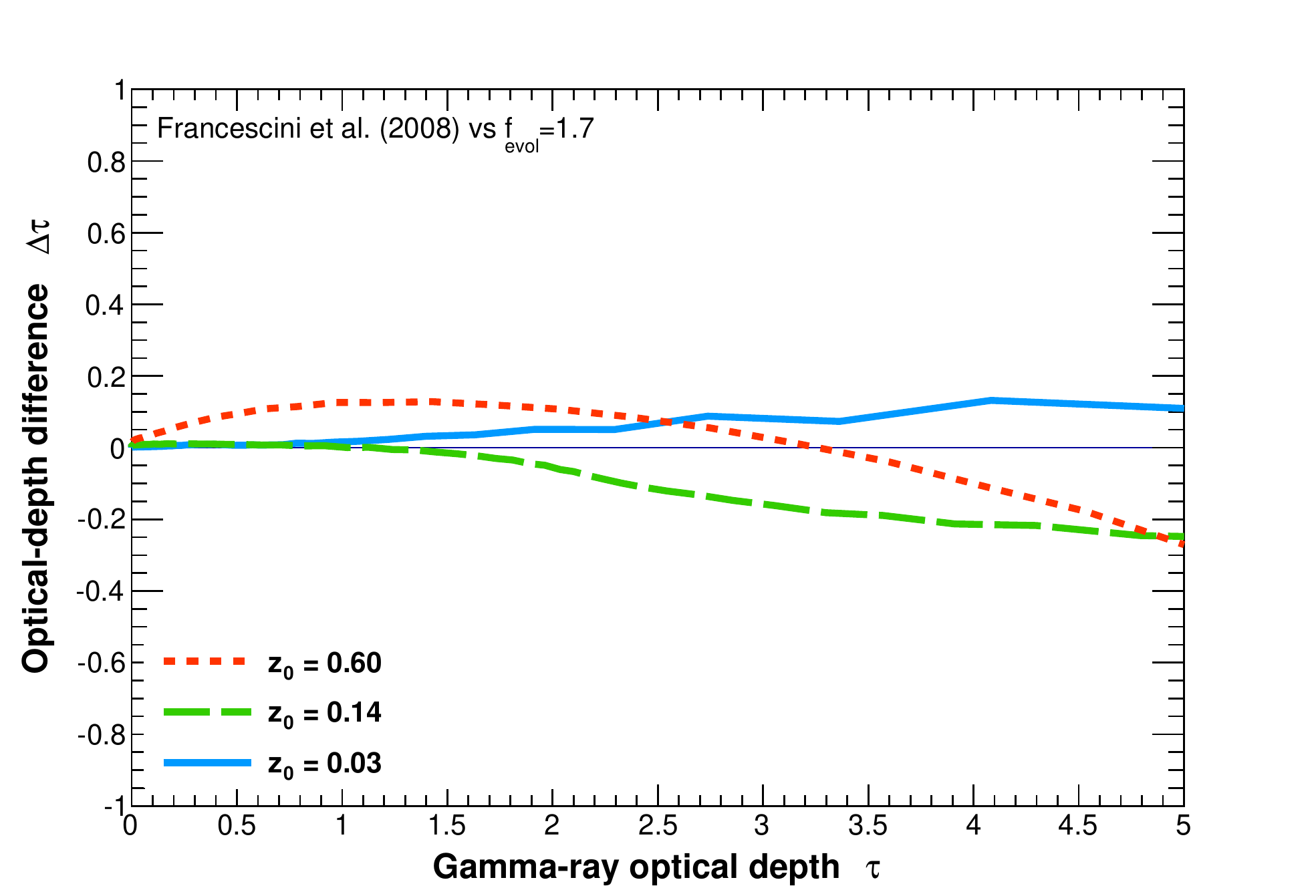}
\hspace{-0.2cm} \includegraphics[width=0.535\textwidth]{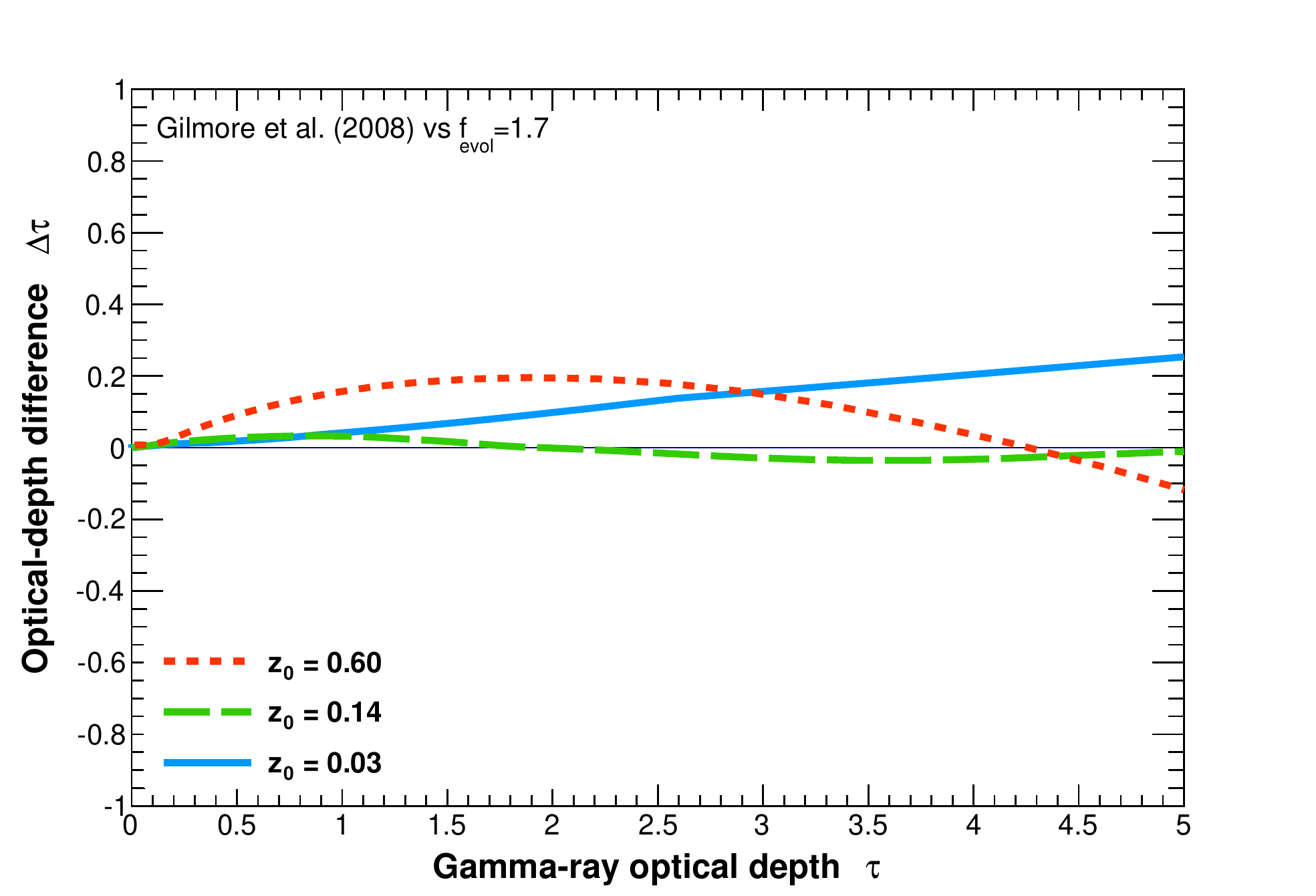}
\hspace{-0.2cm} \includegraphics[width=0.535\textwidth]{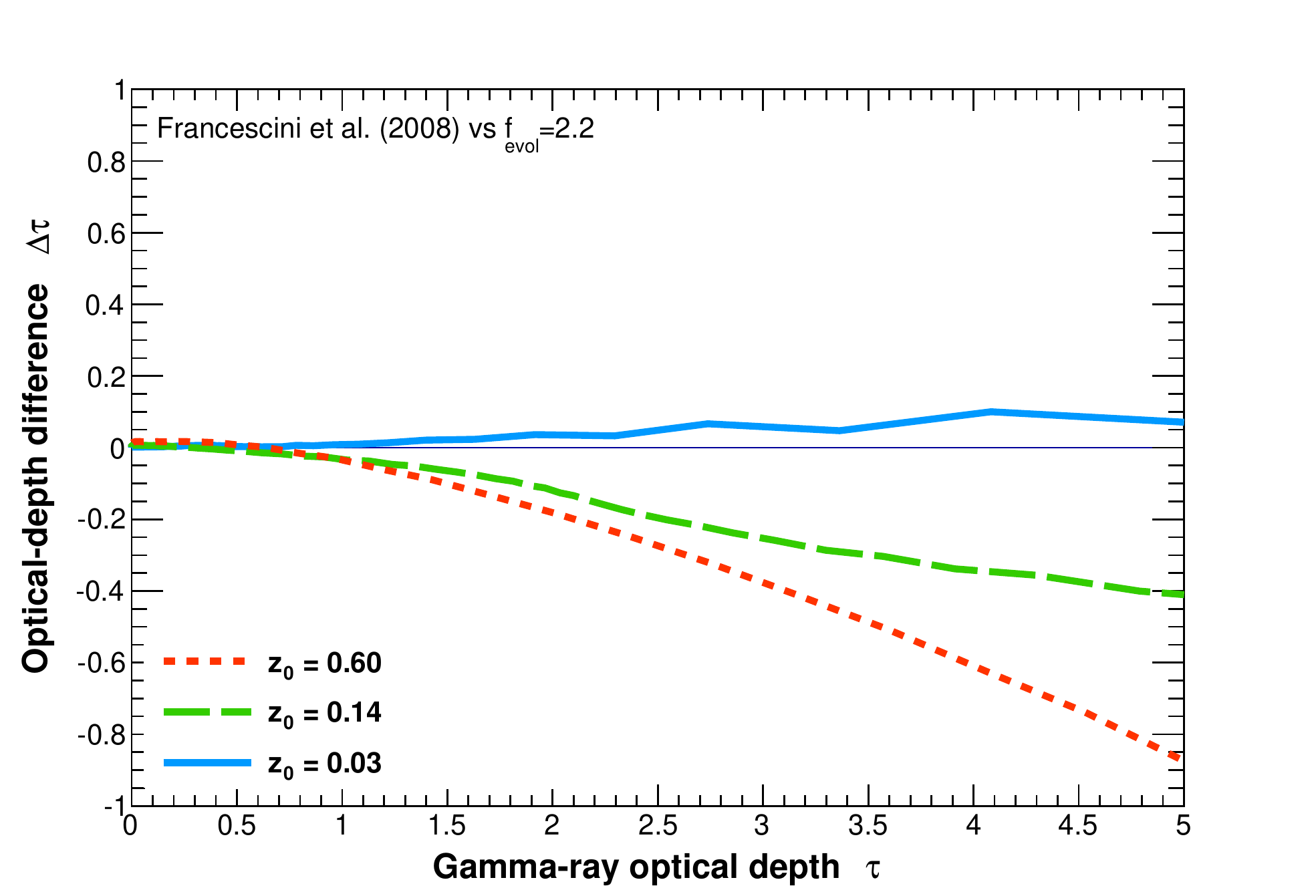}
\hspace{-0.2cm} \includegraphics[width=0.535\textwidth]{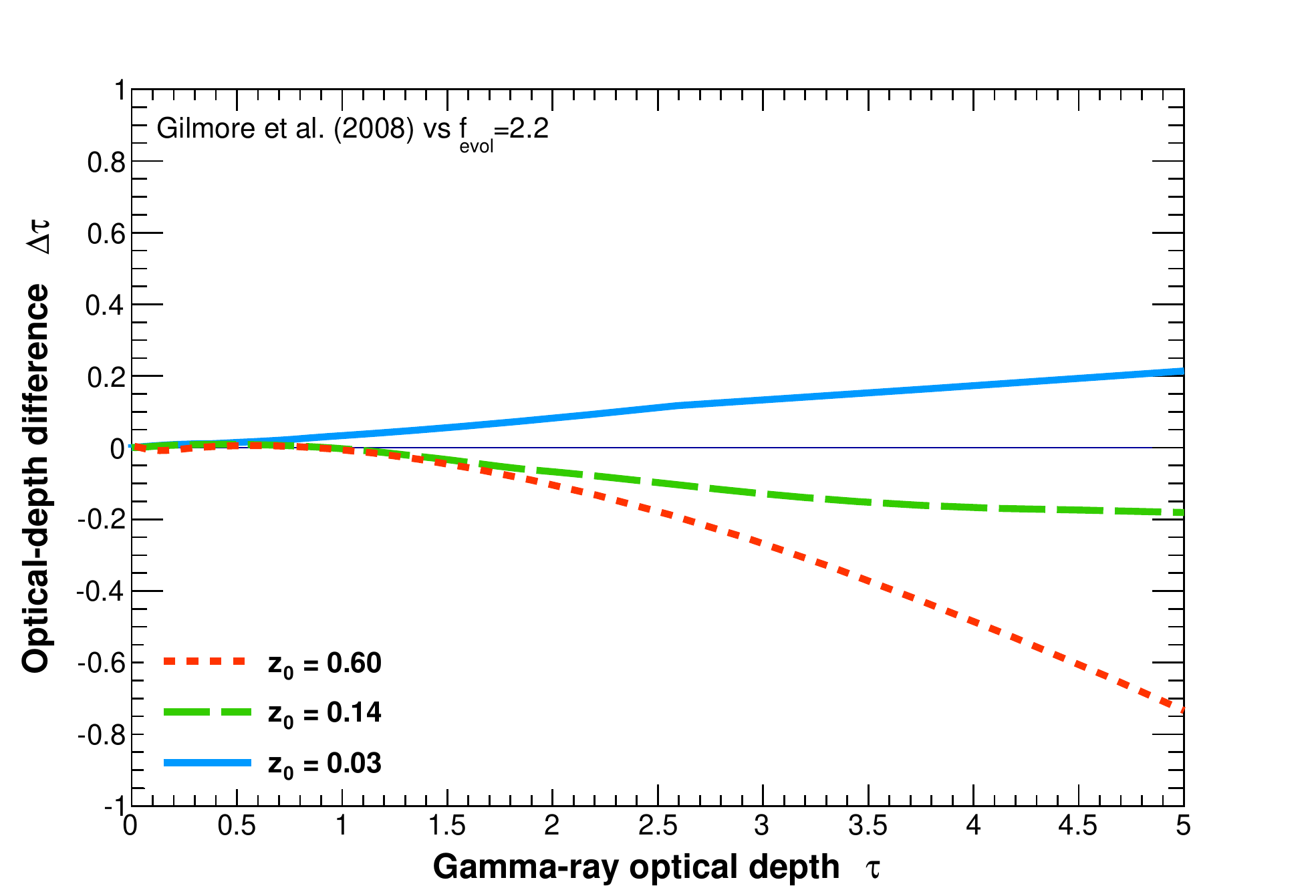}
\caption{{\it Left:} Difference between optical depths obtained with a template evolution and with the model published by \cite{FR08}. {\it Right:} Difference between optical depths obtained with a template evolution and with the model published by \cite{G12}. From {\it top} to {\it bottom}, the evolution parameter is $f_{\rm evol} = 1.2$, $f_{\rm evol} = 1.7$, and $f_{\rm evol} = 2.2$. The intermediate evolution with $f_{\rm evol} = 1.7$ is adopted.}
\label{fig:Evol}
\end{figure*}

We compare, in Fig.~\ref{fig:Tau17F}, gamma-ray optical depths published by \cite{FR08} and \cite{G12} with optical depths derived from Eq.~\ref{Eq:Opt3}, assuming the $z=0$ specific intensities as given in each paper with a template evolution with $f_{\rm evol} = 1.7$. The published attenuation curves and those derived from Eq.~\ref{Eq:Opt3} are in very good agreement, which supports the analytical approach in Sec.~\ref{Sec:Theory}. The differences between published and template optical depths, $\Delta\tau$, are shown in Fig.~\ref{fig:Evol}. The value of $f_{\rm evol} = 1.7$ is adopted (intermediate panels) and can be compared to softer and harder evolution in the bottom and top panels, respectively. We note that the evolution used by \cite{2008IJMPD..17.1515R}, $f_{\rm evol} = 1.2$, results in significant deviations at large redshifts with respect to the models. Similarly, the published optical depths are underestimated by the template approach at large redshifts for a soft evolution with $f_{\rm evol} = 2.2$.

The template evolution with $f_{\rm evol} = 1.7$ results in an optical depth difference on the order of $0.1$ on average, which is comparable with the difference in evolution between the models of \cite{FR08} and \cite{G12} themselves. As far as EBL evolution is concerned, assuming an energy-redshift decoupling in the local universe has then a similar impact on the absorption to using one or another state-of-the-art model. 

Below an optical depth of 3, the deviation in the absorption factor remains smaller than $15\%$, which is below the typical systematic errors on gamma-ray fluxes measured by current-generation ground-based instruments. Another reference point for the difference in optical depth can be provided noting that a $5\%$ deviation in $H_0$, roughly the difference between $H_0 = \unit[70]{km\ s^{-1}\ Mpc^{-1}}$ and the best-fit value from \cite{2014A&A...571A...1P}, results in a $5\%$ deviation in optical depth, which corresponds to $\Delta\tau=0.15$ ($15\%$ error on the absorption) for $\tau=3$. Thus the template approach that we use in this publication introduces errors in the EBL no larger than those resulting from the uncertainties on $H_0$ or from the differences between state-of-the-art models.

For reference, the integrations in Eq.~\ref{Eq:Opt3} and Eq.~\ref{Eq:PPKernel2}, which enable the computation of the optical depth, are performed using the trapezoidal rule, with uniforms steps in redshift of $\delta z = 10^{-3}$ and in logarithmic reduced photon energy, $\delta e_0 = 5\times10^{-3}$, from the threshold of the EBL kernel, $e_0 = -2\ln(1+z_0)$, up to $e_0 = 10^3$. We checked that above $e_0 > 10^3$ the tail of the kernel in Fig.~\ref{fig:Kernel} has a negligible contribution to the convolution product in Eq.~\ref{Eq:Opt3}. The overall uncertainty on the optical depth $\tau(E_0,z_0)$ due to the numerical integration is on the order of 0.01-0.03, mildly depending on the redshift of the gamma-ray source.

\subsection{Gaussian-sum approximation}\label{App:GaussSum}

\begin{figure*}
\centering
\includegraphics[width=0.535\textwidth]{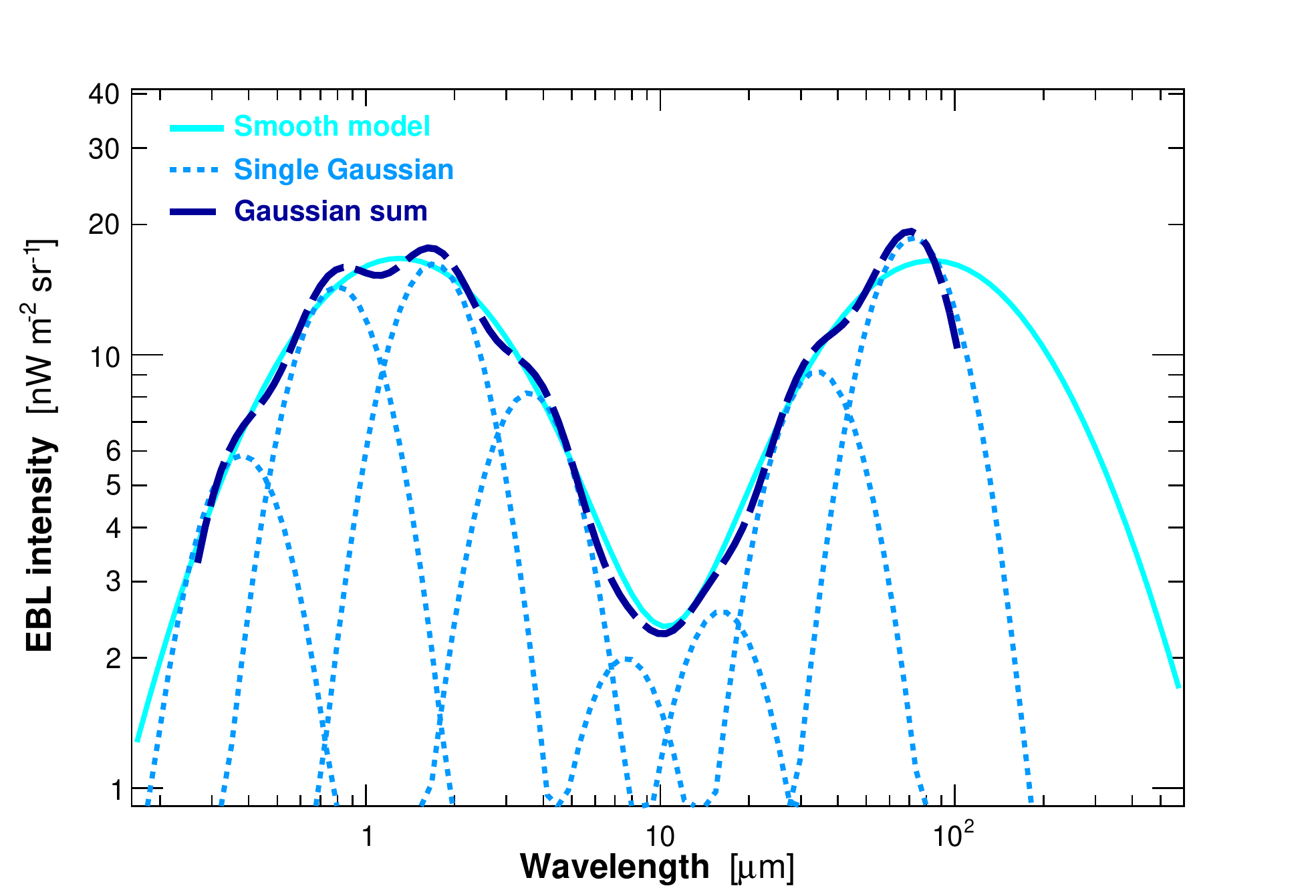}
\caption{Smooth EBL intensity (sum of two log parabolas, solid cyan curve) approximated between $\unit[0.27-105]{\mu m}$ by a sum of Gaussians of width $\Delta_l = 0.75$ (dotted light blue). The resulting intensity is shown as a dark-blue dashed line.}
\label{fig:GaussSum}
\end{figure*}

Besides the evolution of the EBL, our second source of systematic error comes from the approximation of the true spectrum of the EBL by a sum of Gaussians, as in Eq.~\ref{Eq:GaussianModel}. We show in Fig.~\ref{fig:GaussSum} the approximation of a ``smooth" model $\nu I_\nu(\lambda)$, namely a sum of two log parabolas, by a Gaussian sum for a binning $\Delta_l=0.75$, between $\unit[0.26-105]{\mu m}$, as in Sec.~\ref{Sec:DataAn} and \ref{Sec:Results}. The weights of the Gaussian-sum model are obtained by numerically inverting Eq.~\ref{Eq:MatrixRelation}, where we set $\nu I_\nu^i = \nu I_\nu(\lambda_i)$.

The EBL kernel in Eq.~\ref{Eq:Opt3} smooths the EBL density over a wide range of EBL wavelengths. The small deviations in intensity arising from the Gaussian-sum approximation then result in even milder optical-depth deviations, typically $\delta\tau\leq0.1$, $0.01$, and $10^{-3}$ for $\Delta_l = 1.0$, $0.5$, and $0.1$ respectively.

\subsection{Quantifying the systematic errors}\label{App:FulSys}

To estimate the systematic uncertainties arising from the modeling of the EBL, i.e. from the template evolution and the Gaussian-sum approximation, we compare the optical depths derived by \cite{FR08} and \cite{G12} to optical depths obtained with template evolution and approximating the \cite{FR08} or \cite{G12} SEDs at $z=0$ by Gaussian sums. 

We weight the contributions of the different optical depths based on the uncertainties on the spectral points included in the analysis. The measured optical depth depends on the measured flux, $\phi$, as $\ln \phi = \ln \phi_{\rm int} - \tau$, so that the maximum uncertainty on the optical depth, obtained by fixing the intrinsic model, scales as $\sigma_\phi/\phi$. Then, one can estimate the EBL normalization factor $\alpha$ that accounts for the change from the model, of optical depth $\tau_{\rm model}$, to the template approach, of optical depth $\tau_{\rm template}$, by minimizing:
\begin{equation}
\chi^2(\alpha) = \sum \frac{(\tau_{\rm template, i}-\alpha\times\tau_{\rm model, i})^2}{(\sigma_{\phi, i}/\phi_i)^2}
\end{equation}
which gives
\begin{equation}
\alpha = \frac{\sum \tau_{\rm model, i}\tau_{\rm template, i}\times (\phi_i/\sigma_{\phi, i})^2}{\sum \tau_{\rm model, i}^2\times (\phi_i/\sigma_{\phi, i})^2}
\label{Eq:SysEBL}
\end{equation}

Since the optical depth is proportional to the EBL intensity and inversely proportional to the Hubble constant, the systematic relative bias of the optical depth, $\alpha-1$, contributes equally the systematic uncertainty of the EBL intensity and Hubble constant. Using the gamma-ray cosmology sample, and assuming a binning of $\Delta_l = 0.75$, we find an average bias of $\unit[0.5]{\%}$ for the model of \cite{FR08} and $\unit[1.8]{\%}$ for the model of \cite{G12}.

\begin{deluxetable*}{lcc}
\tablecolumns{3} 
\tabletypesize{\scriptsize}
\tablecaption{Systematic uncertainties on the EBL specific intensity estimated with two different models.\label{Tab:Sys}}
\tablehead{\colhead{} & \colhead{\cite{FR08}} & \colhead{\cite{G12}}}
\startdata
EBL modeling & $\unit[0.5]{\%}$ & $\unit[1.8]{\%}$\\
Intrinsic model & $\unit[2.3]{\%}$ & $\unit[5.2]{\%}$\\
Energy scale & $\unit[6.2]{\%}$ & $\unit[6.0]{\%}$\\
 &  & \\
Total & $\unit[6.6]{\%}$ & $\unit[8.1]{\%}$
\enddata
\end{deluxetable*}

This source of systematic uncertainty is compared with the two other principal sources that have been identified: the gamma-ray energy scale and the choice of intrinsic model for each spectrum. The EBL scaling factor of the models of \cite{FR08} and \cite{G12} are estimated as in Sec.~\ref{Sec:EBLSpec}, additionally shifting the energy scale of the Cherenkov experiment within $\pm10\%$ on one hand, and using exclusively log-parabolas or exponential cut-off power laws for the intrinsic spectra on the other hand. A systematic energy bias of $\pm10\%$ is a conservative estimate with respect to the value half as large found in \cite{2010A&A...523A...2M} by studying the marginal mismatch between HE and VHE observations of the Crab nebula, but is comparable to the systematic error typically quoted by the observers. The energy scale impacts the EBL intensity at the $\sim\unit[6]{\%}$ level. Changing the gamma-ray spectral models impacts the EBL normalization at the $\unit[2-5]{\%}$ level. The total systematic uncertainty on the EBL flux level and on the Hubble constant are estimated in Table~\ref{Tab:Sys} by summing in quadrature the various sources of systematic errors to be about $\unit[7-8]{\%}$.

\begin{figure*}
\centering
\includegraphics[width=0.535\textwidth]{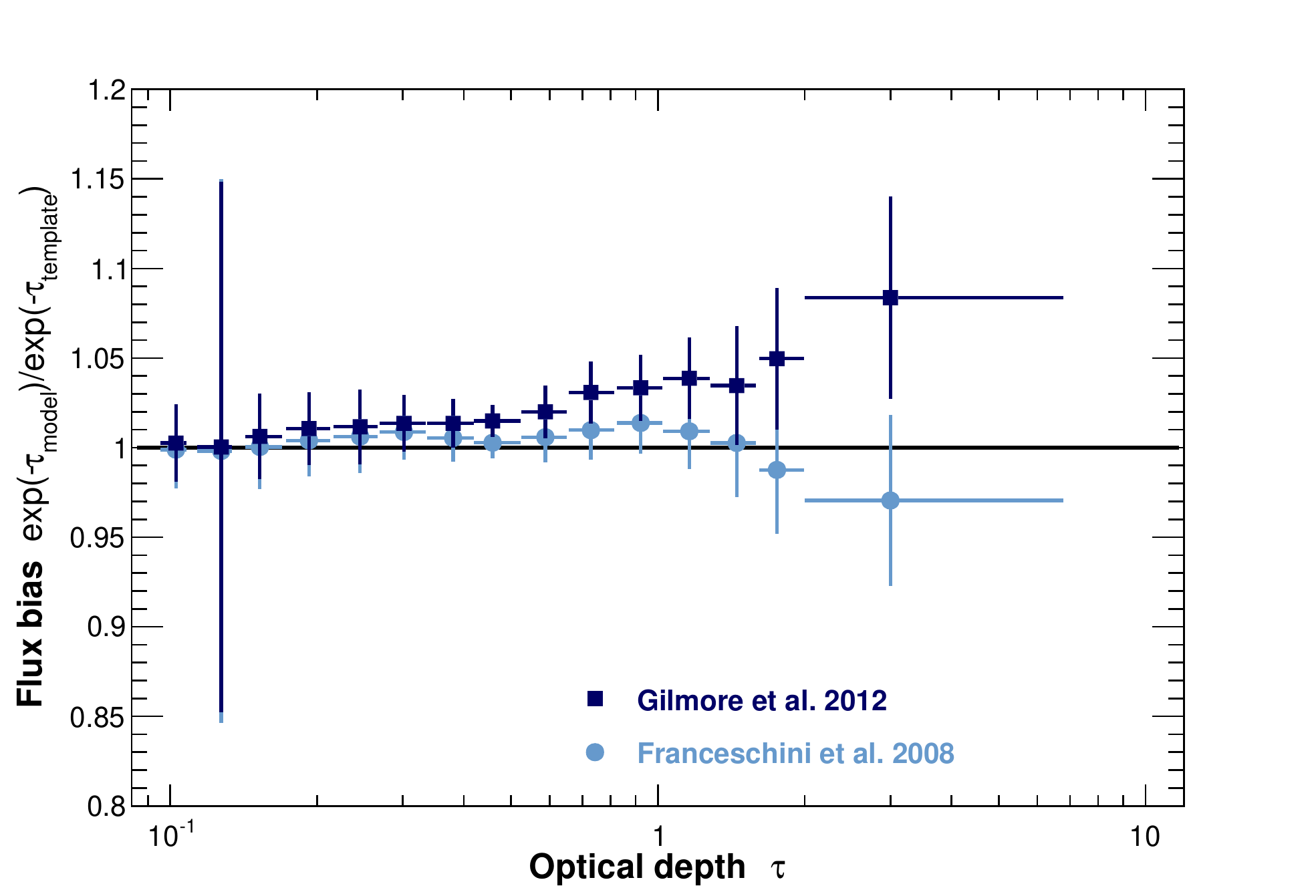}
\caption{Ratio of gamma-ray attenuation from the model of \cite{FR08} (light blue circles) and \cite{G12} (dark blue squares) to the value obtained with our templates, using the full sample of gamma-ray spectra.\vspace{1cm}}
\label{fig:bias}
\end{figure*}

We also estimate the systematic bias on the flux enhancement discussed in Sec.~\ref{Sec:ALP}. Using the full sample, we compute the average flux bias, $FB$, as the flux enhancement resulting from using the template approach instead of the model:
\begin{align}
FB &= \exp[(\alpha-1)\times\tau_{\rm model}]
\end{align}
where $\alpha$ is computed as in Eq.~\ref{Eq:SysEBL}. The flux bias is shown as a function of optical depth in Fig.~\ref{fig:bias}, for the full sample. Using the gamma-ray cosmology sample yields similar average values for each of the two models, though with a rather constant behavior across the whole optical depth range. An underestimation of the absorption of about $\unit[10]{\%}$ at optical depth $\tau>2$ is visible in Fig.~\ref{fig:bias} for the model of \cite{G12}, although the large uncertainties do not allow a firm conclusion. For the sake of argument, we consider $\unit[10]{\%}$ as the systematic uncertainty on the flux enhancement at large optical depths. Improved statistics will result in a more accurate estimate of this systematic uncertainty in the future.

\section{Best-fit parameters and covariance matrices}\label{App:BestFitAndCov}
In the following, we describe the best-fit EBL spectrum derived using gamma-ray data only (4 parameters) and gamma-ray data together with local constraints (8 parameters). The EBL intensities and associated uncertainties in each wavelength bin can be computed from Eq.~\ref{Eq:MatrixRelation}. The best-fit parameters of the gamma-ray data only are shown in Table~\ref{Tab:BF4}, with covariance matrix shown in Table~\ref{Tab:COV4}. The best-fit parameters and covariance matrix for the gamma-ray data and local constraints are shown in Tables~\ref{Tab:BF6} and \ref{Tab:COV6}. The optical depths derived with the eight-point spectrum between $\unit[50]{GeV}$ and  $\unit[20]{TeV}$ are shown in Tables~\ref{Tab:OptDepth1} and \ref{Tab:OptDepth2}. We limit the results to optical depths less than $5$, where we have good control of the systematic errors, as discussed in Appendix~\ref{App:Sys}.
\vfill
\begin{deluxetable*}{cc}
\tablecolumns{2} 
\tabletypesize{\scriptsize}
\tablecaption{Parameters of the best-fit EBL spectrum\\ using gamma-ray data only.\label{Tab:BF4}}
\tablehead{\colhead{$\lambda$} & \colhead{$a_i$}\\ \colhead{$\mu$m} & \colhead{\ nW m$^{-2}$ sr$^{-1}$} }
\startdata
$	0.55	$ & $	18.1	$ \\
$	2.47	$ & $	8.05	$ \\
$	11.1	$ & $	0.91	$ \\
$	49.7	$ & $	4.51	$ 
\enddata
\end{deluxetable*}
\newpage
\begin{deluxetable*}{cccc}
\tablecolumns{4} 
\tabletypesize{\scriptsize}
\tablecaption{Covariance matrix of the best-fit EBL parameters using gamma-ray data only.\label{Tab:COV4}}
\tablehead{\colhead{1} & \colhead{2} & \colhead{3} & \colhead{4}}
\startdata
$	18.66	$ & $	4.58	$ & $	1.62	$ & $	0.38$\\
	\nodata	 & $	1.83	$ & $	0.52	$ & $	0.17$\\
	\nodata&\nodata	& $	0.35	$ & $	0.04	$\\
	\nodata&\nodata&\nodata	 & $	0.21	$
\enddata
\end{deluxetable*}

\begin{deluxetable*}{cc}
\tablecolumns{2} 
\tabletypesize{\scriptsize}
\tablecaption{Parameters of the best-fit EBL spectrum using gamma-ray data and local constraints.\label{Tab:BF6}}
\tablehead{\colhead{$\lambda$} & \colhead{$a_i$}\\ \colhead{$\mu$m} & \colhead{\ nW m$^{-2}$ sr$^{-1}$} }
\startdata
$	0.38	$ & $	7.12	$ \\
$	0.80	$ & $	12.6	$ \\
$	1.70	$ & $	10.3	$ \\
$	3.60	$ & $	6.79	$ \\
$	7.62	$ & $	-0.61	$ \\
$	16.1	$ & $	4.56	$ \\
$	34.1	$ & $	1.30	$ \\
$	72.3	$ & $	11.9	$ 
\enddata
\end{deluxetable*}

\begin{deluxetable*}{cccccccc}
\tablecolumns{8} 
\tabletypesize{\scriptsize}
\tablecaption{Covariance matrix of the best-fit EBL parameters using gamma-ray data and local constraints.\label{Tab:COV6}}
\tablehead{\colhead{1} & \colhead{2} & \colhead{3} & \colhead{4} & \colhead{5} & \colhead{6} & \colhead{7} & \colhead{8}}
\startdata
$21.710	$ & $	-3.450	$ & $	0.659	$ & $	-0.001	$ & $	-0.054	$ & $	-0.007	$ & $	-0.006	$ & $	-0.020	$\\
\nodata	& $	6.153	$ & $	-0.435	$ & $	1.244	$ & $	0.076	$ & $	0.076	$ & $	-0.001	$ & $	0.019	$\\
\nodata&\nodata	& $	0.802	$ & $	-0.108	$ & $	0.232	$ & $	-0.004	$ & $	0.012	$ & $	0.031	$\\
\nodata&\nodata&\nodata	& $	0.827	$ & $	-0.069	$ & $	0.066	$ & $	-0.008	$ & $	0.004	$\\
\nodata&\nodata&\nodata&\nodata	& $	0.434	$ & $	-0.035	$ & $	0.032	$ & $	0.052	$\\
\nodata&\nodata&\nodata&\nodata&\nodata	& $	0.109	$ & $	-0.030	$ & $	0.051	$\\
\nodata&\nodata&\nodata&\nodata&\nodata&\nodata	& $	0.081	$ & $	-0.084	$\\
\nodata&\nodata&\nodata&\nodata&\nodata&\nodata&\nodata	& $	1.743	$
\enddata
\end{deluxetable*}

\begin{deluxetable*}{crrrrrrrrrrrrrrrr}
\tablecolumns{17} 
\tabletypesize{\scriptsize}
\tablecaption{Optical depth between $\unit[50]{GeV}$ and $\unit[20]{TeV}$ for sources at redshift between $0.01$ and $0.31$.\label{Tab:OptDepth1}}
\tablehead{\colhead{TeV / z}&\colhead{0.01}&\colhead{0.03}&\colhead{0.05}&\colhead{0.07}&\colhead{0.09}&\colhead{0.11}&\colhead{0.13}&\colhead{0.15}&\colhead{0.17}&\colhead{0.19}&\colhead{0.21}&\colhead{0.23}&\colhead{0.25}&\colhead{0.27}&\colhead{0.29}&\colhead{0.31} }
\startdata																																	
0.050&0.00&0.00&0.00&0.00&0.00&0.00&0.00&0.00&0.00&0.00&0.00&0.01&0.01&0.01&0.01&0.01	\\
0.061&0.00&0.00&0.00&0.00&0.00&0.00&0.00&0.01&0.01&0.01&0.01&0.01&0.02&0.02&0.02&0.03	\\
0.075&0.00&0.00&0.00&0.01&0.01&0.01&0.01&0.02&0.02&0.02&0.03&0.03&0.04&0.05&0.05&0.06	\\
0.091&0.00&0.00&0.01&0.01&0.02&0.02&0.03&0.04&0.04&0.05&0.06&0.07&0.08&0.09&0.10&0.12	\\
0.111&0.00&0.01&0.02&0.02&0.03&0.04&0.05&0.07&0.08&0.09&0.11&0.12&0.14&0.16&0.18&0.20	\\
0.136&0.01&0.02&0.03&0.04&0.06&0.07&0.09&0.11&0.13&0.15&0.18&0.20&0.23&0.26&0.29&0.32	\\
0.166&0.01&0.03&0.05&0.07&0.09&0.12&0.14&0.17&0.21&0.24&0.27&0.31&0.35&0.39&0.44&0.48	\\
0.202&0.01&0.04&0.07&0.11&0.14&0.18&0.22&0.26&0.31&0.36&0.41&0.46&0.51&0.57&0.63&0.69	\\
0.247&0.02&0.06&0.11&0.15&0.21&0.26&0.32&0.37&0.44&0.50&0.57&0.64&0.71&0.79&0.87&0.95	\\
0.302&0.03&0.09&0.15&0.21&0.28&0.35&0.43&0.51&0.59&0.68&0.76&0.86&0.95&1.05&1.15&1.25	\\
0.368&0.04&0.12&0.20&0.28&0.37&0.47&0.57&0.67&0.77&0.88&0.99&1.11&1.23&1.35&1.48&1.61	\\
0.450&0.05&0.15&0.26&0.37&0.48&0.60&0.72&0.85&0.98&1.11&1.25&1.40&1.54&1.69&1.84&2.00	\\
0.549&0.06&0.19&0.32&0.46&0.60&0.74&0.90&1.05&1.21&1.37&1.54&1.71&1.88&2.06&2.24&2.43	\\
0.671&0.08&0.23&0.39&0.56&0.73&0.90&1.08&1.27&1.46&1.65&1.85&2.05&2.25&2.46&2.67&2.88	\\
0.819&0.09&0.28&0.47&0.66&0.86&1.07&1.28&1.50&1.72&1.94&2.17&2.40&2.63&2.87&3.11&3.35	\\
1.00&0.11&0.32&0.55&0.77&1.01&1.24&1.48&1.73&1.98&2.23&2.48&2.73&2.99&3.25&3.51&3.77	\\
1.22&0.12&0.37&0.62&0.88&1.14&1.40&1.66&1.93&2.20&2.47&2.74&3.01&3.28&3.55&3.82&4.08	\\
1.49&0.13&0.41&0.68&0.96&1.23&1.51&1.79&2.07&2.35&2.63&2.90&3.18&3.45&3.72&4.00&4.27	\\
1.82&0.14&0.43&0.71&1.00&1.28&1.57&1.85&2.13&2.41&2.69&2.97&3.25&3.52&3.80&4.07&4.34	\\
2.22&0.14&0.43&0.72&1.01&1.30&1.58&1.87&2.15&2.43&2.72&3.00&3.29&3.57&3.86&4.15&4.44	\\
2.71&0.15&0.44&0.73&1.02&1.31&1.60&1.90&2.19&2.49&2.80&3.11&3.42&3.73&4.06&4.38&4.72	\\
3.31&0.15&0.45&0.76&1.07&1.38&1.70&2.03&2.36&2.70&3.05&3.41&3.77&4.14&4.52&4.91&\nodata	\\
4.05&0.17&0.50&0.85&1.20&1.56&1.93&2.32&2.71&3.11&3.52&3.95&4.38&4.82&\nodata&\nodata&\nodata	\\
4.94&0.19&0.59&0.99&1.41&1.84&2.29&2.74&3.21&3.69&4.18&4.68&\nodata&\nodata&\nodata&\nodata&\nodata	\\
6.03&0.23&0.70&1.18&1.68&2.19&2.71&3.25&3.80&4.36&4.94&\nodata&\nodata&\nodata&\nodata&\nodata&\nodata	\\
7.37&0.27&0.82&1.39&1.98&2.57&3.19&3.82&4.46&\nodata&\nodata&\nodata&\nodata&\nodata&\nodata&\nodata&\nodata	\\
9.00&0.32&0.97&1.63&2.32&3.03&3.77&4.53&\nodata&\nodata&\nodata&\nodata&\nodata&\nodata&\nodata&\nodata&\nodata	\\
11.0&0.38&1.17&2.00&2.86&3.77&4.73&\nodata&\nodata&\nodata&\nodata&\nodata&\nodata&\nodata&\nodata&\nodata&\nodata	\\
13.4&0.50&1.56&2.69&3.89&\nodata&\nodata&\nodata&\nodata&\nodata&\nodata&\nodata&\nodata&\nodata&\nodata&\nodata&\nodata	\\
16.4&0.73&2.29&3.98&\nodata&\nodata&\nodata&\nodata&\nodata&\nodata&\nodata&\nodata&\nodata&\nodata&\nodata&\nodata&\nodata	\\
20.0&1.11&3.46&\nodata&\nodata&\nodata&\nodata&\nodata&\nodata&\nodata&\nodata&\nodata&\nodata&\nodata&\nodata&\nodata&\nodata	
\enddata
\end{deluxetable*}

\begin{deluxetable*}{crrrrrrrrrrrrrrr}
\tablecolumns{16} 
\tabletypesize{\scriptsize}
\tablecaption{Optical depth between $\unit[50]{GeV}$ and $\unit[6]{TeV}$ for sources at redshift between $0.33$ and $0.61$.\label{Tab:OptDepth2}}
\tablehead{\colhead{TeV / z}&\colhead{0.33}&\colhead{0.35}&\colhead{0.37}&\colhead{0.39}&\colhead{0.41}&\colhead{0.43}&\colhead{0.45}&\colhead{0.47}&\colhead{0.49}&\colhead{0.51}&\colhead{0.53}&\colhead{0.55}&\colhead{0.57}&\colhead{0.59}&\colhead{0.61}}
\startdata
0.050&0.01&0.02&0.02&0.02&0.02&0.03&0.03&0.04&0.04&0.05&0.05&0.06&0.06&0.07&0.08\\
0.061&0.03&0.04&0.04&0.05&0.06&0.06&0.07&0.08&0.09&0.10&0.11&0.12&0.13&0.14&0.15\\
0.075&0.07&0.08&0.09&0.10&0.11&0.12&0.13&0.15&0.16&0.18&0.19&0.21&0.23&0.24&0.26\\
0.091&0.13&0.14&0.16&0.18&0.20&0.21&0.23&0.26&0.28&0.30&0.33&0.35&0.38&0.41&0.43\\
0.111&0.22&0.24&0.27&0.29&0.32&0.35&0.38&0.41&0.45&0.48&0.52&0.55&0.59&0.63&0.68\\
0.136&0.35&0.39&0.42&0.46&0.50&0.54&0.59&0.63&0.68&0.73&0.78&0.83&0.88&0.94&0.99\\
0.166&0.53&0.58&0.63&0.68&0.74&0.80&0.85&0.92&0.98&1.04&1.11&1.18&1.25&1.32&1.40\\
0.202&0.76&0.82&0.89&0.96&1.04&1.11&1.19&1.27&1.35&1.44&1.52&1.61&1.70&1.80&1.89\\
0.247&1.03&1.12&1.21&1.30&1.40&1.49&1.59&1.70&1.80&1.91&2.02&2.13&2.24&2.36&2.47\\
0.302&1.36&1.47&1.58&1.70&1.82&1.94&2.06&2.19&2.32&2.45&2.58&2.72&2.86&3.00&3.14\\
0.368&1.74&1.87&2.01&2.15&2.30&2.45&2.60&2.75&2.90&3.06&3.22&3.38&3.54&3.71&3.88\\
0.450&2.16&2.32&2.49&2.66&2.83&3.00&3.18&3.36&3.54&3.73&3.91&4.10&4.29&4.48&4.68\\
0.549&2.62&2.81&3.00&3.20&3.40&3.60&3.81&4.02&4.23&4.44&4.65&4.86&\nodata&\nodata&\nodata\\
0.671&3.10&3.32&3.55&3.77&4.00&4.22&4.45&4.69&4.92&\nodata&\nodata&\nodata&\nodata&\nodata&\nodata\\
0.819&3.59&3.83&4.08&4.32&4.57&4.82&\nodata&\nodata&\nodata&\nodata&\nodata&\nodata&\nodata&\nodata&\nodata\\
1.00&4.03&4.29&4.54&4.80&\nodata&\nodata&\nodata&\nodata&\nodata&\nodata&\nodata&\nodata&\nodata&\nodata&\nodata\\
1.22&4.35&4.62&4.88&\nodata&\nodata&\nodata&\nodata&\nodata&\nodata&\nodata&\nodata&\nodata&\nodata&\nodata&\nodata\\
1.49&4.53&4.80&\nodata&\nodata&\nodata&\nodata&\nodata&\nodata&\nodata&\nodata&\nodata&\nodata&\nodata&\nodata&\nodata\\
1.82&4.62&4.89&\nodata&\nodata&\nodata&\nodata&\nodata&\nodata&\nodata&\nodata&\nodata&\nodata&\nodata&\nodata&\nodata\\
2.22&4.73&\nodata&\nodata&\nodata&\nodata&\nodata&\nodata&\nodata&\nodata&\nodata&\nodata&\nodata&\nodata&\nodata&\nodata
\enddata
\end{deluxetable*}

\section{Individual spectra}\label{App:IndivSpectra}

The 106 VHE spectra studied in this paper are shown in Fig.~\ref{fig:spec12}-\ref{fig:spec7}. The publications from which these data are extracted can be found in Table~\ref{Tab:GammaData}. Observed VHE spectral points are shown in dark blue and the models best fitting these data are shown as a blue-gray solid line. The intrinsic spectral points are shown in light blue, after deabsorption with the best-fit EBL spectrum. The associated butterfly (one sigma envelope of the best-fit intrinsic model) is shown as a cyan contour. The butterflies associated with {\it Fermi}-LAT data are shown in dashed gray for non-contemporaneous HE and VHE observations, and in dashed black whenever the HE and VHE data are quasi-contemporaneous. The border of each spectral panel is colored as in Fig.~\ref{fig:Ez}, light blue indicating the gamma-ray cosmology sample, dark red the sources with under-constrained redshifts, and orange the FSRQs.

The contemporaneity criterion is rather loose, and the HE and VHE integration windows differ in duration, albeit having a temporal overlap. For this reason, some VHE spectra show a flux level slightly larger than their quasi-contemporaneous counterpart (2 of 33) while some show a somewhat smaller flux level (4 of 33). No conclusion can be drawn from these mild discrepancies given the astrophysical uncertainties in the mechanism responsible for the variability of blazars and the differences between the HE and VHE exposures.

The constraints from the hardness of the intrinsic spectra based on quasi-contemporaneous HE observations (see Eq.~\ref{Eq:IntPrior}) seem {\it  a posteriori} rather weak. Nine spectra show non-zero $\chi^2$ contributions, the most important one being obtained for {\tt Mkn501\_ARGO-YBJ\_flare2011}, with $\chi^2_{\rm HE-VHE}=3.3\times 10^{-2}$. The remaining eight show $\chi^2$ values smaller than $10^
{-2}$.

\begin{figure*}
\centering
\includegraphics[width=0.535\textwidth]{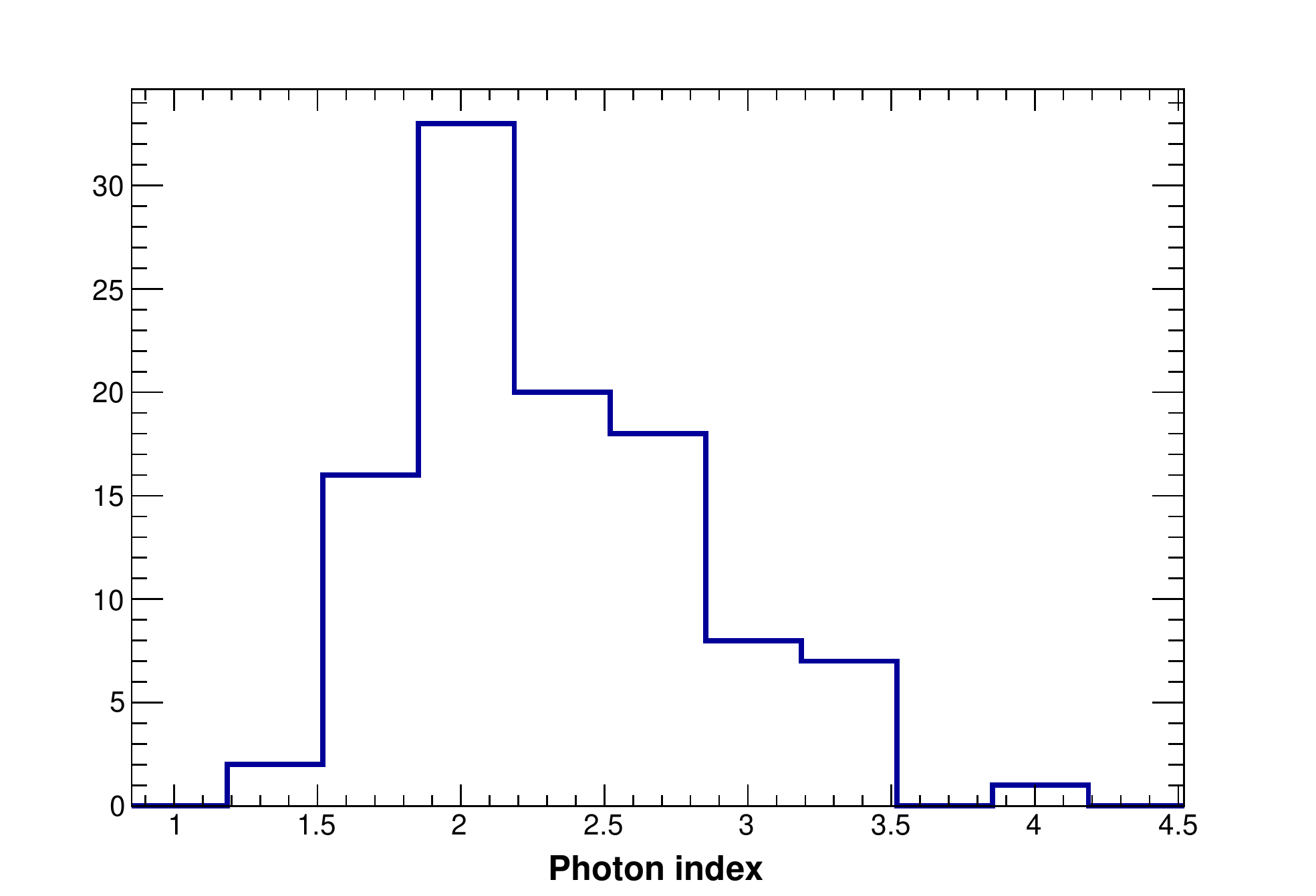}
\caption{Distribution of the intrinsic photon indices.}
\label{fig:distribindex}
\end{figure*}

\begin{figure*}
\hspace{-0.2cm} \includegraphics[width=\textwidth]{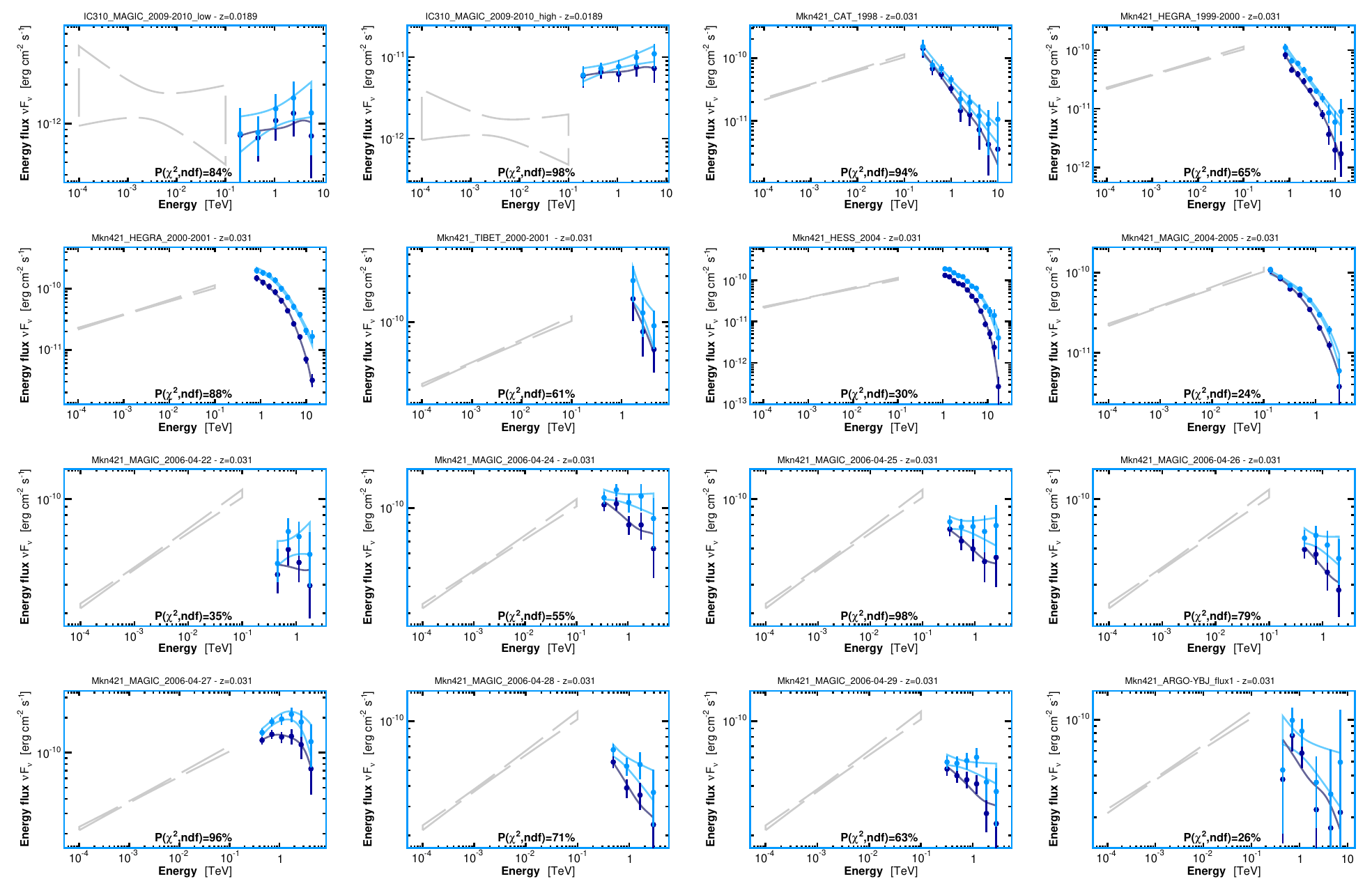}
\caption{Best-fit spectra 1 of 4. See text for more details.}
\label{fig:spec12}
\end{figure*}

\begin{figure*}
\hspace{-0.2cm} \includegraphics[width=\textwidth]{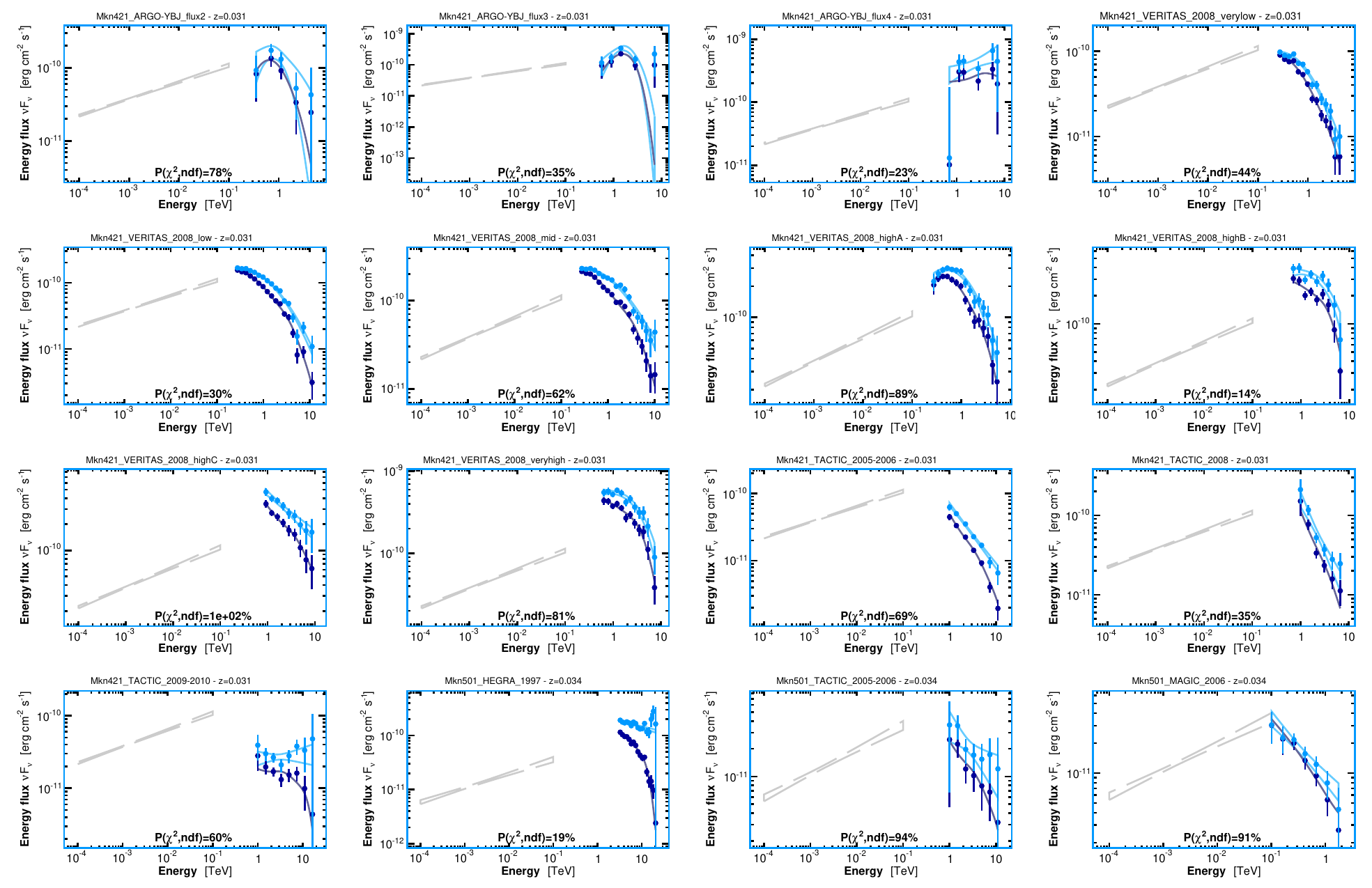}
\hspace{-0.2cm} \includegraphics[width=\textwidth]{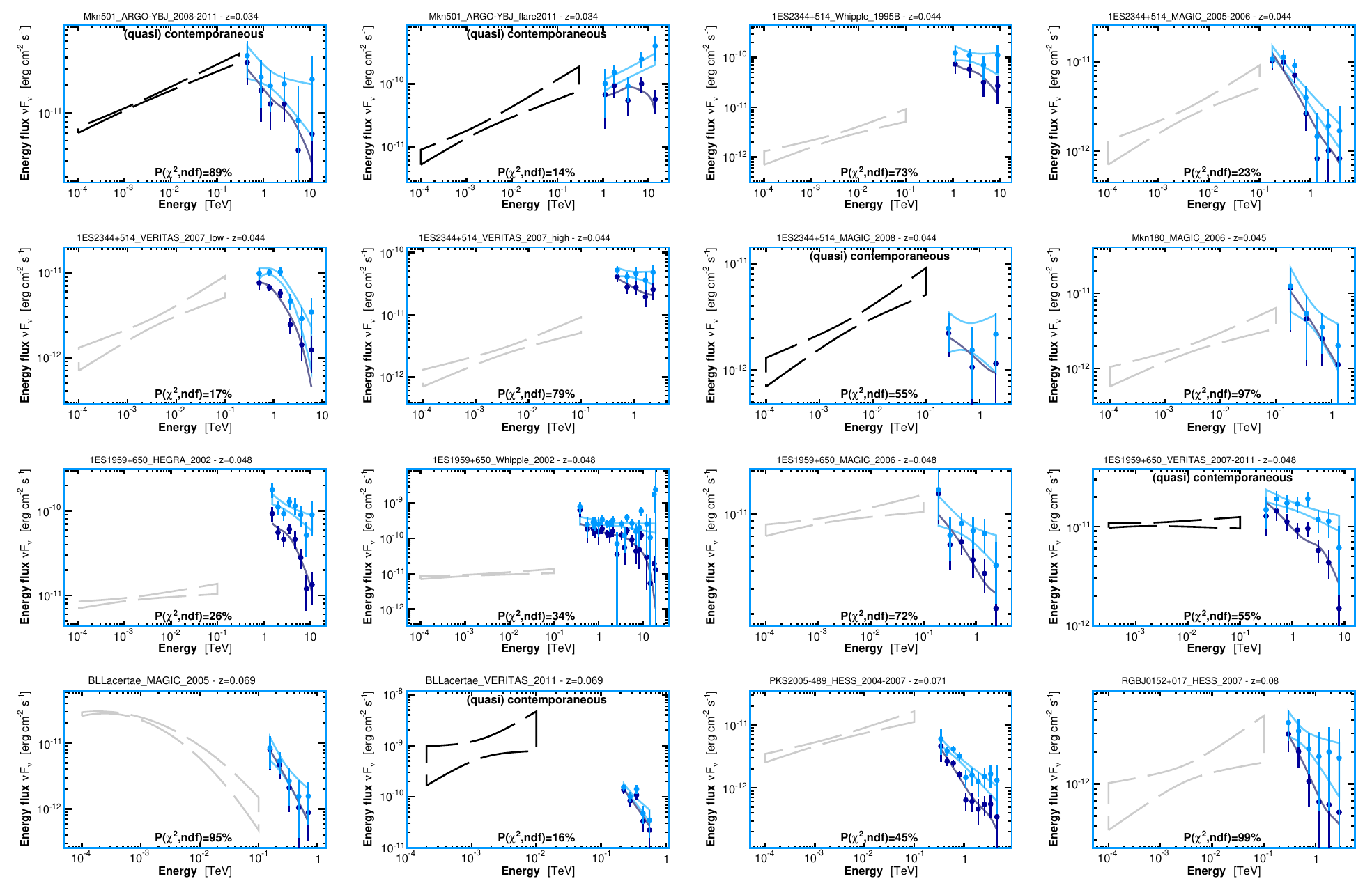}
\caption{Best-fit spectra 2 of 4. See text for more details.}
\label{fig:spec34}
\end{figure*}

\begin{figure*}
\hspace{-0.2cm} \includegraphics[width=\textwidth]{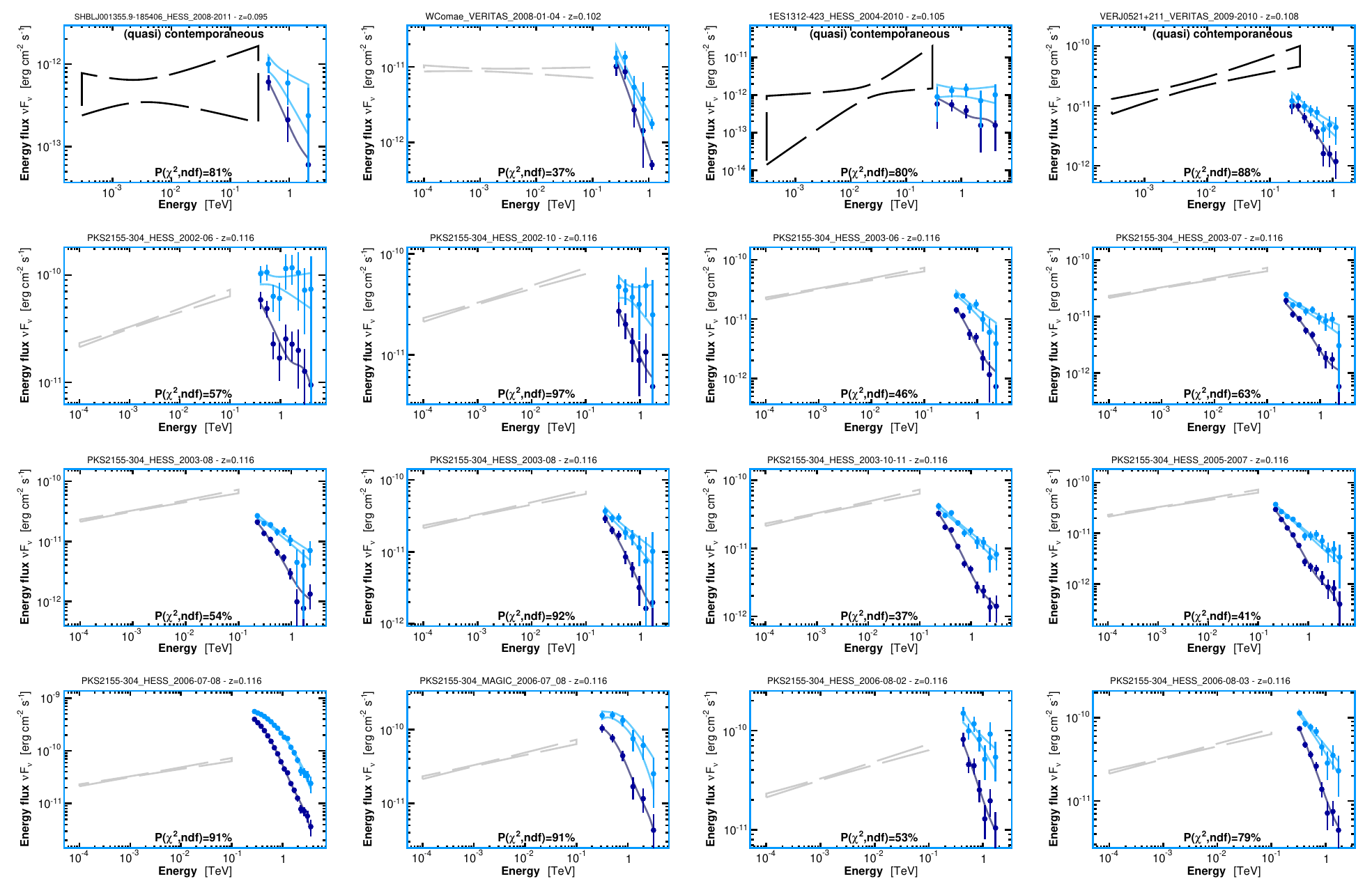}
\hspace{-0.2cm} \includegraphics[width=\textwidth]{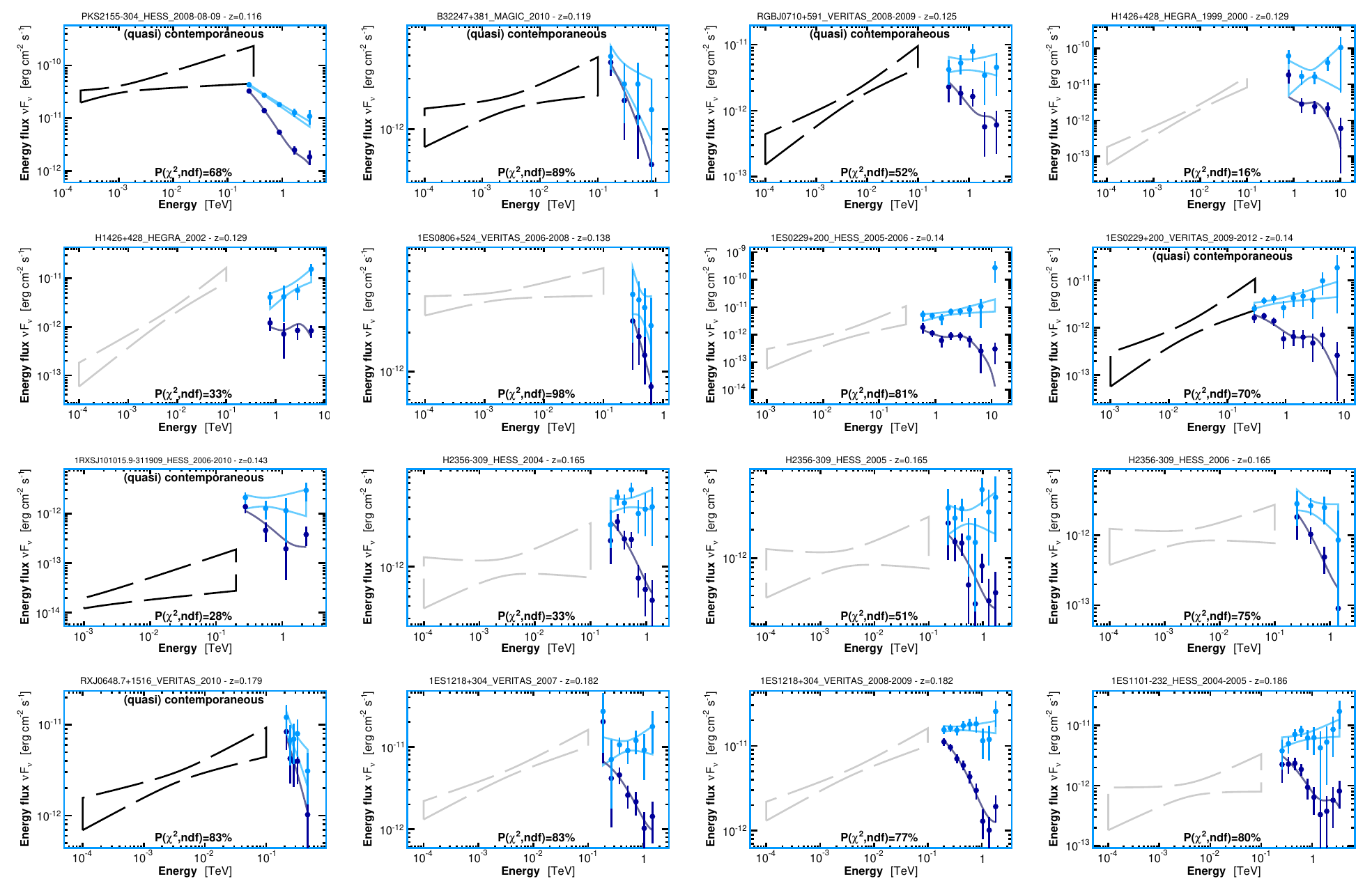}
\caption{Best-fit spectra 3 of 4. See text for more details.}
\label{fig:spec56}
\end{figure*}

\begin{figure*}
\hspace{-0.2cm} \includegraphics[width=\textwidth]{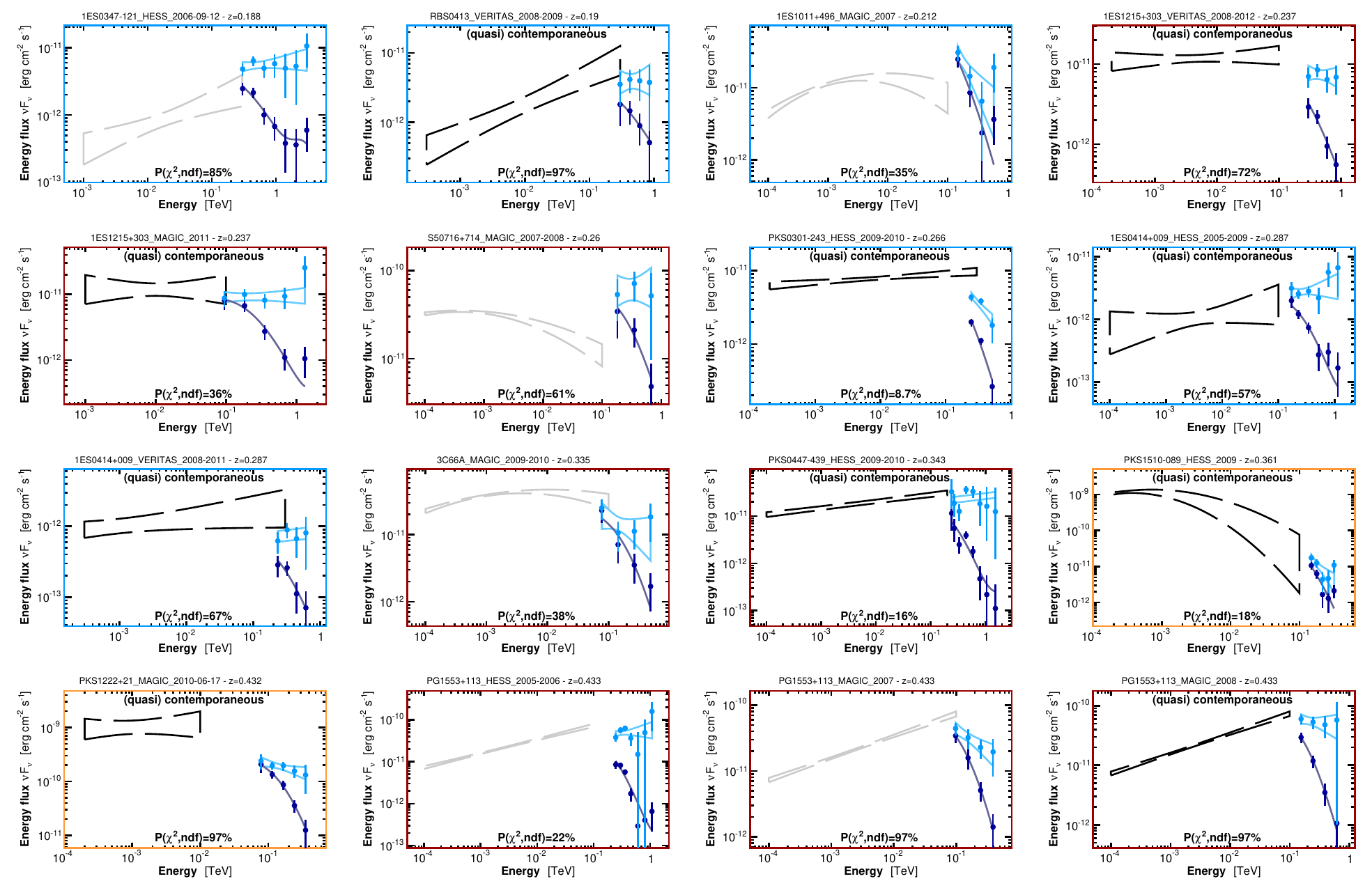}
\hspace{-0.2cm} \includegraphics[width=\textwidth]{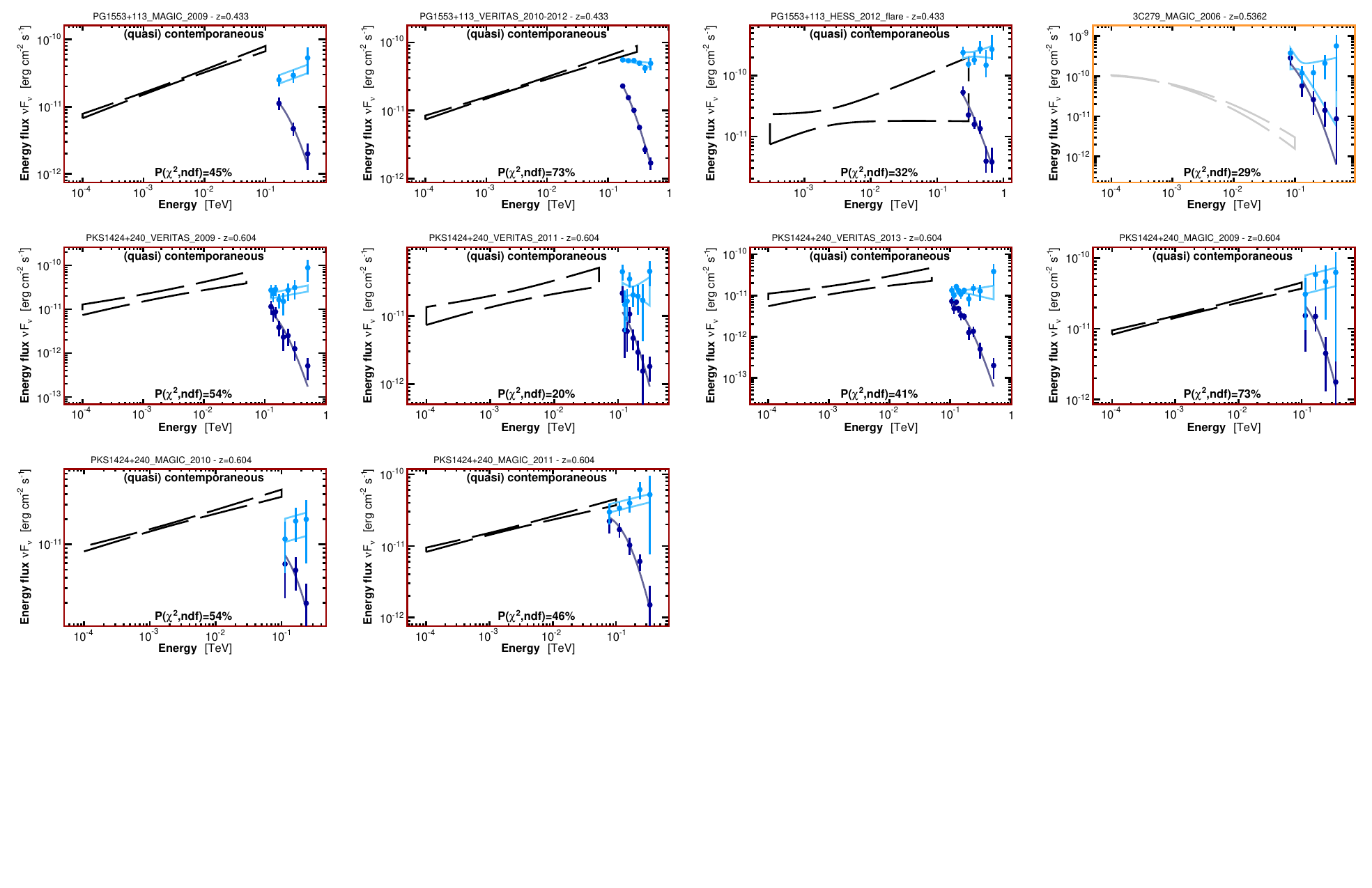}
\caption{Best-fit spectra 4 of 4. See text for more details.}
\label{fig:spec7}
\end{figure*}

The distribution of the intrinsic photon indices is shown in Fig.~\ref{fig:distribindex}. For the spectra described by log parabolas and exponential cut-off power laws, the indices are computed at the decorrelation energy. The minimum indices are $\Gamma = 1.35\pm0.24$ and $\Gamma = 1.37\pm0.30$, obtained for {\tt Mkn421\_MAGIC\_2006-04-27} and {\tt H1426+428\_HEGRA\_2002} respectively, indicating that all of these spectra are compatible with the maximum hardness, $\Gamma>1.5$, expected within a standard synchrotron self-Compton model for an electron index of $2$.  

All the spectra show good fits, with 103 of them having $\chi^2$ probabilities larger than $\unit[15]{\%}$. The three poorest fits are {\tt PKS0301-243\_HESS\_2009-2010}, {\tt Mkn501\_ARGO-YBJ\_flare2011}, and {\tt Mkn421\_VERITAS\_2008\_highB} with $\chi^2$ probabilities of $\unit[9]{\%}$, $\unit[14]{\%}$, and $\unit[14]{\%}$ respectively.

Finally, we would like to stress a natural bias arising from the visual scanning of these spectra. It appears that some VHE spectra show important upward going fluctuations with respect to their best-fit model, see e.g. the last spectral points of {\tt Mkn421\_ARGO-YBJ\_flux3} and {\tt 1ES0229+200\_HESS\_2005-2006}. These cases are particularly striking to the eye with flux enhancements of $\sim1600$ and $\sim23$, respectively. In practice, such deviations are of rather small amplitude when normalized to the uncertainty on the flux, with deviations of only $\unit[1.2]{\sigma}$ and $\unit[1.3]{\sigma}$ for these two examples. The logarithmic scale of the plot is of course responsible for the visual bias that artificially amplifies upward going fluctuations.

\end{document}